\documentclass[tosem, manuscript]{acmart}
\AtBeginDocument{%
  }

\setcopyright{acmlicensed}

\copyrightyear{2025}
\acmYear{2025}
\acmJournal{TOSEM}
\acmVolume{XX}
\acmNumber{X}
\acmArticle{XXX}
\acmDOI{XXXXXXX.XXXXXXX}
\usepackage{longtable}

\usepackage{svg}
\usepackage{subcaption} 
\usepackage{booktabs}
\usepackage{multirow}
\usepackage{makecell} \usepackage{xcolor}
\usepackage[none]{hyphenat}
\usepackage[most]{tcolorbox}
\usepackage{array}
\newtcolorbox{myframe}[1][]{
  enhanced,
  arc=0pt,
  outer arc=0pt,
  colback=white,
  boxrule=0.8pt,
  #1
}
\newcommand{\sectopic}[1]{\vspace{0.2em}\par\noindent{\textit{\bfseries #1}}}
\acmISBN{978-1-4503-XXXX-X/2018/06}

\begin{document}

\title{Monitoring Machine Learning Systems: A Multivocal Literature Review}

\author{Hira Naveed}
\email{hira.naveed@monash.edu}
\orcid{1234-5678-9012}
\affiliation{%
  \institution{Monash University}
  \country{Australia}
}

\author{Scott Barnett}
\affiliation{%
  \institution{Deakin University}
  \country{Australia}}
  \email{scott.barnett@deakin.edu.au}

\author{Chetan Arora}
\orcid{0000-0003-1466-7386}
\affiliation{%
  \institution{Monash University}
  \country{Australia}
}
  \email{chetan.arora@monash.edu}

\author{John Grundy}
\orcid{0000-0003-4928-7076}
\affiliation{%
 \institution{Monash University}
 \country{Australia}}
  \email{john.grundy@monash.edu}

\author{Hourieh Khalajzadeh}
\affiliation{%
  \institution{Deakin University}
  \country{Australia}}
  \email{hkhalajzadeh@deakin.edu.au}

\author{Omar Haggag}
\orcid{0000-0003-2346-3131}
\affiliation{%
  \institution{Monash University}
  \country{Australia}}
\email{omar.haggag@monash.edu}

\renewcommand{\shortauthors}{Naveed et al.}

\begin{abstract}
  \textbf{Context:} Dynamic production environments make it challenging to maintain reliable machine learning (ML) systems. Runtime issues, such as changes in data patterns or operating contexts, that degrade model performance are a common occurrence in production settings. Monitoring enables early detection and mitigation of these runtime issues, helping maintain users' trust and prevent unwanted consequences for organizations.   
  \textbf{Aim:} This study aims to provide a comprehensive overview of the ML monitoring literature. 
  \textbf{Method:} We conducted a multivocal literature review (MLR) following the well-established guidelines by Garousi to investigate various aspects of ML monitoring approaches in 136 papers.
  \textbf{Results:} We analyzed selected studies based on four key areas: (1) the motivations, goals, and context; (2) the monitored aspects, specific techniques, metrics, and tools; (3) the contributions and benefits; and (4) the current limitations. We also discuss several insights found in the studies, their implications, and recommendations for future research and practice. \textbf{Conclusion:} Our MLR identifies and summarizes ML monitoring practices and gaps, emphasizing similarities and disconnects between formal and gray literature. Our study is valuable for both academics and practitioners, as it helps select appropriate solutions, highlights limitations in current approaches, and provides future directions for research and tool development. 
\end{abstract}



\received{20 February 2007}
\received[revised]{12 March 2009}
\received[accepted]{5 June 2009}

\maketitle
\section{Introduction}
The ability of machine learning (ML) to learn from previous data and solve complex problems without being explicitly programmed has been valuable for numerous use cases~\cite{khomh2018software, zhang2003machine, awoyemi2017credit, khanal2020systematic, carvalho2019systematic, adamopoulou2020chatbots}. 
According to ML statistics from 2024~\cite{saasworthy2024mlstats}, nearly 83\% of global organizations rely on ML in some capacity.
However, for organizations to achieve long-term benefits from ML systems, they need to be deployed and maintained in a production environment, handling inference requests at scale from real users. This is challenging since production environments are dynamic and ML systems are sensitive to change~\cite{ovadia2019can}. Shifts in data patterns and novel situations that were not part of the training set are a common occurrence in production~\cite{klein2021mlops}. Even subtle changes in the operating context can lead to unexpected and unwanted behavior, such as reduced accuracy or delayed inference~\cite{8804457, klein2021mlops}. Despite quality assurance activities prior to deployment, the non-deterministic nature of ML makes it impossible to exhaustively assess all possible scenarios~\cite{grieser2020assuring, martinez2021developing}. In this situation, monitoring becomes necessary. Monitoring involves continuously observing an ML system in production by collecting and analyzing relevant data to ensure operational stability and prediction quality~\cite{schroder2022monitoring, eken2024multivocal}. This proactive approach helps identify and mitigate runtime issues in a timely manner, thereby reducing technical debt, maintaining users' trust, and preventing organizations from financial and reputational damage~\cite{schroder2022monitoring, breck2017ml}. 

Given the importance of monitoring for ensuring system reliability~\cite{breck2017ml}, several approaches and tools exist in the literature~\cite{hardt2021amazon, lewis2022augur, azeem2024monitizer}. This includes peer-reviewed, formal literature and unpublished gray literature such as blogs, white papers, and website articles.  Existing reviews on monitoring ML systems are not comprehensive, and most of them only examine peer-reviewed literature~\cite{protschky2025gets, andersen2024monitoring, jain2023survey, rahman2021run, ferreira2024safety, schroder2022monitoring, karval2023catching, pattan2020survey}. This lack of industrial perspectives and comprehensive reviews limits both researchers and practitioners from fully understanding current practices, ongoing challenges, and potential future directions. With the growing volume of work on monitoring ML systems, especially gray literature, overlooking these sources hampers advances that come from solving pressing industry needs.

To address this gap, a systematic review combining both formal and gray literature is essential to provide a holistic overview of monitoring approaches for ML systems. The unified insights from this review have several benefits: 1) they provide a more comprehensive view of monitoring ML systems than either source can offer alone; 2) the findings are more diverse and generalizable, as they draw on a wider range of contexts and evidence; 3) the findings are more practically relevant, as they incorporate ongoing practices from real world deployments; and 4) the trends and gaps in academia and industry become evident. This review can guide future researchers in aligning their work to bridge existing gaps. It can assist practitioners in adopting appropriate solutions, given the ongoing challenges in the practical implementation of monitoring ML systems\cite{shergadwala2022human, klein2021mlops}. Further, it can also provide valuable guidance for tool developers to improve the design and effectiveness of monitoring tools for ML systems.


In this paper, we present a Multivocal Literature Review (MLR) on monitoring ML systems following the guidelines of Garousi et al.~\cite{garousi2019guidelines}. The MLR provides a holistic overview of the ML monitoring landscape, including both formal and gray literature. Our goal was to identify the motivations and context of monitoring ML systems, including the application domain and ML techniques. Further, we explored the various monitored aspects, monitoring approaches, metrics, and tools. We also analyzed the contributions and benefits of the selected studies and discussed current limitations and their implications. Our analysis reveals evident differences between formal and gray literature in motivations, goals, monitoring focus, and contributions. Technical aspects, such as model performance and drift, receive significantly greater attention and benefit from more mature monitoring solutions than responsible ML aspects. Challenges faced by practitioners are often overlooked, and automated and customizable monitoring solutions are scarce. Overall, the key aspects monitored in ML systems are data, model behavior, operations and infrastructure, and responsible use of ML. 

Our findings present a collective summary of formal and gray literature that can help practitioners and academics efficiently identify current trends and limitations of monitoring approaches for ML systems. We also outline several directions for future work to advance ML monitoring. We believe this MLR will be valuable for understanding the ML monitoring landscape, selecting appropriate solutions, and identifying opportunities for future research and practice.

The main contributions of this MLR include:
\begin{itemize}
    \item identification, analysis, data extraction, and synthesis of 136 studies (100 formal and 36 gray literature) highly relevant to monitoring ML systems;
    \item a comprehensive overview of current trends in ML monitoring, including a categorization of monitored aspects, corresponding approaches, metrics, and popular tools;
    \item A comparison of formal and gray literature on monitoring ML systems, identifying alignments and disconnects;
    \item An analysis of  key limitations in ML monitoring research and practice work to date; and
    \item A set of recommendations for future research and practice.
\end{itemize}

The rest of the paper is structured as follows: Section 2 provides an
overview of the background and related work on monitoring ML systems. Section 3 presents our research methodology. Section 4 reports our findings from the selected studies. Section 5 describes the
threats to validity and mitigation strategies. Section 6 details the recommendations for future work. Finally, Section 7 concludes the paper.

\section{Background and Related Work}
In this section, we describe the key background concepts for this review and the related work.
\subsection{Machine Learning Systems}
Machine Learning (ML) is a branch of Artificial Intelligence (AI) that leverages data and models to make decisions. An ML model is a mathematical function that has the capability to learn patterns from data during training and then make predictions on unseen data during inference~\cite{mueller2021machine}.
Similar to traditional software systems, which consist of multiple components to perform specific functions~\cite{ieee1990ieee}, ML systems also comprise several components, one of which is an ML model~\cite{schroder2022monitoring}. 

\subsubsection{Machine Learning Techniques} Depending on the use case, an ML system can use models that are based on different learning techniques~\cite{mueller2021machine}. These techniques are the methods through which an ML model learns to map inputs to outputs using data. ML techniques are divided into three broad categories: supervised learning, unsupervised learning, and reinforcement learning.

In \textit{supervised learning}, a model learns from labeled data, i.e., input data with the corresponding correct outputs~\cite{mueller2021machine, lee2020machine}. For instance, a house price prediction model trained on houses' features (inputs) and their sale prices (outputs). The two main types of supervised learning are classification and regression~\cite{mueller2021machine}. Classification aims to categorize data into distinct classes (e.g., spam vs. not spam), whereas regression is used to predict continuous numerical values (e.g., house prices). Some popular supervised learning algorithms are logistic regression, linear regression, decision tree, random forest, support vector machines (SVM), and artificial neural networks (ANNs)~\cite{lee2020machine}. Supervised learning is widely used in fraud detection, disease diagnosis, and recommender systems~\cite{mueller2021machine}.

In \textit{unsupervised learning}, a model learns from unlabeled data by identifying patterns and grouping similar data~\cite{mueller2021machine, lee2020machine}. For example, an anomaly detection model learns the patterns of normal data and then detects deviations, such as outliers, that do not conform to these patterns. Common unsupervised learning algorithms are K-means clustering, principal component analysis, and isolation forest~\cite{lee2020machine}. Use cases for unsupervised learning include customer segmentation and anomaly detection. While supervised learning works well when labels are available, unsupervised learning is beneficial in scenarios when labels are not available. A hybrid of these two techniques is \textit{semi-supervised learning}, in which the model learns from both labeled and unlabeled data~\cite{van2020survey}. This hybrid learning approach is particularly valuable when there is a large volume of unlabeled data and manual labeling is expensive or time-consuming~\cite{van2020survey, lee2020machine}. Applications of semi-supervised learning include sentiment analysis, image classification, and audio transcription~\cite{van2020survey}.

In \textit{reinforcement learning}, the model learns through trial and error by receiving rewards or penalties, to guide the behavior towards an optimal outcome~\cite{mueller2021machine,lee2020machine}. Reinforcement learning does not require any labeled data for training; instead, the model learns to maximize reward through experience in a simulated or real-world environment~\cite{mueller2021machine}. For example, a reinforcement learning agent learns to play a video game by interacting with the game environment and improving its strategy over time. Well-known algorithms for reinforcement learning are Q-learning, policy gradient, and temporal difference learning~\cite{lee2020machine}. Common use cases include robotics, financial bots, and game agents~\cite{lee2020machine}.

\textit{Deep learning} (DL) is a widely adopted subset of ML that leverages neural networks (NNs) with multiple layers to learn complex data patterns~\cite{lee2020machine}. For example, an image classification model can learn to recognize objects like cats or cars by processing thousands of labeled images, automatically detecting shapes and textures through its layered architecture. DL is a cross-cutting approach that can be applied to all three major ML techniques: supervised, unsupervised, and reinforcement learning~\cite{lee2020machine}. Compared to traditional ML models, DL models are more complex, require larger datasets, and often achieve higher prediction accuracy~\cite{lee2020machine}. Prominent DL architectures include Deep Neural Networks (DNNs), Convolutional Neural Networks (CNNs), and Recurrent Neural Networks (RNNs)~\cite{dong2021survey}. DL is often used in medical imaging, facial recognition, and autonomous vehicles~\cite{dong2021survey}.

\subsubsection{Software Engineering for ML}
Software engineering for ML systems (SE4ML) is different from traditional software engineering~\cite{amershi2019software}. SE4ML consists of two main phases: \textit{development} and \textit{operations}. The \textit{development} phase begins by gathering requirements, formulating the problem to be solved, and defining the goals. Once the goal is clear, relevant data is collected and prepared by cleaning and transforming it into a suitable format. Then, ML model development is carried out by running experiments, selecting an appropriate model, training it, and testing its performance~\cite{amershi2019software}. The \textit{operations} phase comprises deployment, monitoring, and maintenance. Deployment involves making the trained model available to end-users by integrating it into a production environment and setting up the necessary infrastructure and APIs. The production environment is sometimes also referred to as \textit{runtime}. The deployed ML system is then monitored to detect any issues in the production environment and alert engineers~\cite{amershi2019software}. Finally, maintenance is performed to address these issues and ensure optimal performance.

Since ML systems do not need explicit programming, they are increasingly being adopted by businesses to solve complex problems where traditional algorithms are ineffective~\cite{geron2022hands}. According to a 2023 survey~\cite{rackspace2023report}, organizations reported that ML increased innovation by 17\% and employee efficiency by 20\%, while reducing expenses by 16\%. The surge in ML systems can be seen across all domains, from social media and e-commerce to critical domains like engineering, finance, healthcare, and autonomous vehicles. 

\subsection{Runtime Issues in Machine Learning Systems}
Runtime issues refer to unwanted behaviors that occur when ML models are deployed in production. A common example is a drop in model accuracy, which can arise due to unexpected changes in the production environment or unintentional mistakes during development.

One major runtime issue is \textit{model performance decay}, a gradual decline in how well the model performs over time~\cite{lewis2022augur}. This decay often results from \textit{data drift} (also called covariate drift), where the distribution of input data shifts away from the training data~\cite{moreno2012unifying, lewis2022augur}. In addition, changes in the relationship between inputs and predictions, known as \textit{concept drift}, can further degrade performance~\cite{moreno2012unifying}. Similarly, shifts in model predictions themselves, referred to as \textit{label drift} or prior probability shift, can also impact accuracy~\cite{lipton2018detecting, moreno2012unifying, zamani2021machine}. Label drift rarely occurs in isolation and is often accompanied by other distributional changes. Since ground truth labels are often unavailable at runtime, data drift and label drift are used as proxies to estimate model performance~\cite{chen2022estimating}.
In addition to distributional changes, \textit{data quality issues} such as incorrect formats, missing values, or corrupted data can lead to errors in predictions and degrade model effectiveness~\cite{ehrlinger2019daql}. These issues may arise from missing constraints on input data or bugs in the ML pipeline, the automated process that handles everything from data collection and preparation to training, evaluation, deployment, and monitoring~\cite{shankar2022towards}. Failures within this pipeline can also cause runtime problems~\cite{shankar2022towards}.

Another major cause of runtime issues is \textit{out-of-distribution} (OOD) data. ML models tend to perform well on data within their training data distribution but not on data significantly OOD~\cite{song2023deeplens}. OOD data includes novel inputs, anomalies, random or corrupted inputs (noise), drifted data, and malicious inputs (adversarial attacks)~\cite{ferreira2021benchmarking}. In addition to purely technical issues, runtime issues can also have a human-centric aspect. For example, \textit{safety violations}, which occur when an ML system does not adhere to safety constraints and operates in a risky and harmful manner~\cite{ferreira2024safety}. This is especially dangerous in critical domains such as healthcare and autonomous vehicles. Another concern is \textit{fairness violations}, which occur when the model predictions are unfair for certain groups based on sensitive attributes such as gender, age, or race ~\cite{ghosh2022faircanary}. \textit{Privacy} is also critical in ML systems; data leakage and model theft can cause serious violations at runtime~\cite{ko2023privmon}.

\subsection{Monitoring Machine Learning Systems}
\begin{figure}
    \centering
    \includegraphics[width=0.4\linewidth]{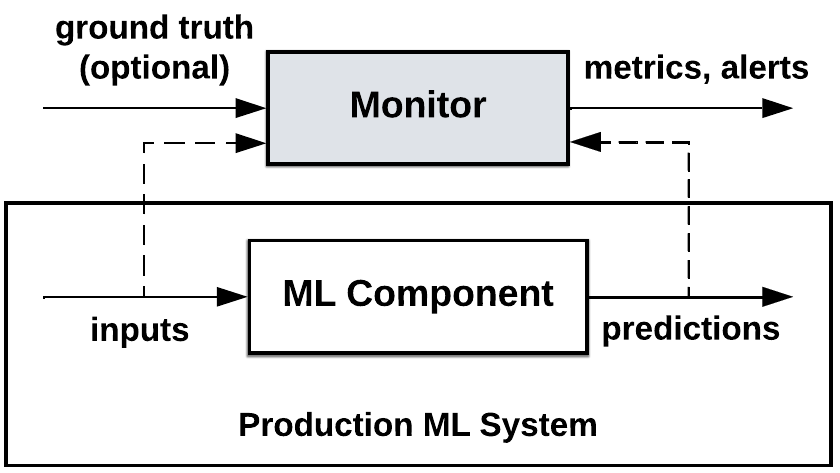 }
    \caption{ML Monitoring Architecture}
    \label{fig:monitoring architecture}
\end{figure}

Monitoring is a key step in the operations phase, focused on continuously observing the behavior of production ML systems to detect unwanted behavior~\cite{karval2023catching}. It involves instrumenting the system's source code, collecting relevant data, and analyzing it to detect runtime issues. Figure \ref{fig:monitoring architecture} shows a basic ML monitoring architecture. The monitor captures input data and predictions of an ML component and calculates metrics to detect deviations from expected behavior.

Due to the non-deterministic and brittle nature of ML, even subtle changes in the data or operating environment can significantly affect the performance of an ML system~\cite{breck2017ml, karval2023catching}. Quality assurance activities like testing have limitations and cannot assess all possible scenarios. Monitoring can be viewed as an extension of testing that evaluates the system at runtime and ensures desired outcomes.~\cite{breck2017ml} describes monitoring as a primary consideration for ensuring the production readiness of an ML system, improving reliability, reducing technical debt, and decreasing maintenance costs. 

The monitoring approach can be manual, semi-automated, or fully automated. \textit{Manual monitoring} involves engineers periodically inspecting system operations or relying on user feedback~\cite{jayalath2022enhancing}. In contrast, \textit{semi-automated monitoring} combines automated steps with human involvement, i.e., human-in-the-loop~\cite{ginart2022mldemon}. On the other hand, \textit{fully automated monitoring} operates independently, detecting issues and raising alerts without any human intervention~\cite{kourouklidis2023domain}.

Monitoring enables early detection and mitigation of runtime issues in ML systems, such as data drift, performance degradation, delayed predictions, fairness violations, safety violations, etc~\cite{schroder2022monitoring}. Which aspect to monitor depends on the use case, domain, and context of the ML system. A monitor can detect one or more aspects by calculating relevant metrics using the collected data. Larger ML systems may have multiple monitors for different aspects of the system.

\subsection{Related Work} \label{RW}
Several secondary studies have been conducted to summarize the ML monitoring literature. We found systematic literature reviews (SLRs), a systematic mapping study (SMS), a MLR, a scoping review, and various survey papers. 

Schroder et al. conducted an SLR~\cite{schroder2022monitoring} to collect and categorize the current challenges and methods for monitoring production ML systems. The authors proposed a taxonomy of ML monitoring methods with three categories (data, models, and non-technical), and listed real-world incidents corresponding to each category. Another SLR~\cite{karval2023catching} summarized the existing monitoring and explainability methods for ML, monitoring methods were categorized under data drift, outlier detection, and adversarial detection. Leest et al. present an SMS~\cite{leest2025tea} of contextual information in ML monitoring. They classify contextual information in framework with three dimensions - system, aspect, and representation - and outline common patterns in ML monitoring context. A MLR combined with an interview study was conducted by~\cite{protschky2025gets} to identify the characteristics of production ML systems and the steps for successfully monitoring and mitigating production issues. The authors classify these steps as define, measure, assess, act, and control. A scoping review~\cite{andersen2024monitoring} was performed on monitoring methods for AI in healthcare, the focus was on performance monitoring methods and rationales for selecting certain monitoring strategies. The authors highlight a scarcity of evidence and practical guidance for implementing performance monitoring of AI in healthcare settings. 

Pattan et al. performed a survey~\cite{pattan2020survey} and collected information about the factors that affect the behavior of production ML models, monitoring challenges, monitoring metrics, methods, and tools. Jain et al.~\cite{jain2023survey} conducted a survey on the importance of model monitoring and the techniques and tools for monitoring Natural Language Processing (NLP) applications. Rahman et al.~\cite{rahman2021run} surveyed ML monitoring for robotic perception (a subset of computer vision) to identify trends and approaches to monitor the performance and reliability of perception systems. The authors mapped monitoring approaches to stages in the perception pipeline and categorized them based on past experiences, inconsistencies during inference, and uncertainty estimation and confidence. Another survey on safety monitoring for ML perception functions was presented in~\cite{ferreira2024safety}. The authors provided a comprehensive review of the potential threats in perception systems, requirements elicitation for safety monitors, safety monitoring strategies, techniques to recover from failure, and methods to evaluate monitors.

A common property of existing studies~\cite{andersen2024monitoring, jain2023survey, rahman2021run, ferreira2024safety} is their limited scope; they tend to focus on specific domains such as healthcare, NLP, or perception systems. Other studies~\cite{schroder2022monitoring, karval2023catching, pattan2020survey, leest2025tea,protschky2025gets} with a broader scope focus only on specific aspects such as monitoring context, steps, methods, tools, or challenges, and do not provide a comprehensive review of the existing literature. \textbf{None of these studies focused on the goals, ecosystem of tools, benefits, and limitations of ML monitoring solutions while also incorporating gray literature. The only study that considered gray literature focused narrowly on monitoring steps, leaving these broader dimensions unexplored.}

\section{Research Methodology} 
We conducted a Multivocal Literature Review (MLR) on monitoring approaches for ML systems. An MLR is a type of Systematic Literature Review (SLR) that includes gray literature such as websites, blogs, and unpublished academic articles in addition to formal literature such as conference and journal papers. This review was motivated by the growing interest of practitioners and academics, as evident by the surge in ML monitoring tools and publications over the past five years. This review aims to provide a holistic and systematic overview of ML monitoring literature to guide future research and practice. This review has been conducted according to the guidelines for multivocal literature reviews in software engineering by Garousi et al.~\cite{garousi2019guidelines}.

\begin{figure}
    \centering
    \includegraphics[width=1.0\linewidth]{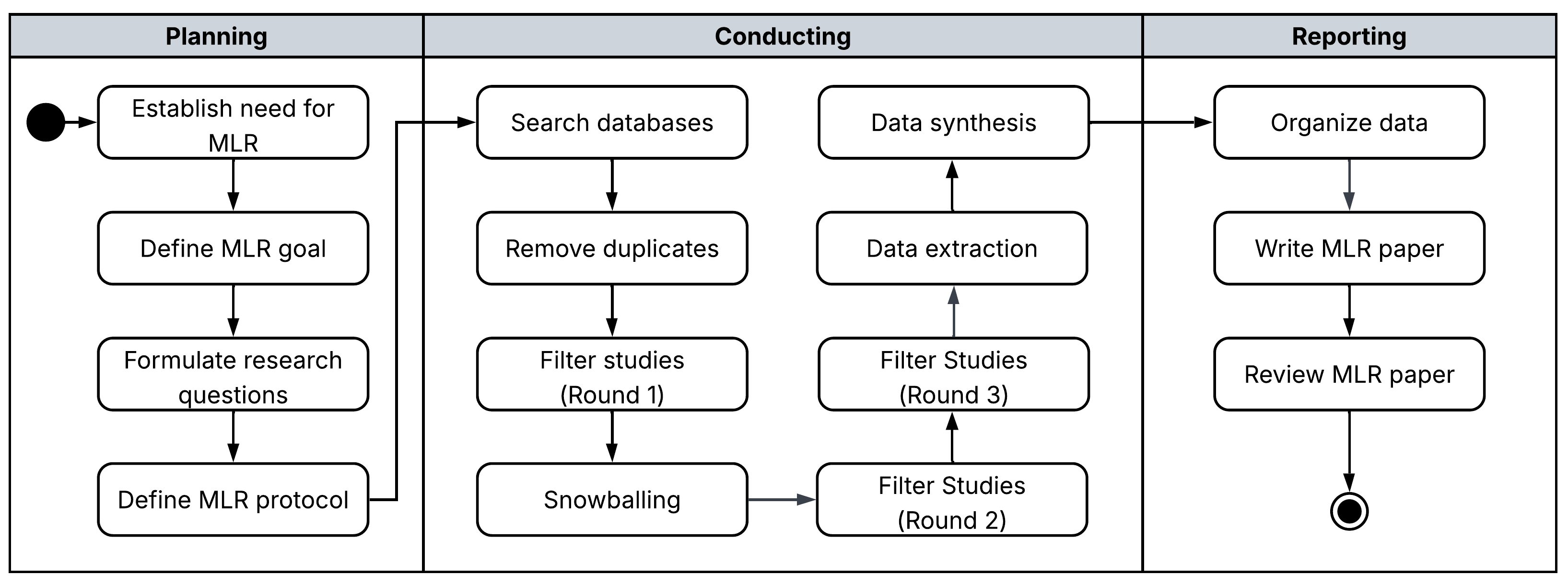}
    \caption{Multivocal Literature Review Process}
    \label{fig:MLR process}
\end{figure}

Fig. \ref{fig:MLR process} shows an overview of our review process consisting of three stages; planning, conducting, and reporting. The \textit{planning stage} began by establishing a need for the MLR, followed by defining the goal, formulating the research questions, and defining the MLR protocol, according to~\cite{garousi2019guidelines}. During the \textit{conducting stage}, we searched six databases and created an initial pool of studies by filtering the results based on predefined criteria. We performed snowballing on the initial pool to find additional relevant studies that may have been missed in the database search. Two more rounds of filtering were performed, which resulted in a final pool of 136 studies. Next, we extracted and synthesized data for the 136 selected studies, listed in Appendix \ref{StudiesList}. In the \textit{reporting stage}, we organized the findings in tables and figures, drafted the MLR paper, and refined the final manuscript.

\subsection{Motivation}
Monitoring ML systems is challenging~\cite{bernardi2019150, yao2017complexity}, but crucial for businesses to ensure that these systems are working as expected~\cite{bosch2021engineering}. This continuous oversight helps build stakeholder trust and protects organizations from potential reputational and financial harm~\cite{schroder2022monitoring}. While organizations are increasingly recognizing this, many still struggle with monitoring~\cite{bosch2021engineering}. Engineers responsible for ML monitoring need to carefully evaluate several factors before implementing a solution~\cite{shergadwala2022human}. These include identifying the specific aspects of the system that need to be monitored, such as data quality, prediction quality, or response time. They must also determine a suitable monitoring approach, metrics, and tools. Tool selection is based on the features offered, infrastructure requirements, and cost. Lastly, the benefits and limitations of each approach need to be weighed out before making a selection. Due to the growing interest of academics and practitioners, there is a wide range of ML monitoring solutions available, yet finding the appropriate one is still challenging for engineers~\cite{shergadwala2022human}.

We found a lack of a systematic and holistic overview of the ML monitoring literature. Existing SLRs and surveys have limitations as discussed in section \ref{RW}. To structure existing knowledge and provide clarity to practitioners and academics, a comprehensive multivocal literature review on monitoring ML systems was needed. We believe that a collective summary and comparison of formal and gray literature would be useful for understanding the ML monitoring landscape, selecting appropriate solutions, and identifying gaps for future research or tool development.

\subsection{Research Questions}
We defined three research questions to explore the goals, solutions, contributions, and importance of existing ML monitoring approaches.

\textbf{RQ1. What are the motivations and goals of monitoring approaches for ML systems?} RQ1 explored the motivations and goals for monitoring ML systems, and further examined the application domain(s) of the monitored techniques.

\textbf{RQ2. What monitoring solutions have been developed for ML systems?} RQ2 explored the monitoring techniques and metrics proposed for ML systems, along with the tools used, ground truth dependencies and logging capabilities of ML monitoring solutions.

\textbf{RQ3. What are the benefits of the proposed monitoring solutions for ML systems?} RQ3 examined the contributions, study type, and benefits of the ML monitoring solutions.

\textbf{RQ4. What are the limitations of the proposed monitoring solutions for ML systems?} RQ4 identified the limitations of ML monitoring solutions.

\subsection{Study Selection}

\begin{table}
    \caption{Inclusion and Exclusion Criteria}
    \centering
    \small
    \renewcommand{\arraystretch}{1.2}
    \begin{tabular}{p{2cm}p{12cm}}
     \hline
     \textbf{Criteria ID} & \textbf{Criterion} \\
     \hline
      I01   & Studies that focus on monitoring machine learning systems
      \begin{enumerate}
          \item Monitoring solutions for machine learning systems
          \item Monitoring guidelines for machine learning systems
          \item Monitoring challenges in machine learning systems
      \end{enumerate}\\
      I02   & Studies with full text accessible\\
      I03 & Studies in English language \\
    \hline
    \hline
      
      E01   & Studies that do not focus on monitoring machine learning systems
      \begin{enumerate}
          \item Studies about monitoring software systems that do not have machine learning components
          \item  Studies about testing and evaluation of machine learning systems
          \item Studies about MLOps with minimal focus on monitoring 
          \item Studies that only provide technical details of drift or out of distribution detection methods
          \item Studies about adapting machine learning systems to mitigate runtime issues
      \end{enumerate}\\
  
        E02 & Studies shorter than 4 pages\\
        E03 & Secondary and tertiary studies\\
        E04 & Papers with inadequate information to extract\\
        E05 & Personal blogs, videos, and presentations\\
        
        E06 & Studies with an extended version available \\
        E07 &  Gray literature sources with terms that restrict research use or competitor comparisons without permission\\
        E08 & Studies on monitoring foundation models such as large language models  \\
    \hline
    \end{tabular}
     \renewcommand{\arraystretch}{1}
    \label{tab:selection criteria}
\end{table}

The search process for systematic reviews begins with an initial pool of studies found from searching databases with a predefined search string. This initial pool often has several irrelevant studies that need to be filtered out before progressing to the next stage. We developed comprehensive inclusion and exclusion criteria as part of our MLR protocol (shown in Table \ref{tab:selection criteria}) that we used to filter studies. For this review, we selected studies that were in English language, had full text accessible, and focused on monitoring ML systems. We excluded studies that did not focus on ML monitoring, such as studies about general software monitoring, ML testing, MLOps studies that have insufficient monitoring details, and studies on updating ML systems to mitigate runtime issues. During the search, we found many studies that only presented a new drift or out-of-distribution (OOD) detection method. These studies were excluded from our review because they lacked integrations with the broader ML monitoring process. Additionally, we excluded studies with insufficient information to answer our research questions, that were shorter than four pages, were secondary or tertiary reviews, or had an extended version (e.g., a conference paper with a corresponding journal version or gray literature with a peer-reviewed counterpart). For gray literature, we checked the terms and conditions of websites and excluded any with copyright restrictions. Personal blogs, videos, and presentations were also excluded from this review since their inclusion would require ethics approval from our university. To limit the scope of this review, we did not include studies about monitoring large language models (LLMs) or any other type of foundation models.

\subsection{Search Process}
\begin{figure}[ht]
    \centering
    \includegraphics[width=0.9\linewidth]{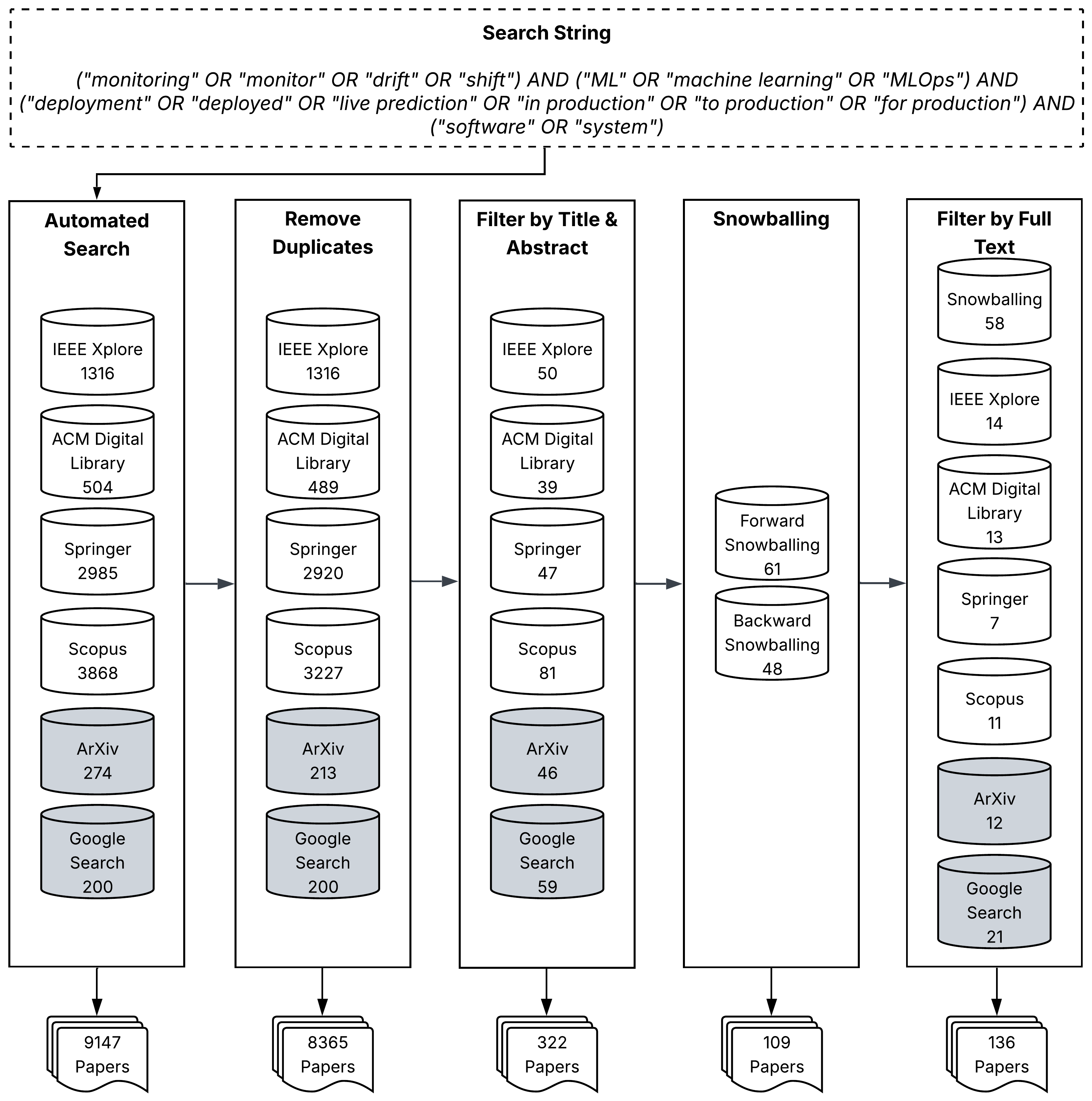}
    \caption{Search Process}
    \label{fig:search process}
\end{figure}

Fig. \ref{fig:search process} provides an overview of the search process followed in this MLR. 

For automated search in online databases, we formulated a search string with monitoring and ML-related keywords. The search string was tested on multiple databases and refined to balance relevance with a feasible number of search results. The final search string is shown in Fig.~\ref{fig:search process}.
The search was conducted on six online databases, four for formal literature (IEEE Xplore, ACM Digital Library, Springer, and Scopus) and two for gray literature (ArXiv and Google Search Engine). Since the results from the Google Search Engine were too many to evaluate manually, our stopping criteria were effort-bound. We relied on Google's page rank algorithm and selected the top 200 results as recommended~\cite{garousi2019guidelines}. 

The search resulted in an initial pool of 9147 papers from all six databases. The results were exported from the databases to Google Sheets and duplicates were removed based on matching titles. After duplicate removal, we were left with 8365 papers for further filtering. The first round of filtering was done by checking the title and abstract of each paper against the predefined inclusion and exclusion criteria (\ref{tab:selection criteria}). The first round of filtering yielded 322 potentially relevant papers. This significant reduction was because the database search was applied to full texts, which retrieved several papers with similar keywords but irrelevant for this review (Exclusion criteria 1).

Automated database search is not sufficient to gather all relevant papers for a review, so to complement the automated search results, we conducted snowballing, a manual search process following the guidelines of Wohlin~\cite{wohlin2014guidelines}. Backward snowballing (checking references) and forward snowballing (checking citations) were performed for all 322 papers from the previous round. We performed three iterations of snowballing until no new papers were found. 109 new potentially relevant papers were discovered through all iterations of snowballing. Further filtering using the selection criteria was done for the 431 (322 +109) papers, first by skimming through the full text of each paper and then by reading in detail. We carefully selected a final pool of 136 papers for this review. To ensure accuracy during the selection process, multiple authors reviewed and discussed borderline papers before making a decision.

\subsection{Data Extraction and Synthesis}

\begin{figure}
    \centering
    \includegraphics[width=1\linewidth]{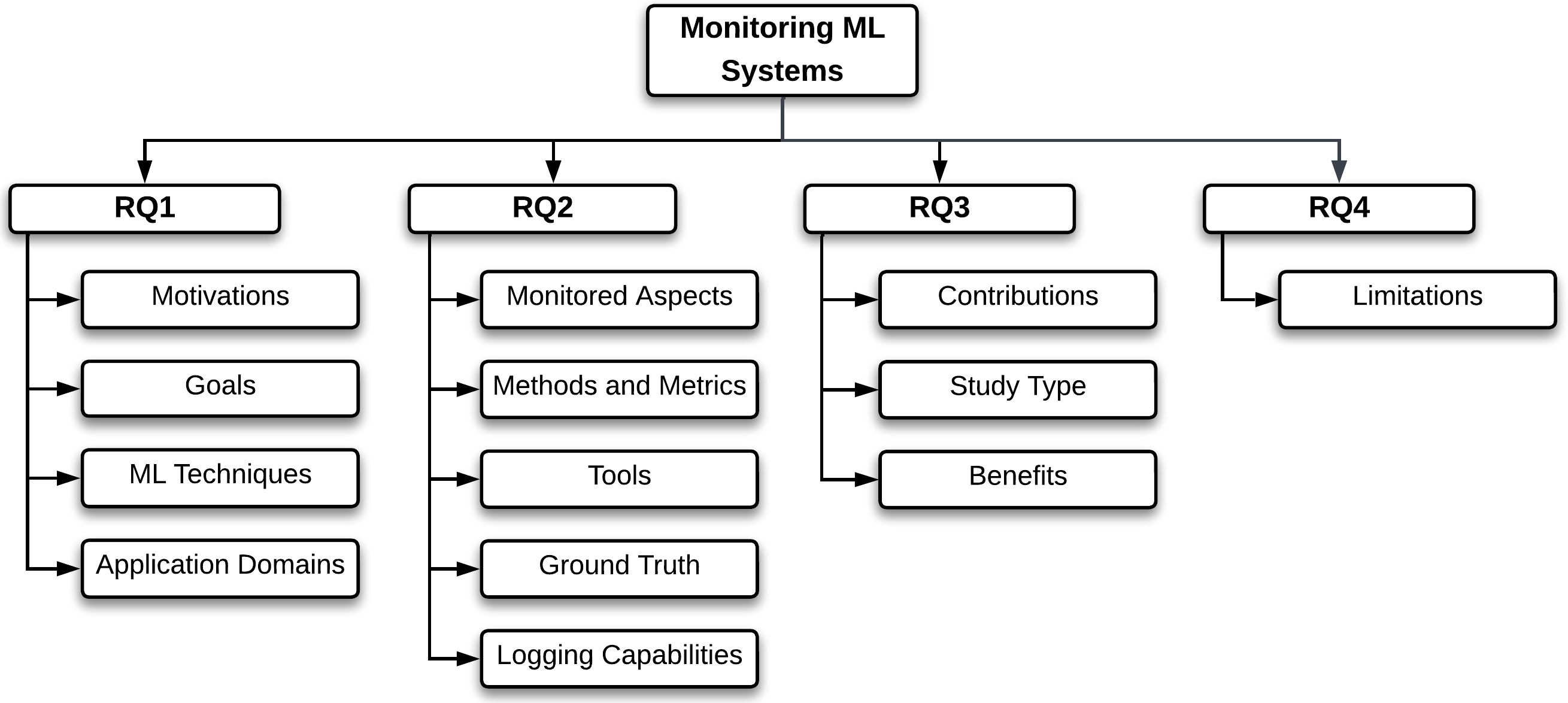}
    \caption{Data Extracted}
    \label{fig:data extracted}
\end{figure}

We created a Google form for data extraction with questions corresponding to each research question. The data extracted corresponding to each research question is shown in Fig.~\ref{fig:data extracted}. To maintain traceability between papers and the extracted data we collected study identification data such as title, authors, year, source, and study type (formal or gray literature). Our form had a total of 25 questions including questions for demographic information. We used a combination of short answers, long answers, checkboxes, and radio buttons depending on the data to be extracted. For quality assessment, we ran pilot tests; the first author extracted data for nine papers and compared them with the data extracted by other authors. A close match was found during the pilot tests, after which data for the remaining papers were extracted.

The extracted data consisted of quantitative data and qualitative data. To synthesize quantitative data, we leveraged descriptive statistics (frequency and percentages), visualizations through charts, and tabular comparisons. To synthesize qualitative data, we applied thematic analysis~\cite{clarke2017thematic}. The steps we followed for thematic analysis were familiarizing ourselves with the data, assigning codes, grouping similar codes, assigning themes to code groups, and refining the themes. Through the synthesis of collected data, we identified interesting patterns and trends that are shared in the following sections of this review.

\section{Results}
This section presents the findings from our MLR on monitoring ML systems. We organize these results based on the three research questions described earlier.
\subsection{Publication Trends}

\begin{figure}[h]
    \centering
    \begin{subfigure}{0.48\textwidth}
        
        \centering
        \includegraphics[width=\linewidth]{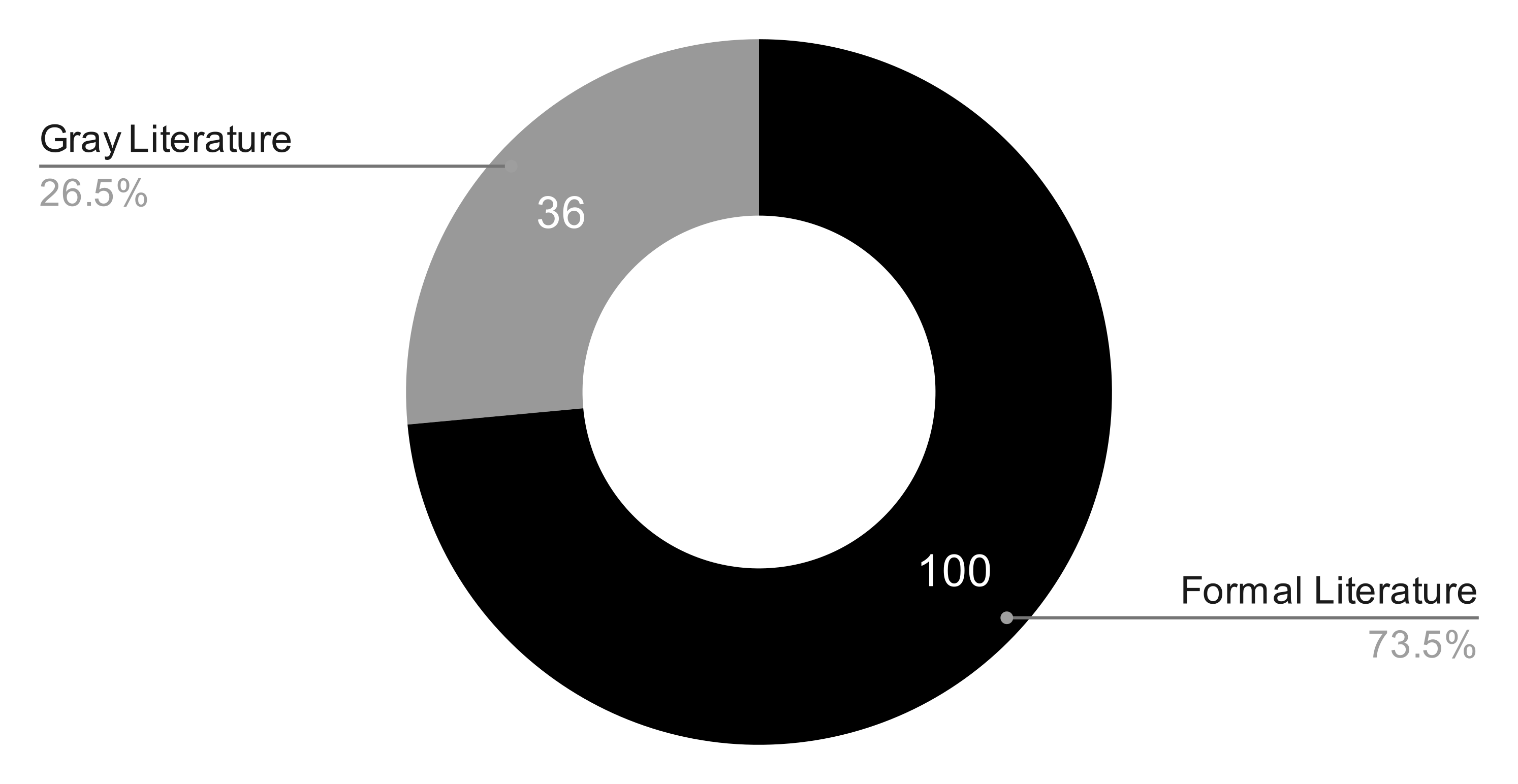}
        \vspace{4mm}
        \caption{Study Distribution by Literature Type}
        \label{fig:distribution by literature type}
        
    \end{subfigure}
    \hfill
    \begin{subfigure}{0.48\textwidth}
        \centering
        \includegraphics[width=\linewidth]{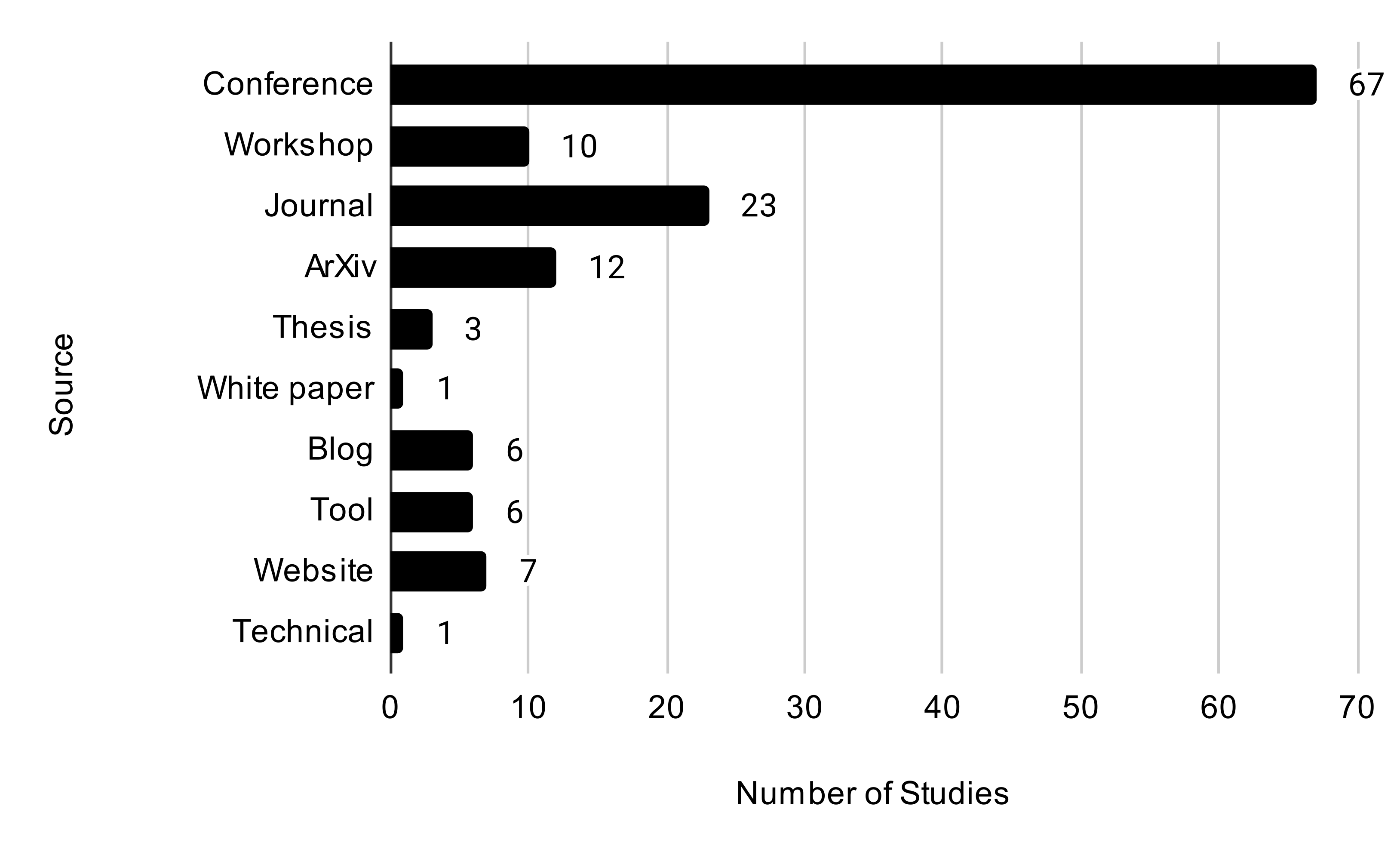}
        \caption{Study Distribution by Source Type}
        \label{fig:sources}
    \end{subfigure}
    \caption{Literature Distribution}
\end{figure}

The search process resulted in 136 studies in the final pool. Out of these, 100 were formal literature studies and 36 were gray literature studies. Although much more relevant gray literature exists, most of those studies met one or more exclusion criteria and were therefore excluded from this MLR. Fig. \ref{fig:sources}, shows the specific sources of studies; among formal literature, we found conference papers, journal papers, and workshop papers; among gray literature, we found ArXiv papers, theses, white papers, blogs, tool documentations, websites, and technical reports. The majority of formal literature studies found were conference papers and journal articles, contributing to 49\% and 17\% of the total studies, followed by workshop papers contributing 7\%. For gray literature, most of the studies were ArXiv papers (9\%), websites (5\%), blogs (4.5\%), and tool documentations (4.5\%). The remaining few (2\% or less) studies consist of theses, white papers, and technical reports.

\begin{figure}[h]
    \centering
    
    \begin{subfigure}{0.48\textwidth}
    
        \centering
        \includegraphics[width=\linewidth]{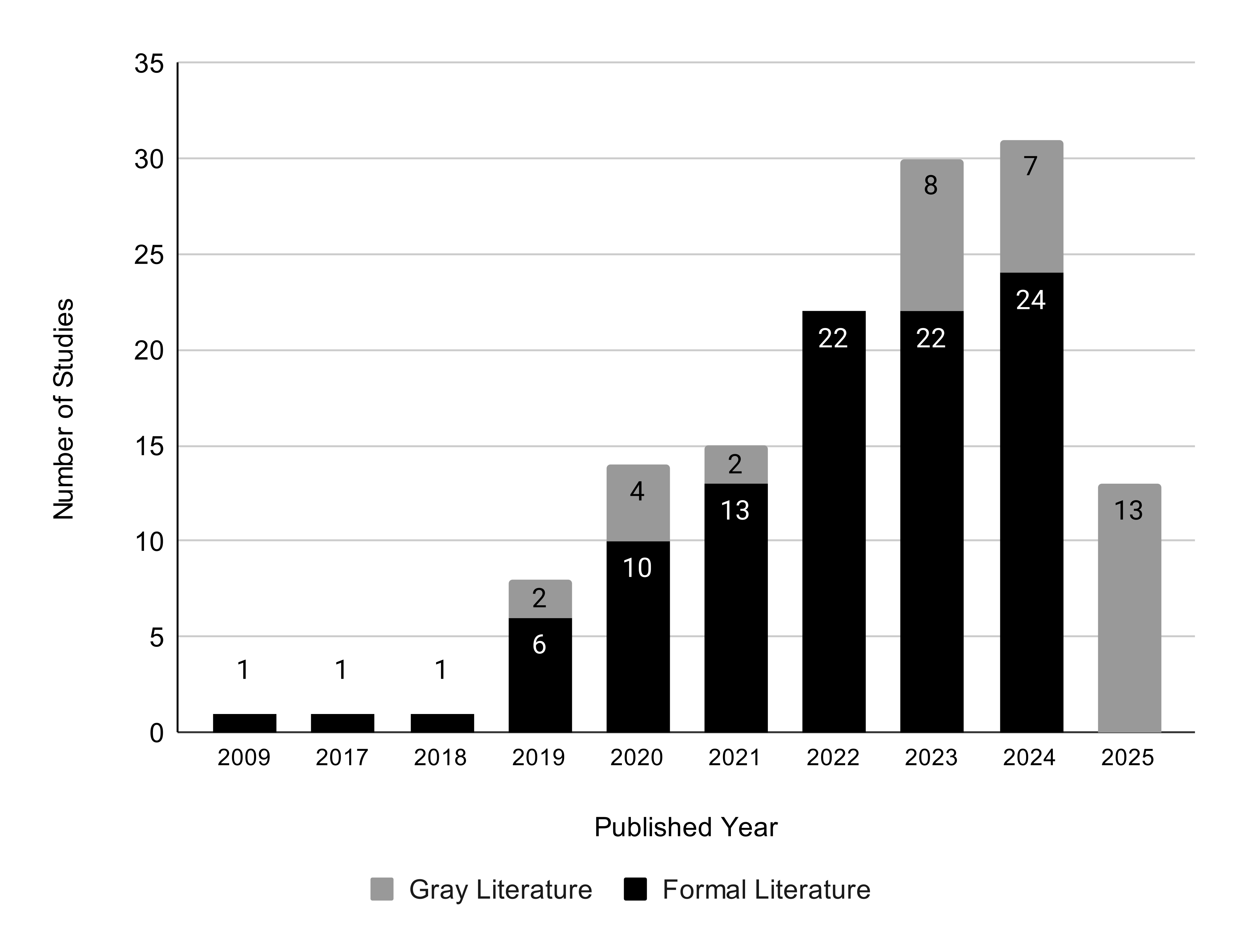}
       
        \caption{Study Distribution by Year}
        \label{fig:year}
        
    \end{subfigure}
    \hfill
    \begin{subfigure}{0.48\textwidth}
        \centering
        \includegraphics[width=\linewidth]{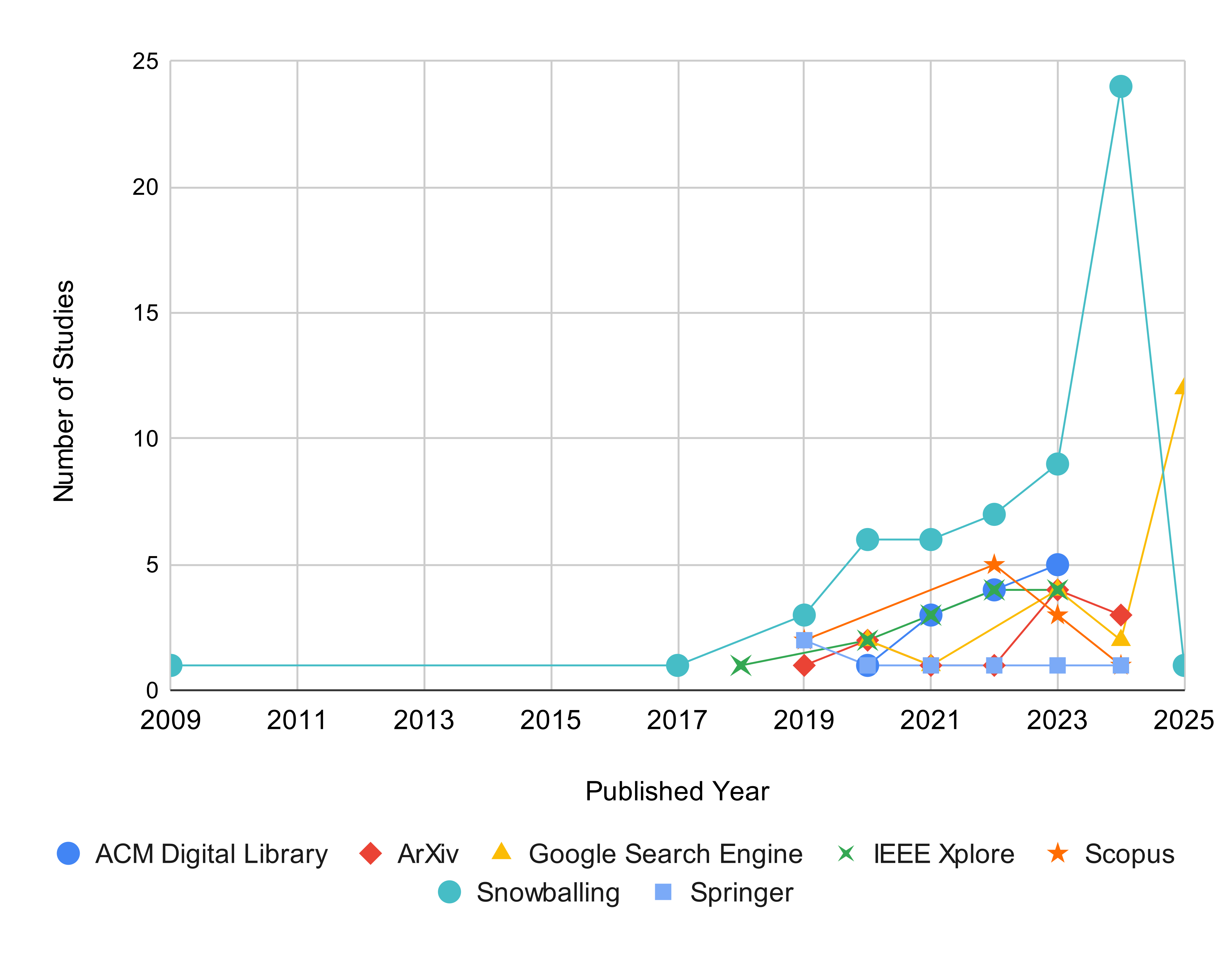}
        \caption{Study Distribution by Year and Source}
        \label{fig:year and source}
    \end{subfigure}
    \caption{Publication trends}
\end{figure}

Fig. \ref{fig:year} illustrates the distribution of studies over a period of 17 years, from 2009 to 2025. A rise in ML monitoring studies is observed from 2019. This could be due to several popular AI incidents around that time, like the biased Amazon recruitment tool~\cite{dastin2022amazon} and the erroneous cancer treatment recommender by IBM~\cite{strickland2019ibm}. Since 2019, there has been a steady increase in studies, with the largest number appearing in 2024. Gray literature studies increased significantly since 2023, with the highest number in 2025. The trend suggests that studies in 2025 would be higher than previous years, but since the search process for this review ended in early 2025, we may not have captured all relevant studies published later in the year. Fig. \ref{fig:year and source} shows the studies identified from various databases and through snowballing over the years. Overall, the rising number of ML monitoring studies since 2019 highlights the relevance of the area and an increasing interest among practitioners and academics.

\subsection{RQ1. What are the motivations and goals of monitoring approaches for ML systems?}
\subsubsection{Motivations}
\begin{figure}
    \centering
    \includegraphics[width=1\linewidth, clip, trim=0 1 0 0]{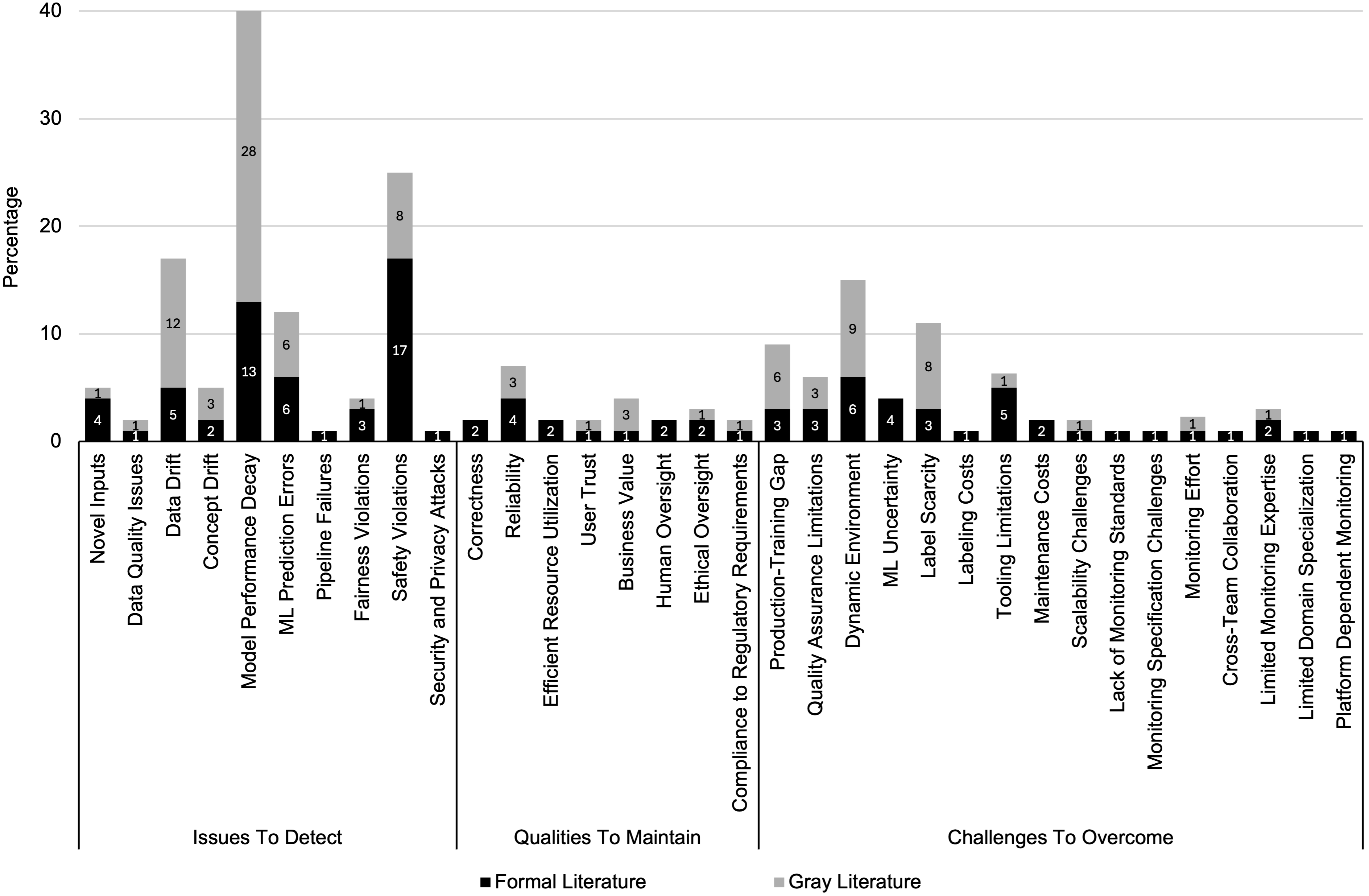}
    \caption{Monitoring Motivations}
    \label{fig:motivations}
\end{figure}

This section explores the reasons presented in the included studies for monitoring ML systems. Fig. \ref{fig:motivations} shows the specific motivations; we divided these motivations into three categories: \textit{issues to detect}, \textit{qualities to maintain}, and \textit{challenges to overcome}. These categories are not mutually exclusive, and many studies fall into multiple categories.

\sectopic{Issues To Detect} category is overall the most frequently mentioned among the motivations for monitoring ML systems. \textit{Model performance decay} is the most cited motivation in this category, with greater emphasis in gray literature (28\%) than in formal literature (13\%). This is followed by \textit{safety violations}, 17\% in formal literature and only 8\% in gray literature. \textit{Data drift} and \textit{ML prediction errors} also received a moderate level of attention. Interestingly, we found that formal literature and gray literature both emphasize different issues. For instance, the issues of declining model performance and data drift are mentioned nearly twice as much in gray literature than in formal literature (P121, P122). We expect this might be due to the greater importance of these issues in industrial ML systems. The opposite is observed for safety violations, which are significantly more cited in formal literature, with many studies focusing on safety monitoring for cyber-physical systems (P13). Other runtime issues like \textit{novel inputs}, \textit{data quality issues}, \textit{concept drift}, \textit{pipeline failures}, \textit{fairness violations}, and \textit{security and privacy attacks} received minimal attention (5\% or less). Except for data quality issues and ML prediction errors, an overall difference in priorities is evident. These differences highlight a potential disconnect between industrial and academic motivations for monitoring ML systems.

\sectopic{Qualities To Maintain} category received the least attention among all studies, possibly because the authors focused more on describing runtime issues to motivate their work. Our results show that while less cited overall, \textit{Reliability} stands out as an important quality of ML systems in both literatures, 4\% in formal literature and 3\% in gray literature, due to challenging production environments (P81). Maintaining \textit{business value} has the second highest occurrence in this category, with greater importance in gray literature (3\%) given its significance in organizational contexts. In contrast, \textit{Ethical oversight}, which refers to aspects of responsible ML, and human-centric qualities such as \textit{user trust}, \textit{compliance to regulatory requirements}, and monitoring with \textit{human oversight} received limited attention in both literatures (3\% or less), possibly due to the challenges associated with their practical implementation (P135). Only a few studies motivated their work with the aim of maintaining qualities such as \textit{correctness} (2\%) and \textit{efficient resource utilization} (2\%).

\sectopic{Challenges To Overcome} category was adequately represented in selected studies, particularly challenges that make monitoring ML systems necessary. Among these challenges, monitoring due to the \textit{dynamic environment} of production ML systems was most prominent, 6\% in formal literature and 9\% in gray literature. We found that dynamic production environments due to drifts and anomalies appear to be a major challenge in both literatures (P10, P74). \textit{Label scarcity} is another frequently cited motivation for monitoring approaches. We observed that label scarcity is mentioned nearly three times as much in gray literature (8\%) than in formal literature (3\%). This discrepancy may stem from two reasons: first, formal literature often assumes the availability of labeled data (P17, P44); second, due to the existence of several studies that address the label scarcity issue, it may no longer be considered an important challenge (P22, P86). Similarly, the challenges due to a mismatch in production and training, i.e., the \textit{production-training gap} is also mentioned twice as much in gray literature (6\%) than formal literature (3\%). Surprisingly, \textit{limitations of monitoring tools} were cited significantly more in formal literature. Other challenges mentioned in limited studies (3\% or less) include \textit{limited monitoring expertise}, \textit{monitoring effort}, and \textit{scalability challenges}. More nuanced challenges, such as \textit{ML uncertainty}, \textit{labeling costs}, \textit{maintenance costs}, \textit{lack of monitoring standards}, \textit{monitoring requirements specification challenges}, \textit{cross-team collaboration}, \textit{limited domain specialization}, and \textit{platform dependent monitoring}, were underrepresented (4\% or less) and appeared only in formal literature.

\subsubsection{Goals} This section investigates the goals for monitoring ML systems described in the included studies. Table \ref{tab:monitoring-goals} summarizes our findings into five comprehensive themes explained ahead.

\begin{table}[htbp]
\centering
\small
\caption{Monitoring Goals}
\renewcommand{\arraystretch}{1.2}
\begin{tabular}{>{\raggedright\arraybackslash}p{2.5cm}>{\raggedright\arraybackslash}p{3.5cm}p{3cm}p{3cm}c}
\hline
\textbf{Themes} & \textbf{Monitoring Goals} & \textbf{Formal Literature } & \textbf{Gray Literature } & \textbf{Percentage} \\
\hline
\multirow{3}{*}{\parbox{2.5cm}{\raggedright\textbf{Model Performance and Quality Assurance} }} 
& Detect Model Performance Degradation & P1, P2, P3, P6, P10, P14, P15, P17, P19, P21, P25, P26, P33, P37, P42, P46, P69, P70, P71, P80, P82, P85, P87, P92, P95, P99, P102, P104, P106, P112 & P52, P65, P74, P84, P110, P120, P121, P122, P123, P125, P127, P128, P132, P134, P136 & 18.8\% \\

& Continuous Quality Evaluation of System & P14, P18, P21, P29, P30, P34, P40, P47, P56, P57, P59, P66, P73, P79, P88, P91, P96, P97, P101, P105, P111 & P55, P113, P114, P115, P116, P118, P119, P120, P129, P130, P131, P132, P133, P134 & 14.6\% \\

& Early Issue Detection and Resolution & P1, P8, P9, P13, P16, P26, P29, P70, P78, P91, P92, P98, P103, P106 & P53, P54, P74, P122, P128 & 7.9\% \\

& Issue diagnosis & P10, P16, P18, P21, P26, P37, P38, P63 & P65, P120, P130, P133 & 5.0\% \\

& Evaluate Monitoring Methods & P58, P90 & - & 0.8\% \\

\hline
\multirow{2}{*}{\parbox{2.5cm}{\raggedright\textbf{Ensuring Responsible ML} }} 
& Ensure Safe and Reliable Operation & P9, P12, P23, P24, P29, P32, P61, P72, P76, P77, P83, P85, P86, P87, P94, P95, P98, P100, P104, P107, P108, P135 & P51, P67, P109, P129, P130 & 11.3\% \\

& Ensure Fairness & P18, P21, P22, P46, P48, P80 & P120, P134 & 3.3\% \\

& Incorporate Human Feedback in Monitoring & P13, P20, P98, P103 & - & 1.7\% \\
& Ensure Trustworthiness & P21, P28, P60 & - & 1.3\% \\

& Ensure Privacy & P39, P80 & P120 & 1.3\% \\

& Facilitate Responsible ML & P2, P135 & - & 0.8\% \\
\hline
\multirow{2}{*}{\parbox{2.5cm}{\raggedright\textbf{Efficient Workflows and Operations}}} 
& Integrate Monitoring in ML Workflows & P3, P4, P5, P8, P11, P15, P16, P17, P27, P28, P41, P44, P59, P60, P63, P101 & P93, P136 & 7.5\% \\

& Provide a Monitoring Platform & P17 & P113, P114, P115, P116, P118, P119, P121, P123, P128, P129, P130 & 5.0\% \\

& Support Engineers & P1, P5, P15, P34, P35, P38 & - & 2.5\% \\

& Automate Monitor Creation & P5, P75, P79, P80 & - & 1.7\% \\

& Platform Agnostic Monitoring & P39 & P114 & 0.8\% \\
& Prevent Alert Fatigue & - & P84 & 0.4\% \\
\hline
\multirow{3}{*}{\parbox{2.5cm}{\raggedright\textbf{Challenges, Guidelines, and Best practices} }} 
& Monitoring Guidelines & P64, P89, P100, P135 & P50, P51, P117, P124, P126 & 3.8\% \\

& Monitoring Challenges & P35, P43, P49, P62, P68, P89 & P50 & 2.9\% \\

& Consider Tradeoffs between Monitoring Requirements & P6, P7, P27, P36, P45, P68, P88, P111 & - & 2.5\% \\

& Industry Best Practices & P43, P49, P62, P64, P81 & - & 2.0\% \\

& Standardize Monitoring Requirements & P81 & - & 0.4\% \\

& Monitoring Limitations  & P31 & - & 0.4\% \\
\hline

\multirow{1}{*}{\parbox{2.5cm}{\raggedright\textbf{Label Scarcity} }} 
& Monitoring without Labels & P22, P25 & P54, P65, P110, P134 & 2.5\% \\

& Monitoring with Limited Labels & P36, P46 & - & 0.8\% \\
\hline
\end{tabular}
\renewcommand{\arraystretch}{1}
\label{tab:monitoring-goals}
\end{table}

\sectopic{Model Performance and Quality Assurance} is the most prominent theme among all. Some of the key monitoring goals in this theme are \textit{detecting model performance degradation}, such as decreasing accuracy in production, and \textit{continuous quality evaluation of the system}, which refers to maintaining the overall quality of the ML system beyond model performance. The strong emphasis on these two goals (> 33\%) shows the importance of predictive performance and overall system stability in both literatures. Additionally, \textit{early issue detection and mitigation} and \textit{issue diagnosis} were other important goals in this category, highlighting efforts to proactively detect, understand, and mitigate runtime issues. In comparison, very few studies considered the \textit{evaluation of monitoring methods} as a primary goal of their work.

\sectopic{Ensuring Responsible ML} is the second most prominent theme, encompassing critical goals such as \textit{ensuring safe and reliable operation}, \textit{fairness}, \textit{trustworthiness}, and \textit{privacy}. A small portion of studies \textit{facilitate responsible ML} by operationalizing multiple aspects. For use cases where automated monitoring cannot be completely relied upon, some studies aim to \textit{incorporate human feedback in monitoring}. Monitoring goals related to responsible ML were more common in formal literature (> 80\%) than in gray literature, possibly due to the difficulties of implementing it in practice (P135).

\sectopic{Efficient Workflows and Operations} is the third most common theme, consisting of monitoring goals related to the creation and integration of monitors. While formal literature tends to focus on goals such as \textit{integrate monitoring in ML workflows}, and \textit{automate monitor creation}, gray literature, in contrast, focuses more on \textit{providing a monitoring platform}. Given the industrial context, the greater emphasis on a platform seems justified. However, \textit{supporting engineers} with better monitoring solutions is completely absent in gray literature, which was unexpected. Only a few studies aimed to create \textit{platform agnostic monitoring} solutions and \textit{prevent alert fatigue} among engineers.

\sectopic{Challenges, Guidelines, and Best practices} is the fourth theme. While comparatively fewer studies focused on this area, it covers important aspects of the ML monitoring landscape. A notable goal in this category is providing \textit{monitoring guidelines}, which are available from both formal and gray literature perspectives. \textit{Industry best practices}, in particular, are shared by a small portion of formal literature. Additional objectives, mentioned in a few formal and even fewer gray literature studies, included highlighting \textit{monitoring challenges}, \textit{considering tradeoffs between monitoring requirements}, such as accuracy and scalability, \textit{standardizing monitoring requirements}, and discussing the \textit{limitations of monitoring}.

\sectopic{Label Scarcity} is the fifth theme, covering an important challenge in model performance monitoring. Since measuring model metrics such as accuracy and precision is not possible without labeled data, the goal of some studies was to specifically address this challenge and present monitoring solutions that can \textit{monitor with limited labels} or even \textit{monitor without labels}. While other studies may also address this problem, it was less commonly (3.3\%) stated as an explicit goal.

\subsubsection{Machine Learning Techniques Monitored} An ML technique refers to the algorithm that enables a model to learn patterns in data and make predictions. Since monitoring approaches are often dedicated to specific ML techniques, this section looks into the various ML techniques monitored in the literature. Our findings are presented in Table \ref{tab:ml_techniques}.
\begin{table}[htbp]
\centering
\small
\renewcommand{\arraystretch}{1.2}
\caption{Machine Learning Techniques Monitored}
\begin{tabular}{>{\raggedright\arraybackslash}p{3cm}>{\raggedright\arraybackslash}p{2cm}>{\raggedright\arraybackslash}p{3.5cm}>{\raggedright\arraybackslash}p{3.5cm}c}
\hline
\textbf{Machine Learning Technique} & \textbf{Subtype}  & \textbf{Formal Literature } & \textbf{Gray Literature } & \textbf{Percentage}\\
\hline
\multirow{1}{*}{\textbf{Supervised Learning}} 
 & Deep Learning & P6, P11, P13, P14, P26, P29, P38, P41, P47, P57, P58, P60, P61, P66, P69, P72, P73, P75, P77, P78, P79, P85, P86, P87, P88, P90, P91, P92, P94, P95, P96, P97, P99, P102, P103, P104, P105, P106, P112 & P54, P67, P74, 84, P93, P109, P136 & 32.4\% \\

 & Traditional & P1, P2, P20, P26, P36, P37, P38, P42, P44, P56, P57, P60, P68, P80, P82 & P65, P84, P110, P122, P132 & 14.0\% \\
 
 & Not Specified & P5, P9, P12, P21, P25, P33, P34, P45, P46, P59, P62, P64, P81 & P114, P116, P117, P118, P119, P123, P124, P125, P127, P130, P134 & 17.0\%\\
\hline
\textbf{Semi Supervised Learning} & - & P34 & - & 0.7\%\\
\hline
\textbf{Unsupervised Learning} & Deep Learning  & P71 & - & 0.7\%\\
\hline
\multirow{1}{*}{\textbf{Reinforcement Learning}} & Deep Learning  & P76, P98, P107, P108 & - & 2.8\%\\
 & -  & P83 & - & 0.7\%\\

\hline

\textbf{Model Agnostic} & -  & P32, P48 & P53, P113, P115 & 3.5\% \\
\hline
\textbf{Not Mentioned} & -  & P3, P4, P7, P8, P10, P15, P16, P17, P18, P19, P22, P23, P24, P27, P28, P30, P31, P35, P39, P40, P43, P49, P63, P70, P89, P100, P101, P111, P135 & P50, P51, P52, P55, P120, P121, P126, P128, P129, P131, P133 & 28.2\% \\
\hline
\end{tabular}
 \renewcommand{\arraystretch}{1}
\label{tab:ml_techniques}
\end{table}

Overall, \textit{supervised learning} techniques, such as classification and regression, were the most popular, particularly those leveraging \textit{deep learning}. The substantial volume of formal literature focusing on deep learning can be attributed to the widespread use of deep neural networks (DNNs) in cyber-physical systems - the application domain for 26\% of formal literature studies. The inherent uncertainty and large input space of DNNs require high levels of safety monitoring (P72, P73). Another commonly monitored technique was \textit{traditional} supervised learning, which includes algorithms such as logistic regression (P36) and random forest (P20). Many studies did \textit{not specify} the type of supervised learning technique being monitored (P5). Monitoring \textit{semi-supervised learning}, \textit{unsupervised learning}, and \textit{reinforcement learning} was rare and exclusively implemented in formal literature. A limited number of formal and gray literature studies presented monitoring approaches that were \textit{model agnostic}, i.e., could be used to monitor any ML technique (P53). Whereas, a significant portion of studies from both literatures did not mention the type of ML technique monitored. This might be because these approaches are model agnostic; however, it was not explicitly mentioned in the papers.

\subsubsection{Application Domains} 
\begin{figure}
    \centering
    \includegraphics[width=0.8\linewidth, clip, trim=2 4 2 4]{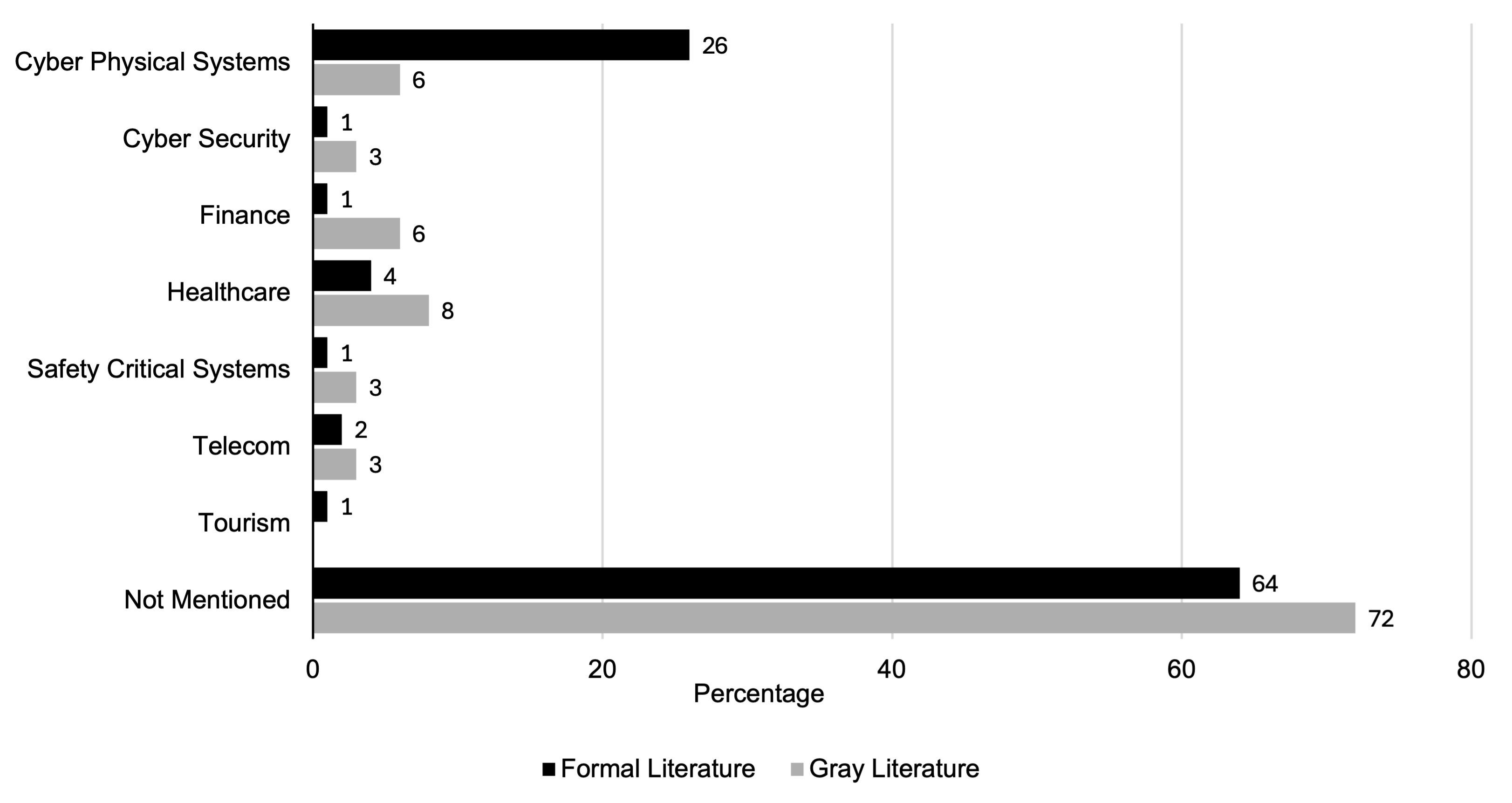}
    \caption{Application Domains}
    \label{fig:domain}
\end{figure}

Fig. \ref{fig:domain} presents the main application domains of the included studies. The domain is a crucial characteristic of the monitoring context; therefore, we discuss this further in this section. 

The majority of studies, 64\% formal and 72\% gray, do \textit{not mention} the application domain, as they intend to be generic and applicable to any domain (P23, P114). \textit{Cyber physical systems} is the most common domain in monitoring literature, particularly formal literature (26\%), due to the importance of reliability and safety monitoring for such systems (P90, P98). Few studies focused on monitoring ML systems in \textit{healthcare} (P44) and \textit{finance} (P65), whereas monitoring approaches for other domains such as \textit{cyber security} (P10), \textit{safety critical systems}, \textit{telecom} (P26), and \textit{tourism} (P57) were rare (5\% or less). 

\begin{center}
\begin{myframe}[width=48em,top=5pt,bottom=5pt,left=5pt,right=5pt,arc=10pt,auto outer arc,title=\centering\textbf{RQ1 Answer Summary}]
The primary motivation for monitoring ML systems is to detect runtime issues such as model performance decay, safety violations, and data drift. Challenges in ML monitoring, such as dynamic production environments and label scarcity, are also mentioned as rationale by some studies. Relatedly, the monitoring goal for most studies is to maintain model performance and overall quality of the ML system. Considerably fewer studies aim to monitor ML systems to ensure responsible ML and improve workflows. Among ML techniques monitored, supervised learning is most prominent, particularly deep learning based classification. A fair portion of studies do not specify the ML technique being monitored, whereas monitoring approaches for semi-supervised learning, unsupervised learning, and reinforcement learning are rare. Additionally, most monitoring approaches tend to be generic, while only a few are domain-specific.
    \end{myframe}
\end{center}

\subsection{RQ2. What monitoring solutions have been developed for ML systems?}
\subsubsection{Monitored Aspects} 
\begin{figure}[h]
    \centering
    \begin{subfigure}{0.49\textwidth}
        \centering
        \includegraphics[width=1\linewidth]{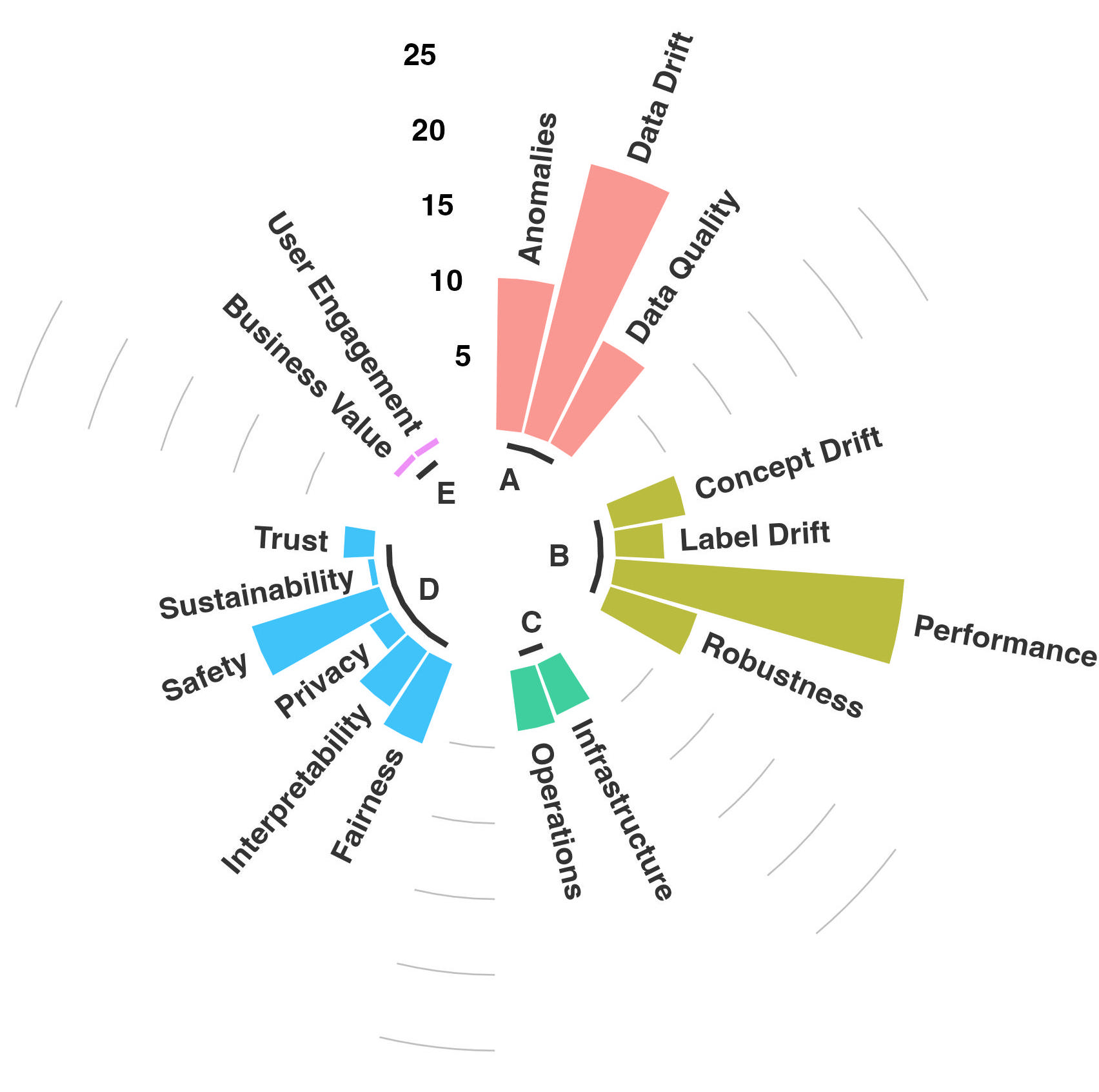}
        \caption{Formal Literature}
        \label{fig:FL-aspects}
    \end{subfigure}
    \hfill
    \begin{subfigure}{0.49\textwidth}
        \centering
         \includegraphics[width=0.9\linewidth]{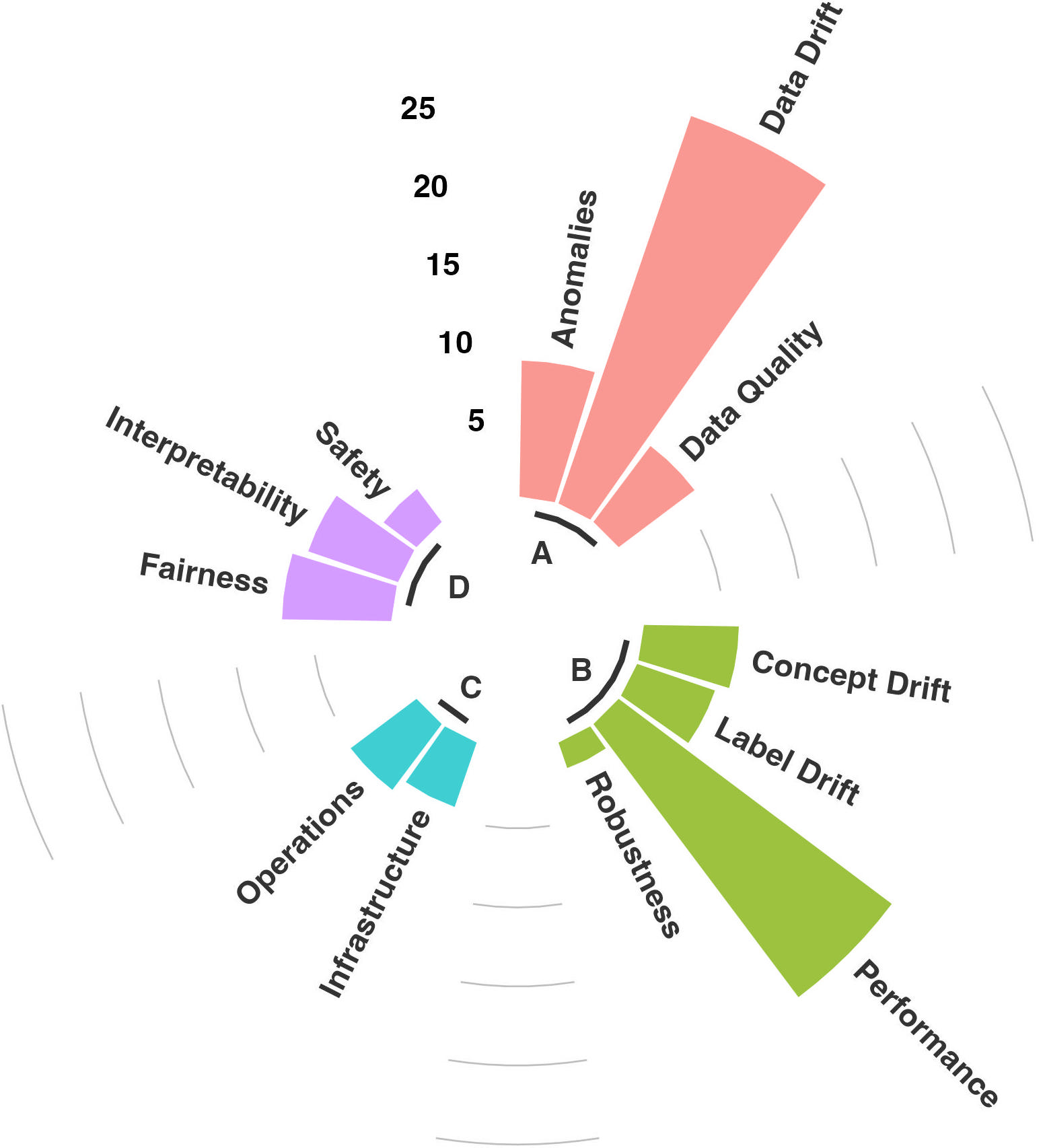}
        \caption{Gray Literature}
        \label{fig:GL-aspects}
    \end{subfigure}
    
    \caption{Monitored Aspects}
    \begin{center}{\footnotesize A) Data B) Model Behavior C) Operations \& Infrastructure D) Responsible ML E) Business } \end{center}
    
\end{figure}

This section presents our findings on \textit{what} was being monitored in the selected studies. Fig. \ref{fig:FL-aspects} and \ref{fig:GL-aspects} show the percentage of monitored aspects in formal and gray literature, divided into six categories described below.

\sectopic{Data} category consists of monitored aspects related to the input data of an ML model. \textit{Data drift} occurred most frequently in this category, more so in gray literature (26\%) than formal literature (18\%). Since evolving data is a primary cause of performance degradation in ML models, many studies monitor data distribution changes (P16, P118). Other moderately monitored aspects include \textit{data quality} and \textit{anomalies}. Monitoring \textit{data quality} involves looking for issues such as missing, corrupted, and incorrectly formatted data that can negatively impact model performance (P25, P117). \textit{Anomalies}, on the other hand, are rare, unusual inputs that significantly deviate from expected data patterns and often lead to inaccurate predictions (P11, P16). 

\sectopic{Model Behavior} refers to the functionality of the ML model in an ML system. In this category, model \textit{performance} was a highly monitored aspect among both literatures (P1, P44). This was expected since accurate predictions are the most important quality of an ML model. The second highest was \textit{concept drift}, closely followed by \textit{label drift}. Both of these drift types provide insights into the underlying relationship between model inputs and outputs that may impact model performance, hence their importance in literature (P6, P56). The distribution of the top three monitored aspects in this category, \textit{performance}, \textit{concept drift}, and \textit{label drift}, is similar across formal and gray literature, indicating a similar level of interest. Monitoring for more nuanced aspects of model behavior, such as \textit{robustness}, was found significantly more in formal literature (6\%) than in gray literature (2\%). 

\sectopic{Operations and Infrastructure} related aspects of the ML system are covered under this category. A decent portion of studies monitor for system and model \textit{operations} such as inference latency, throughput, and ML service health (P19, P63). In comparison to formal literature, more gray literature studies monitored operational aspects. We expect that this might be due to the relative simplicity of monitoring operational aspects and their greater importance in industrial applications. Comparatively, fewer studies monitor for \textit{infrastructure} aspects such as CPU and memory utilization (P7, P101). 

\sectopic{Responsible ML} category covers monitoring ethical and human-centric characteristics of ML systems. \textit{Fairness}, including related concepts such as bias, is a commonly monitored aspect (P18, P114). Within the field of responsible ML, \textit{fairness} has received the most attention, possibly due to several high-profile incidents~\cite{dastin2022amazon, dressel2018accuracy}. \textit{Safety} was also frequently monitored, especially in formal literature, since many studies focus on cyber-physical systems (P32, P76). Monitoring \textit{interpretability}, which refers to explanations for model decisions (P37, P50), was less common among studies.   
Relatively few and only formal literature studies monitored \textit{privacy}, \textit{trust}, and \textit{sustainability} of ML systems.

\sectopic{Business} captures aspects of an ML system that directly affect organizational performance. While these aspects are critical for success, monitoring for such aspects is extremely rare; we found only one paper in formal literature that monitors \textit{business value} and \textit{user engagement} (P135).

\subsubsection{Monitoring Techniques and Metrics}
This section provides a comprehensive overview of the monitoring techniques and metrics applied in studies to monitor ML systems. The findings have been categorized in terms of monitoring focus into data, ML model behavior, operations and infrastructure, responsible ML, and business. 
\sectopic{Data} monitoring in ML systems involves the detection of anomalies, identification of data drift, and assessment of data quality as shown in Table \ref{tab:data_monitoring_techniques}. 

\begin{table}[htbp]
\centering
\footnotesize
\renewcommand{\arraystretch}{1.1}
\caption{Data Monitoring Techniques and Metrics}
\begin{tabular}{|c|c| >{\raggedright\arraybackslash}p{2.5cm}>{\raggedright\arraybackslash}p{3cm}p{2.7cm}p{2.7cm}|}
\hline
\textbf{} & \textbf{Monitored Aspect}  & \textbf{Monitoring Technique } & \textbf{Metrics} & \textbf{Formal Literature} & \textbf{Gray Literature}\\
\hline
\multirow{61}{*}{\rotatebox{90}{\textbf{DATA}}} 
 & \multirow{10}{*}{\textbf{Anomalies}} 
 & Statistical & Custom & P16, P17, P75, P97, P106 & P117, P121\\
 \cline{3-6}
 &  & Distance-based & Euclidean distance & P73  & - \\\cline{4-6}
 & & & Mahalanobis distance  & P75, P77 & -\\\cline{4-6}
 & & & Earth Mover’s distance & - & P93\\\cline{3-6}
  &  & Rule-based & - & P28, P30, P47, P59, P75, P85, P101 & P117, P121\\
 \cline{3-6}
 &  & Machine learning & - & P11, P23, P25, P77, P87, P90, P101, P112 & P50, P53, P109, P114, P130 \\
 \cline{3-6}
 &  & Deep learning & - & P11, P15, P19, P69, P112, P75, P90, P98 & - \\
 \cline{2-6}
  & \multirow{39}{*}{\textbf{Data Drift}} 
  & Data distribution-based
(Statistical) & Kolmogorov–Smirnov statistic  & P2, P3, P5, P8, P16, P17, P25, P32, P63, P64, P82 & P50, P52, P53, P55, P74, P123, P124, P134\\

    \cline{4-6}
    & & & Z-statistic & P2, P36 & -\\\cline{4-6}
    & & & T-statistic & P17 & -\\\cline{4-6}
    & & & Kruskal-Wallis statistic & P82  & -\\\cline{4-6}
    & & & Maximum Mean Discrepancy & -& P50 \\\cline{4-6}

    & & & Bhattacharyya coefficient & P3 & -\\\cline{4-6}
    & & &  Exponential moving avg. & P5 & -\\\cline{4-6}
    & & &  Chi-square statistic& P64, P82, P135 & P123\\\cline{4-6}
    & & &  Entropy-based & P62, P75 & -\\\cline{4-6}
    & & &  Bayesian-based  & P36, P70 & -\\\cline{4-6}
    & & &  Mean, Min, Max, Std dev, Std diff, Quantiles & P64, P71, P81 & P50, P93, P125 \\\cline{4-6}
    & & & Cumulative Sum Statistic & P100 & -\\\cline{4-6}
    & & &  Custom & P33, P44, P59, P66, P81, P86, P88, P92, P97 & P113, P116, P119, P120, P128, P129, P130, P131 \\\cline{3-6}

    &  &  Data distribution-based
(Distance) & Kullback-Leibler divergence&P2, P63, P75, P111, P135  & P53, P93, P110, P114, P115, P121, P132\\
   \cline{4-6}
    & & &  Hellinger distance & P2, P82 & P121, P132\\\cline{4-6}
    & & & Energy distance  &  P2, P75  & -\\\cline{4-6}
    & & & Jensen-Shannon divergence & - &  P93, P110, P114, P117, P118, P123, P125\\\cline{4-6}
    & & &  Total variation distance & P2 & P121, P132\\\cline{4-6}
    & & &  Probability stability index & P62, P63, P80 & P53, P114, P115, P123, P125, P126\\\cline{4-6}
    & & &  Characteristic stability index & - & P126\\\cline{4-6}
    & & &  L-Infinity distance & P5 & P118\\\cline{4-6}
    & & &   Earth Mover’s distance & P8 & P93, P115\\\cline{4-6}
    & & &   Euclidean distance & P71, P96 & -\\\cline{4-6}
    & & &   Mahalanobis distance & P75, P96 & -\\\cline{4-6}
    
    & & &   Wasserstein distance & P32, P111 & P114, P117, P123, P134\\\cline{4-6}
    & & &   Custom & P13, P105 & -\\\cline{3-6}
    &  & Feature Contribution  & LIME/SHAP scores & P17, P18, P19, P21, P37, P135 & P118, P128\\\cline{4-6}
    & & &   NDCG & P15, P17 & P117, P123\\\cline{3-6}
    & & Rule-based & - & P11, P38, P69 & P54\\
    \cline{3-6}
  & & Machine learning & - & P11, P16, P21, P23, P25, P36, P56, P71, P90, P96 & P53, P109, P124, P133\\
  \cline{3-6}
  & & Deep learning & - & P11, P13, P19, P41, P69, P75 & -\\
 \cline{2-6}
 
 & \multirow{12}{*}{\textbf{Data Quality}} & Statistical &  Mean, Min, Max, Std Dev, Percentiles  & P15, P19, P59, P64, P81, P87 & P93\\\cline{4-6}
 & & &  Missing values percentage & P10, P19, P30, P64, P81 & P115, P117, P128, P131\\\cline{4-6}
 & & &  Missing records percentage & P30 &-\\\cline{4-6}
 & & &  Missing metadata percentage & P10 &-\\\cline{4-6}
 & & &  Zeros/Nulls percentage & P19, P81 & P123\\\cline{4-6}
 & & &  Distinct category count/ratio & P19 &-\\\cline{4-6}
 & & &  Custom & P5, P135 &-\\\cline{3-6}

 & & Metadata profiling &  Amount of data & P59 & - \\\cline{4-6}
 & & &  Invalid format rate & - & P117, P123\\\cline{4-6}
 & & &  Version & P19, P81, P135 &-\\\cline{3-6}
 & & Rule-based &  - & P5, P30 & P117\\\cline{3-6}
   
\hline
\end{tabular}
 \renewcommand{\arraystretch}{1}
\label{tab:data_monitoring_techniques}
\end{table}

\textit{Anomaly detection} is implemented through several techniques, broadly categorized as statistical, distance-based, rule-based, and learning-based. \textit{Statistical} and \textit{distance-based} approaches are relatively straightforward, lightweight and interpretable for detecting anomalies, while \textit{rule-based} approaches are suitable for domain-specific anomaly detection when customizations are needed. \textit{Machine learning} and \textit{deep learning} are the most frequently used methods, particularly in formal literature, for detecting dynamic and complex patterns of anomalies, especially in high-dimensional data. 

Monitoring techniques for \textit{data drift} are primarily \textit{statistical} or \textit{distance-based}, comparing previous and current data distributions to identify deviations. These methods are standardized, straightforward, and computationally efficient. Common statistical test metrics include Kolmogorov–Smirnov statistic, Chi-square statistic, and entropy-based measures, which are used to compare feature distributions over time. Custom statistical metrics were also proposed in several studies. Distance-based metrics such as Kullback–Leibler divergence, Hellinger distance, and Wasserstein distance are effective in quantifying shifts in probability distributions. Additionally, scores computed through LIME and SHAP~\cite{salih2025perspective} have been leveraged to detect feature attribution drift and \textit{feature contribution}. \textit{Machine learning} and \textit{deep learning} methods, in contrast, utilize ML models to detect data drift by learning complex data patterns. These approaches are more adaptive and effective for detecting subtle shifts that statistical techniques might miss. Overall, distance-based methods were cited more in gray literature, while learning-based methods were found more in formal literature. Rule-based methods were rarely used for data drift detection, possibly because they do not adapt well when the data changes.

For \textit{data quality} monitoring, basic \textit{statistical} summaries such as mean, standard deviation, and percentiles are commonly used to assess overall data consistency, while specific metrics like missing value percentages, distinct category ratios, and \textit{metadata profiling} help identify quality issues in more detail. Data quality checks are often combined with \textit{rule-based} constraints to enforce data integrity.

\sectopic{Model Behavior} monitoring ensures that the ML model itself performs correctly, which includes monitoring for concept drift, label drift, model performance, and robustness. Tables \ref{tab:model_monitoring_techniques1} and \ref{tab:model_monitoring_techniques2} summarize the key techniques and metrics found in literature.

\begin{table}[htbp]
\centering
\footnotesize
\renewcommand{\arraystretch}{1.1}
\caption{Model Behavior Monitoring Techniques and Metrics}
\begin{tabular}{|c|c|>{\raggedright\arraybackslash}p{2.5cm}>{\raggedright\arraybackslash}p{3cm}p{2.5cm}p{3cm}|}
\hline
\textbf{} & \textbf{Monitored Aspect}  & \textbf{Monitoring Technique } & \textbf{Metrics} & \textbf{Formal Literature} & \textbf{Gray Literature}\\
 
\hline
\multirow{27}{*}{\rotatebox{90}{\textbf{MODEL BEHAVIOR}}} 
& \multirow{17}{*}{\textbf{Concept Drift}} & Error rate-based & Squared error  & P6, P26, P100 & P114\\\cline{4-6}
& & & Accuracy & P10, P17, P135 & P53, P114\\\cline{4-6}
& & & Precision & P17 & P121 \\\cline{4-6}
& & & Exponential moving average & P5 & -\\\cline{4-6}
& & & Custom & - & P128, P130\\\cline{3-6}
&  & Data distribution-based & Kolmogorov–Smirnov statistic & P26  & P114\\\cline{4-6}

& &(Statistical)& R statistic & P42 & -\\\cline{4-6}
& & & Bayes factor & P70 & -\\\cline{4-6}
& & & Mean, Std dev.& - & P121\\\cline{3-6}
&  & Data distribution-based& Euclidean distance & P71  & P114\\\cline{4-6}
& &(Distance) & Kullback-Leibler divergence & - & P121 \\\cline{4-6}
& & & Hellinger distance & - & P121 \\\cline{4-6}
& & & Total Variation distance & - & P121 \\\cline{4-6}
& & & Jensen-Shannon divergence & - & P65\\\cline{4-6}
& & & Custom & - & P132\\\cline{2-6}

& \multirow{13}{*}{\textbf{Label Drift}} &  Data distribution-based
(Statistical) & Mean, Std. dev, Variance, Quantiles & P15, P100  & - \\\cline{4-6} 
& & & Kolmogorov–Smirnov statistic  & P123 & -\\\cline{4-6}
& & & Chi-square statistic& P123 & -\\\cline{4-6}
& & & Friedman statistic  & P82 & -\\\cline{3-6}
&  & Data distribution-based & Kullback-Leibler divergence & P56  & P53 \\\cline{4-6}
& &(Distance) & Wasserstein distance  & - & P117 \\\cline{4-6}
& &  & Total variation distance & - & P120 \\\cline{4-6}
& &  & Jensen-Shannon divergence  & P63 & P117, P118, P123 \\\cline{4-6}
& &  & Chebyshev distance & - & P123\\\cline{4-6}
& &  & Custom & - & P120, P131 \\\cline{2-6}
\hline
\end{tabular}
 \renewcommand{\arraystretch}{1}
\label{tab:model_monitoring_techniques1}
\end{table}

Monitoring techniques for \textit{concept drift} are categorized as error rate-based or data distribution-based. \textit{Error rate-based} methods track changes in error rate of ML models using metrics such as accuracy and precision~\cite{lu2018learning}. These methods directly capture drops in model performance, but they rely on the availability of ground truth labels to calculate prediction errors. \textit{Data distribution-based} methods measure the difference between distributions of historical and current data by applying statistical tests and distance metrics~\cite{lu2018learning}. Commonly used metrics include Kolmogorov–Smirnov statistic, Euclidean distance, and Kullback-Leibler divergence. Data distribution-based methods do not rely on labels and provide more granular detection; however, the shifts detected may not always impact overall model performance.

\textit{Label drift} monitoring techniques also compare data distributions using \textit{statistical} tests and \textit{distance} measures to quantify changes in the distribution of true labels over time. Statistical metrics include mean, standard deviation, Chi-square statistic, and Friedman statistic, whereas distance metrics consist of Jensen-Shannon divergence, Kullback-Leibler divergence, and Wasserstein distance.

\begin{table}[htbp]
\centering
\footnotesize
\renewcommand{\arraystretch}{1.1}
\caption{Model Behavior Monitoring Techniques and Metrics cont.}
\begin{tabular}{|c|c|>{\raggedright\arraybackslash}p{2.5cm}>{\raggedright\arraybackslash}p{3cm}p{2.5cm}p{3cm}|}
\hline
\textbf{} & \textbf{Monitored Aspect}  & \textbf{Monitoring Technique } & \textbf{Metrics} & \textbf{Formal Literature} & \textbf{Gray Literature}\\
\hline 
\multirow{40}{*}{\rotatebox{90}{\textbf{MODEL BEHAVIOR}}} 

& \multirow{33}{*}{\textbf{Performance}} & Ground truth comparison & Accuracy & P1, P5, P10, P17, P28, P34, P36, P46, P60, P63, P66, P80, P135 & P50, P84, P113, P114, P115, P117, P119, P120, P122, P123, P125, P126, P128, P130, P131\\\cline{4-6}

& &  & Precision & P1, P34, P45, P60, P66, P135 & P50, P115, P117, P120, P121, P123, P125, P126, P128, P130\\\cline{4-6}

& &  & Recall & P1, P10, P34, P44, P66, P135 & P50, P115, P117, P120, P121, P123, P125, P128, P130, P131\\\cline{4-6}

& &  & F1 score & P1, P17, P60, P66, P135  & P113, P115, P117, P120, P122, P125, P128, P130\\\cline{4-6}

& &  & Area under curve & P16, P17, P100 & P113, P120, P125, P130\\\cline{4-6}
& &  & False positive rate & P10, P63 & P120, 128\\\cline{4-6}
& &  & Mean absolute error & P3, P8, P21, P135 & P113, P117, P123, P124, P125, P127, P130, P136\\\cline{4-6}
& &  & Mean squared error & P6, P17, P21, P44, P46, P100, P135 & P113, P117, P120, P123, P124, P125, P127, P130, P136\\\cline{4-6}
& &  & Log loss & P16, P135 & -\\\cline{4-6}
& &  & Average positive predictive value & P45, P68 & - \\\cline{4-6}
& &  & Average negative predictive value & P45, P68 & P121\\\cline{3-6}

&  & Uncertainty-based & Uncertainty estimate & P37  & P134\\\cline{4-6}
 & &  & Prediction confidence score & P72, P99 & P84\\\cline{4-6}
& &  & Max softmax  & P72 & -\\\cline{4-6}
& &  & Softmax entropy & P72 & -\\\cline{4-6}
& &  & Variation ratio & P72 & -\\\cline{4-6}
& &  & Predictive entropy & P72 & -\\\cline{4-6}
& &  & Mutual information & P72 & -\\\cline{4-6}
&  & & Neuron activation path similarity & P105  & - \\\cline{3-6}
&  & Auxiliary model & Predictions difference & - & P110\\\cline{3-6}

&  & Machine learning & - & P95, P112  & P109, P124\\\cline{3-6}
&  & Deep learning & - & P57, P69, P102, P104, P112  & -\\\cline{2-6}
& \multirow{7}{*}{\textbf{Robustness}}
& Auxiliary model & Predictions difference & P62 &\\\cline{3-6}

& &Vulnerability assessment & Custom& P60 & - \\\cline{3-6}
&& Data distribution-based (Statistical) & Custom & P97, P106 &\\\cline{3-6}
& & Formal specification & Custom & P12 & - \\\cline{3-6}

& &  Machine learning & -&P11, P23, P25, P90 &  P50\\\cline{3-6}
& &  Deep learning &- & P11, P69, P75, P90 & \\\cline{2-6}
\hline
\end{tabular}
 \renewcommand{\arraystretch}{1}
\label{tab:model_monitoring_techniques2}
\end{table}

\textit{Model performance} monitoring is primarily done by \textit{comparing predictions with ground truth labels} through standard metrics like accuracy, precision, recall, mean square error, and log loss. These provide a direct measure of model correctness and are frequently reported in both formal and gray literature. In scenarios where labels are not available, uncertainty-based, auxiliary model-based, and learning-based techniques are applied. \textit{Uncertainty-based} techniques track prediction confidence scores, or model internal metrics such as entropy measures and softmax outputs to assess changes in the model’s confidence. Another technique for monitoring performance involves running an additional ML model trained for the same task, referred to as an \textit{auxiliary model}, alongside the primary model, and comparing their predictions to identify potential errors. Lastly, \textit{machine learning} and \textit{deep learning} can also be employed to predict errors in the primary ML model when labels are not available. These learning-based approaches are typically used for monitoring neural networks and were more commonly found in formal literature.

\textit{Robustness} of ML models is essential to maintain performance in the face of unexpected conditions such as adversarial inputs. Traditional monitoring approaches consist of \textit{vulnerability assessments}, \textit{auxiliary model comparisons}, \textit{formal specifications}, and \textit{data distribution comparison} using custom metrics. \textit{Machine learning} and \textit{deep learning} approaches are also often applied to assess robustness of the primary ML model. These approaches are particularly beneficial for neural networks where explicit rules or thresholds are hard to define. 

\sectopic{Operations and Infrastructure} monitoring focuses on tracking the resources and performance of ML systems. Table \ref{tab:ops_monitoring_techniques} shows the prominent techniques and metrics found in literature.

\begin{table}[htbp]
\centering
\footnotesize
\renewcommand{\arraystretch}{1.1}
\caption{Operations and Infrastructure Monitoring Techniques and Metrics}
\begin{tabular}{|c|c|>{\raggedright\arraybackslash}p{2.5cm}>{\raggedright\arraybackslash}p{3cm}p{2.7cm}p{2.7cm}|}
\hline
\textbf{} & \textbf{Monitored Aspect}  & \textbf{Monitoring Technique } & \textbf{Metrics} & \textbf{Formal Literature} & \textbf{Gray Literature}\\
\hline
\multirow{37}{*}{\rotatebox{90} {\textbf{OPERATIONS \& INFRASTRUCTURE}}}
& \multirow{12}{*}{\textbf{Infrastructure}} & System/container  & CPU usage & P4, P7, P63, P66, P101, P135 & P113, P126, P130\\\cline{4-6}
&  &resource tracking & GPU usage & P4, P135 & P113\\\cline{4-6}
& & & Memory usage & P4, P7, P63, P66, P101, P135 & P126, P130\\\cline{4-6}
& & & Memory availability & P14 &P113\\\cline{4-6}
& & & Disk input/output operations & P4, P7, P66, P101, P135 &P126 \\\cline{4-6}
& & & Network traffic & P4, P7, P14, P66, P135 &P126 \\\cline{4-6}
& & & Garbage collector statistics & P66 & -\\\cline{3-6}

& & Model resource tracking & CPU usage & P66 & -\\\cline{4-6}
& & & GPU usage & P66 & -\\\cline{4-6}
& & & Memory usage & P66 & -\\\cline{4-6}
& & & Disk input/output operations & P4, P7 & -\\\cline{4-6}
& & & Network traffic & P4, P7, P14 & -\\\cline{2-6}

& \multirow{25}{*}{\textbf{Operations}}  & System performance tracking & Inference latency & P4, P19, P63, P66, P101, P135 & P113, P116, P119, P120, P132\\\cline{4-6}
& & & Mean inference latency & P66 & -\\\cline{4-6}
& & & Tail inference latency & P66 & -\\\cline{4-6}
& & & Throughput & P63 & P113, P120, P132\\\cline{4-6}
& & & Predictions per second & - &P132\\\cline{4-6}
& & & Predictions per day &- &P116\\\cline{4-6}
& &  & Inference requests per day  & P64 & P116, P119, P120, P128 \\\cline{4-6}
& & & Model serving time& P4, P14 &-\\\cline{4-6}
& &  & Model training time & P4 &-\\\cline{4-6}
& &  & Uptime  & P101, P135 &-\\\cline{4-6}
& &  & Downtime  & P135 &-\\\cline{4-6}
& &  & Service health status & P101 &P130\\\cline{3-6}

& & Release and recovery & Deployment frequency & P135 &-\\\cline{4-6}
&  &  tracking & Mean time to recovery &  P135 &-\\\cline{4-6}
& & & Model retraining frequency &  P135 &-\\\cline{4-6}
& & & Model rollback rate &  P135 &-\\\cline{4-6}
& & & Deployment success rate &  P135 &-\\\cline{3-6}

& & System age tracking & Model age & P81 &-\\\cline{4-6}
& & & Model age at each pipeline stage & P81 & -\\\cline{4-6}
& & & Feature computation tables/processes age & P81 & - \\\cline{3-6}
& & Error tracking & Failed HTTP responses, timeouts & P4, P135 & P113\\\cline{2-6}
\hline

\end{tabular}
 \renewcommand{\arraystretch}{1}
\label{tab:ops_monitoring_techniques}
\end{table}

\textit{Infrastructure monitoring} of an ML system consists of hardware and runtime \textit{resource tracking}, which can be measured at either the system, container, or model level. This includes tracking CPU and GPU usage, memory consumption, disk input/output operations, and network traffic. Some studies also track metrics such as garbage collection and memory availability, especially in high-load environments. While resource tracking is more commonly performed at the system or container level, model level monitoring can provide insights into the resource usage of specific ML workloads. Overall, infrastructure monitoring enables early detection of resource bottlenecks and supports maintaining the health and stability of the entire system.

\textit{Operations monitoring} emphasizes the runtime behavior and reliability of ML systems by \textit{tracking system performance}. Key metrics include inference latency, throughput, prediction rates, inference frequency, requests per day, and uptime. \textit{Tracking release and recovery} cycles of ML systems also provides operational insights such as deployment frequency, model retraining frequency, model rollback rate, and mean recovery time. Additionally, \textit{system age} and \textit{error} tracking cover metrics such as model age, failed requests, and timeouts. System performance tracking, especially inference latency, is frequently reported in studies, while release and recovery, system age, and error tracking were underrepresented.

\sectopic{Responsible ML} monitoring covers ensuring fairness, interpretability, safety, sustainability, and trust in ML systems. While other aspects of responsible ML exist, these were the ones mentioned in the included studies. The main techniques and metrics are summarized in Table \ref{tab:respml_monitoring_techniques}.

\begin{table}[htbp]
\centering
\footnotesize
\renewcommand{\arraystretch}{1.1}
\caption{Responsible ML Monitoring Techniques and Metrics}
\begin{tabular}{|c|c|>{\raggedright\arraybackslash}p{2.5cm}>{\raggedright\arraybackslash}p{3cm}p{2.7cm}p{2.7cm}|}
\hline
\textbf{} & \textbf{Monitored Aspect}  & \textbf{Monitoring Technique } & \textbf{Metrics} & \textbf{Formal Literature} & \textbf{Gray Literature}\\
\hline
\multirow{53}{*}{\rotatebox{90}{\textbf{RESPONSIBLE ML}}}

& \multirow{29}{*}{\textbf{Fairness}} & Outcome parity & Disparate impact &  P18, P62 & P114, P120\\\cline{4-6}
& & & Statistical parity difference& P17, P80 & P120\\\cline{4-6}
& & & Demographic parity difference & P62, P135 &-\\\cline{4-6}
& & & Difference in acceptance/rejection rates & P18 &-\\\cline{4-6}
& & & Difference in positive observed labels  & P18 & -\\\cline{4-6}
& & & Custom  & P48 & -\\\cline{3-6}

& & Performance parity & Equal opportunity & P17, P135 & P120\\\cline{4-6}
& & & Accuracy difference & P18, P46 & P113\\\cline{4-6}
& & & Recall difference & P18 & -\\\cline{4-6}
& & & Area under curve difference & P44 & -\\\cline{4-6}
& & & Equalized odds & P62, P135 & P117, P120\\\cline{4-6}
& & & Treatment equality & P18 & -\\\cline{4-6}
& & & Error rate difference &  & P120\\\cline{4-6}
& & & False discovery rate &  & P120\\\cline{4-6}
& & &Positive predictive performance difference & P45 & P117 \\\cline{4-6}
& & &Negative predictive performance difference & P45 & -\\\cline{4-6}
& & & Difference in conditional acceptance/rejection & P18 &- \\\cline{4-6}
& & & Conditional demographic disparity in labels & P18 &- \\\cline{3-6}

& & Prediction distribution-based & Quantile demographic drift & P22& -\\\cline{3-6}

& & Counterfactuals & Custom & P21& -\\\cline{3-6}
& & Custom & Custom & & P134\\\cline{2-6}
&  \multirow{5}{*}{\textbf{Interpretability}} & Feature attribution & SHAP score& P17, P18, P19, P21, P22, P37 &P50, P114, P118, P128, P133\\\cline{4-6}
& & & LIME score & P17, P19, P135 &  P50, P114\\\cline{4-6}
& & & Feature importance score & P15, P64, P135 &  P114\\\cline{4-6}
& & & Custom & P26 & - \\\cline{3-6}
& & Counterfactuals & Custom & P21, P135 & P50 \\\cline{2-6}

& \multirow{3}{*}{\textbf{Privacy}} & Tracking data exposure risk & Membership privacy risk score & P80 &-\\\cline{3-6}
& & Machine learning & - & P24, P39 & - \\\cline{2-6}

& \multirow{9}{*}{\textbf{Safety}} & Rule-based & - &  P9, P12, P29, P83, P90, P107& P67\\\cline{3-6}
&& Data distribution-based (Statistical) & Kolmogorov–Smirnov statistic & P32 & \\\cline{3-6}
&& Uncertainty-based & Softmax mean & P58 & \\\cline{4-6}
&&& Softmax std. dev & P58 & \\\cline{3-6}
& & Machine learning & - & P11, P23, P24, P76, P77, P79, P90, P95& P109\\\cline{3-6} 
& & Deep learning & - & P11, P61, P90, P91, P94, P98& \\\cline{2-6} 

& \multirow{3}{*}{\textbf{Sustainability}} & Environmental impact & Energy efficiency & P135 & -\\\cline{4-6}
& & tracking & Carbon footprint & P135 & -\\\cline{4-6}
& & & E-waste & P135 & -\\\cline{2-6}

& \multirow{4}{*}{\textbf{Trust}} & Uncertainty-based & Predictive variance & P13 & -\\\cline{3-6}
& & Trust Score Aggregation& Custom & P21, P28 & - \\\cline{3-6}
& & Vulnerability assessment & Custom & P60 & - \\\cline{3-6}
& & Deep learning & - & P78 & - \\
\hline
\end{tabular}
 \renewcommand{\arraystretch}{1}
\label{tab:respml_monitoring_techniques}
\end{table}

For \textit{fairness} monitoring, the most frequently used techniques are outcome parity and performance parity. \textit{Outcome parity} measures the difference in prediction outcomes between groups using metrics such as disparate impact, statistical parity difference, and demographic parity difference. \textit{Performance parity} evaluates whether performance levels are consistent between groups through metrics like equal opportunity, equalized odds, and differences in accuracy, recall, or area under the curve. Performance parity measures require ground truth labels for comparison, whereas outcome parity measures do not. More specialized techniques, such as prediction distribution-based and counterfactuals, are used in a smaller number of studies. \textit{Prediction distribution-based} fairness checks compare the distribution of predictions across demographic groups, and \textit{counterfactual} fairness checks test if a model’s prediction would remain the same if a sensitive attribute was altered while keeping all others constant.

\textit{Interpretability} monitoring often relies on \textit{feature attribution} methods to track changes in model reasoning over time. Common metrics include LIME scores, SHAP scores, and feature importance scores that help identify how the model weighs different features. Custom metrics are also proposed in one study; however, SHAP scores were most frequently used in both formal and gray literature. \textit{Counterfactual} explanations are also used to assess whether small changes in the inputs still produce correct predictions.

\textit{Privacy} monitoring focuses on \textit{tracking data exposure risks}, with membership privacy risk scores used to detect potential leakage of training data in model outputs. \textit{Machine learning-based} detection methods are also employed to analyze model inputs to identify privacy attacks. Privacy monitoring approaches for ML systems were scarce in the literature, as most privacy-preserving measures are implemented at design time.

\textit{Safety} monitoring aims to ensure the safe operation of ML systems through rule-based, data distribution-based, uncertainty-based, and learning-based techniques. \textit{Rule-based} methods enforce predefined constraints on the model outputs, such as triggering brakes when an object is detected within the defined safety distance. These rules can be defined in natural language, code, or formal logic. \textit{Data distribution-based} methods use statistical measures like the Kolmogorov–Smirnov statistic to detect shifts in input distributions that could signal unsafe conditions. \textit{Uncertainty-based} approaches identify low-confidence predictions with metrics such as softmax mean and softmax standard deviation. \textit{Machine learning} and \textit{deep learning} based methods were leveraged frequently to detect unsafe behaviors, particularly in cyber-physical systems with neural networks.   

\textit{Sustainability} monitoring \textit{tracks the environmental impact} of ML systems using metrics such as energy efficiency, carbon footprint, and e-waste. While particularly relevant for large-scale ML systems, such monitoring was rarely reported in the literature.

\textit{Trust} monitoring aims to maintain the confidence of stakeholders in the ML system. Approaches include \textit{uncertainty-based} estimation of predictive variance, \textit{trust score aggregation} from multiple other performance and robustness metrics, and \textit{vulnerability assessment} to identify weaknesses in the system. \textit{Deep learning} based runtime assurance methods are also used to build trust in the ML system.

\sectopic{Business} monitoring aims to measure customer satisfaction, financial performance, and alignment with business objectives of the ML system. The techniques and metrics for business monitoring are shown in Table \ref{tab:bus_monitoring_techniques}. Although business aspects are crucial for organizational success, they were rarely reported in the literature, possibly because their measurement is considered straightforward. 

\begin{table}[ht]
\centering
\footnotesize
\renewcommand{\arraystretch}{1.1}
\caption{Business Monitoring Techniques and Metrics}
\begin{tabular}{|c|c|>{\raggedright\arraybackslash}p{2.5cm}>{\raggedright\arraybackslash}p{3cm}p{2.7cm}p{2.7cm}|}
\hline
\textbf{} & \textbf{Monitored Aspect}  & \textbf{Monitoring Technique } & \textbf{Metrics} & \textbf{Formal Literature} & \textbf{Gray Literature}\\
\hline
\multirow{11}{*}{\rotatebox{90}{\textbf{BUSINESS}}}
& \multirow{8}{*}{\textbf{Business Value}} & Business value tracking & User retention rate & P135  & -\\\cline{4-6}
& & & User conversion rate & P135 & -\\\cline{4-6}
& & & Return on investment & P135 & -\\\cline{4-6}
& & & Total development cost & P135 & -\\\cline{4-6}
& & & Model training cost & P135 & -\\\cline{4-6}
& & & Maintenance cost & P135 & -\\\cline{4-6}
& & & Data acquisition and storage cost & P135 & -\\\cline{2-6}

& \multirow{3}{*}{\textbf{User Engagement}} & User behavior tracking & Usage frequency & P135 & -\\\cline{4-6}
& & & Session length & P135 & -\\\cline{4-6}
& & & User satisfaction score & P135 & -\\\cline{4-6}
\hline
\end{tabular}
 \renewcommand{\arraystretch}{1}
\label{tab:bus_monitoring_techniques}
\end{table}

\textit{Business value} is measured by tracking metrics such as user retention rate, user conversion rate (e.g., from trial to subscription), and return on investment. It also included financial metrics such as total development cost, training cost, maintenance cost, and data storage and acquisition costs. 

\textit{User engagement} is measured by tracking behavioral metrics such as usage frequency, session length, and user satisfaction scores, which provide insights into how actively and positively user interact with the system. 

\subsubsection{Monitoring Tools}
\begin{figure}
    \centering
    \includegraphics[width=0.7\linewidth]{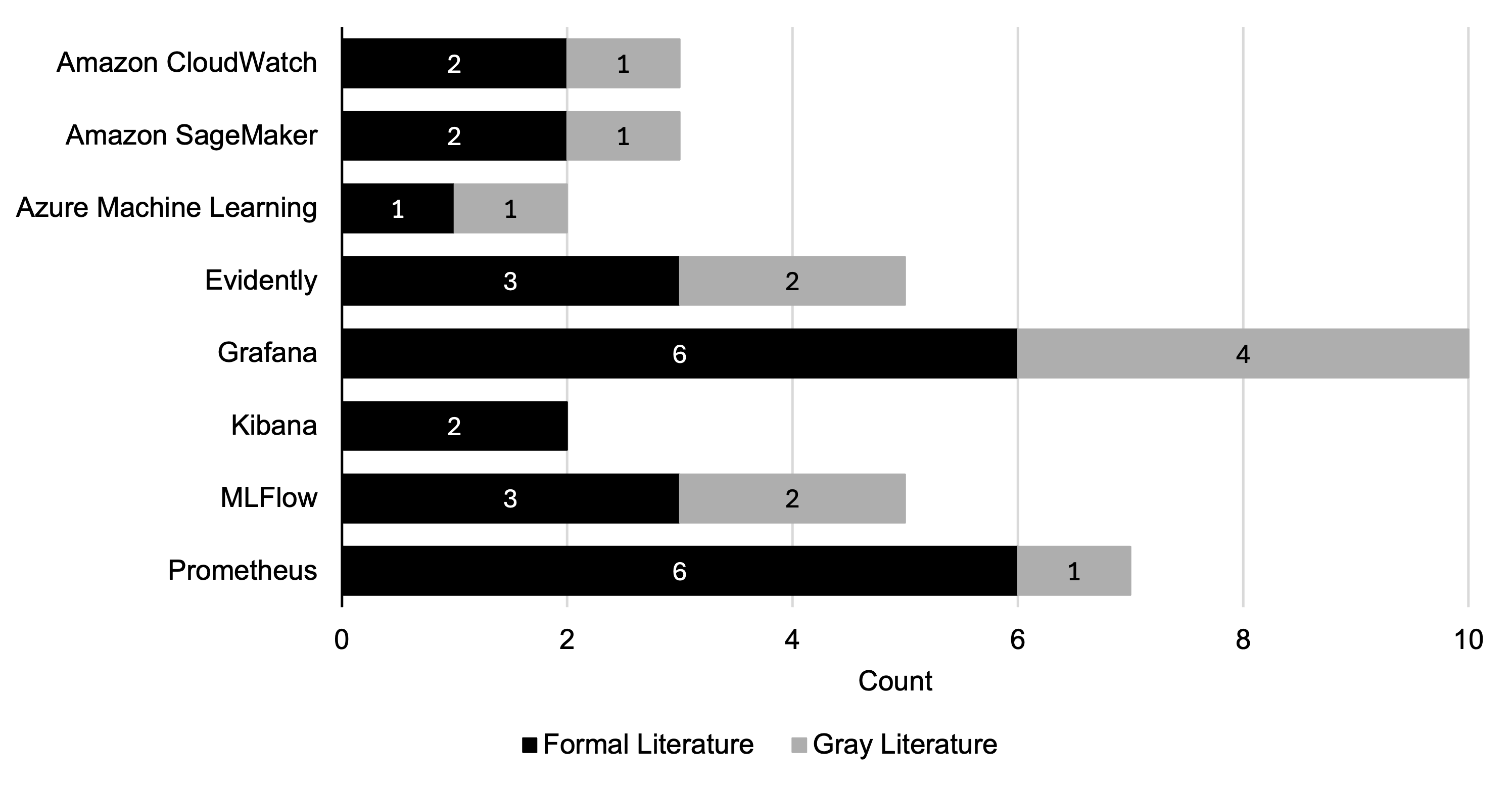}
    \caption{Frequently Mentioned Tools}
    \label{fig:tools}
\end{figure}

Fig. \ref{fig:tools} shows frequently mentioned monitoring tools in the selected studies. Grafana~\cite{grafanaobservability}, Prometheus~\cite{prometheus}, Evidently~\cite{evidently2025monitoring}, and MLflow~\cite{mlflowtracking} emerge as the most popular tools due to their capabilities of interactive visualizations, metrics collection, ML-specific metrics, and ML lifecycle observability, respectively. Grafana is often preferred because it easily integrates with multiple data sources and creates sophisticated, interactive dashboards for visualizing monitoring data (P66, P93). Prometheus provides efficient, reliable, and scalable metrics collection, alerting, and a query language for selecting and aggregating monitoring data in real-time (P8, P49). Evidently, on the other hand, provides monitoring features tailored to ML systems, such as detecting model performance degradation and data drift (P80, P132). Similarly, MLFlow also provides ML-specific monitoring by tracking relevant metrics (P1, P50). A notable characteristic of these four tools is that they are all open source. Another open source tool mentioned is Kibana~\cite{kibana}, which offers capabilities for log analysis, interactive visualizations, and alerting (P66, P101). Monitoring tools offered by cloud providers such as Amazon CloudWatch~\cite{awscloudwatch}, Amazon SageMaker~\cite{awssagemaker}, and Azure Machine Learning~\cite{azuremlmonitoring} were also mentioned for their scalability, integration with cloud ecosystems, and support for alerting, visualization, and logging. 

\begin{table}[htp]
\centering
\caption{Monitoring Tools}
\label{tab:monitoring-tools}
\small

\setlength{\tabcolsep}{3.5pt}  
\begin{tabular}{>{\raggedright\arraybackslash}p{3cm}>{\raggedright\arraybackslash}p{3.5cm}p{1cm}p{1cm}p{1cm}p{1cm}|p{1cm}p{1cm}p{1cm}}
\toprule
\textbf{Distribution Model} & \textbf{Tool} & \rotatebox{90}{\textbf{Data}} & \rotatebox{90}{\parbox{2cm}{\raggedright\textbf{ML Model Behavior}}} & \rotatebox{90}{\parbox{2cm}{\raggedright\textbf{Operations \& Infrastructure}}} & \rotatebox{90}{\textbf{Responsible ML}} & \rotatebox{90}{\textbf{Alerting}} & \rotatebox{90}{\textbf{Visualization}} & \rotatebox{90}{\textbf{Log Processing}} \\
\midrule
\multirow{1}{=}{\textbf{Cloud Provider}} 
& Amazon CloudWatch~\cite{awscloudwatch} & & & $\checkmark$ & & $\checkmark$ & $\checkmark$ & $\checkmark$ \\
& Amazon SageMaker~\cite{awssagemaker} & $\checkmark$ & $\checkmark$ & $\checkmark$ & $\checkmark$ & $\checkmark$ & $\checkmark$ & $\checkmark$ \\
& Azure Machine Learning~\cite{azuremlmonitoring} & $\checkmark$ & $\checkmark$ & $\checkmark$ &  & $\checkmark$ & $\checkmark$ & $\checkmark$ \\
& Azure Monitor~\cite{azure_monitor_overview} &  &  & $\checkmark$ &  & $\checkmark$ & $\checkmark$ & $\checkmark$ \\
& Oracle ML Monitoring~\cite{oracle_ml_monitoring} & $\checkmark$ & $\checkmark$ & &  & $\checkmark$ & $\checkmark$ & \\
& Vertex AI~\cite{google_vertex_ai_model_monitoring} & $\checkmark$ & $\checkmark$ & $\checkmark$ & $\checkmark$ & $\checkmark$ & $\checkmark$ & $\checkmark$ \\
\midrule
\multirow{1}{=}{\textbf{Enterprise}}
& Arthur AI~\cite{arthur_ai_observability} & $\checkmark$ & $\checkmark$ & & $\checkmark$ & $\checkmark$ & $\checkmark$ & \\
& Censius~\cite{censius_ai_monitoring} & $\checkmark$ & $\checkmark$ & & $\checkmark$ & $\checkmark$ & $\checkmark$ & \\
& Comet~\cite{comet_model_production_monitoring} & $\checkmark$ & $\checkmark$ & & $\checkmark$ & $\checkmark$ & $\checkmark$ & \\
& DataDog~\cite{datadog_product_overview} & $\checkmark$ & & $\checkmark$ & & $\checkmark$ & $\checkmark$ & $\checkmark$ \\
& DataRobot~\cite{datarobot_ai_observability} & $\checkmark$ & $\checkmark$ & $\checkmark$ &  & $\checkmark$ & $\checkmark$ & $\checkmark$ \\
& Datatron~\cite{datatron_ai_monitoring_governance} & $\checkmark$ & $\checkmark$ & & $\checkmark$ & $\checkmark$ & $\checkmark$ & $\checkmark$\\
& Great Expectations~\cite{greatexpectations}& $\checkmark$ & & & & $\checkmark$ & & \\
& H2O.ai~\cite{h2o_mlops} & $\checkmark$ & $\checkmark$ & $\checkmark$ & $\checkmark$ & $\checkmark$ & $\checkmark$ & \\

& IBM Watson~\cite{ibm_watson_studio} & $\checkmark$ & $\checkmark$ & & $\checkmark$ & $\checkmark$ & $\checkmark$ & \\
& Iguazio~\cite{iguazio_model_monitoring} & $\checkmark$ & $\checkmark$ &  &  & $\checkmark$ & $\checkmark$ &  \\
& KNIME~\cite{knime2021monitoring} & $\checkmark$ & $\checkmark$ & & & & $\checkmark$ & \\
& Mona~\cite{mona_ml_monitoring} & $\checkmark$ & $\checkmark$ & & $\checkmark$ & $\checkmark$ & $\checkmark$ & $\checkmark$ \\
& NannyML~\cite{nannyml}& $\checkmark$ & $\checkmark$ & & & $\checkmark$  & $\checkmark$ & $\checkmark$\\
& Seldon Core  & $\checkmark$ & $\checkmark$ & $\checkmark$ & $\checkmark$ & $\checkmark$ & $\checkmark$ & \\
& Superwise~\cite{superwise_ml_monitoring} & $\checkmark$ & $\checkmark$ & & $\checkmark$ & $\checkmark$ & $\checkmark$ & $\checkmark$ \\
& Valohai~\cite{valohai_model_monitoring}& & $\checkmark$ & $\checkmark$ & & $\checkmark$ & $\checkmark$ & \\

\midrule
\multirow{1}{=}{\textbf{Open Source}}
& AlerTiger~\cite{xu2023alertiger} & $\checkmark$ & $\checkmark$ & & & $\checkmark$ & $\checkmark$ & \\
& Alibi Detect~\cite{alibidetect}& $\checkmark$ &  & & $\checkmark$ & & & \\
& Apache JMeter~\cite{apachejmeter}& & & $\checkmark$ & &  & $\checkmark$ & \\
& Docker Stats~\cite{dockerstats} & & & $\checkmark$ & & & & \\
& Doubt~\cite{doubt} &  & $\checkmark$ & &  & & & \\
& Evidently~\cite{evidently2025monitoring} & $\checkmark$ & $\checkmark$ & & $\checkmark$ & $\checkmark$ & $\checkmark$ & \\
& Explanation Space~\cite{explanationspace} & & & & $\checkmark$ & & & \\
& Fluentbit~\cite{fluentbit} & & & $\checkmark$ & & & & $\checkmark$ \\
& Google Heapster~\cite{heapster} & & & $\checkmark$ & &  & & \\
& Grafana~\cite{grafanaobservability} & & & & & $\checkmark$ & $\checkmark$ & \\

& Kibana~\cite{kibana} &  & & $\checkmark$ & & $\checkmark$ & $\checkmark$ & $\checkmark$ \\

& Logstash~\cite{logstash} & & & & & & & $\checkmark$ \\
& MLFlow~\cite{mlflowtracking}& & $\checkmark$ & & & & $\checkmark$ & \\
& MLRun~\cite{mlrun} & $\checkmark$ & $\checkmark$ & $\checkmark$ & & $\checkmark$ & $\checkmark$ & \\

& Prometheus~\cite{prometheus} & & & $\checkmark$ & & & $\checkmark$ & \\
& PyOD~\cite{pyod} & $\checkmark$ & & & & & & \\
& Redash~\cite{redash} & & & & &$\checkmark$ & $\checkmark$ & \\
& skshift~\cite{skshift} & $\checkmark$ & & & & & & \\
& strucchange R~\cite{strucchange} &  & $\checkmark$ & & & & & \\
& Telegraph~\cite{telegraf} & & & $\checkmark$ & &  & &  \\
& TensorBoard~\cite{tensorboard} & & $\checkmark$ & & & & $\checkmark$ & \\
& Uncertainty Wizard~\cite{uncertaintywizard} & & $\checkmark$ & & & & & \\
& Verifai~\cite{verifai} & & $\checkmark$ & &  & & & \\
& Vertier~\cite{vetiver2025monitoring} &  & $\checkmark$ & &  & & $\checkmark$ & \\
\bottomrule
\end{tabular}
\renewcommand{\arraystretch}{1}
\end{table}
\normalsize

Table \ref{tab:monitoring-tools} lists all the monitoring tools found in the selected studies along with their monitoring capabilities. The diversity of tools highlights the multifaceted nature of monitoring ML systems. We have categorized these tools in terms of their distribution model, \textit{cloud provider}, \textit{enterprise}, or \textit{open source}. Tools classified as cloud provider are offered by companies whose primary business is cloud infrastructure, and monitoring is a part of their broader cloud ecosystem. Some examples of such tools are Amazon SageMaker (P17, P113) and Vertex AI (P118). Tools under the enterprise category are from companies whose primary business is software solutions for monitoring, such as Arthur AI (P114) and NannyML (P124). Lastly, tools categorized as \textit{open source} are developed and maintained by communities, and are free to use and distribute. For instance, Alibi Detect (P50), MLRun (P132), and Vertier (P127) are open source tools.

Table \ref{tab:monitoring-tools} also indicates which aspects of an ML system are monitored by each tool — data, ML model behavior, operations and infrastructure, and responsible ML—as well as additional monitoring capabilities such as alerting, visualization, and log processing. 

Cloud provider tools offer comprehensive coverage across all monitored aspects and capabilities for ML systems. These tools are integrated within their respective cloud platform ecosystems, enabling seamless monitoring of ML systems deployed on the same infrastructure. Nearly all of these tools support alerting, visualizations, and log processing, while also including monitoring functionalities for data drift, model performance, and system operations. Vertex AI (P118), for example, covers all four aspects by monitoring data drift, label drift, resource usage, and interpretability. It also supports customized alerting, visualizations, and logging, making it a complete monitoring solution for ML systems deployed on Google Cloud~\cite{googlecloud}.

Enterprise tools often focus more on monitoring ML model behavior, data, and responsible ML concerns such as fairness and interpretability. A few tools also monitor operations and infrastructure metrics like DataDog (P49), but most stand out for offering advanced data and model monitoring. For example, tools such as Mona (P129) and Superwise (P131) can monitor data quality issues, anomalies, data drift, label drift, and model performance through a wide range of metrics. Customized alerts and visualization dashboards are another common feature among enterprise tools. Domain-specific monitoring solutions are also offered by some tools, like Censius (P128) provides monitoring solutions tailored to healthcare, finance, manufacturing, and more.

Open source tools vary significantly in monitoring focus and capabilities. Tools such as MLflow (P1, P50) and MLRun (P132) provide model performance tracking, while tools like Prometheus (P8, P49) and Fluentbit (P66) are adopted for operations and infrastructure monitoring of ML systems. Open source libraries like Evidently (P8, P117) and Alibi Detect (P50) offer more specialized capabilities such as data drift and anomaly detection. For interactive visualizations and monitoring dashboards, tools like Grafana (P27, P62) and Kibana (P66) can be paired with others to enhance observability. While open source tools require extra integration effort compared to enterprise or cloud provider tools, they are more flexible, transparent, and customizable. 

A key observation from the table is that operations and infrastructure monitoring is often not available in tools dedicated for ML systems, instead, dedicated resource monitoring tools fill this gap. Another observation is the uneven distribution of support for responsible ML monitoring. While many enterprise solutions offer fairness analysis and explainability, these features are rarely found in cloud providers and open source tools. We also observed that open source tools tend to be narrower in their scope, whereas cloud provider and enterprise tools usually have broader scope. Lastly, log processing was underrepresented, particularly among open source tools.

\begin{figure}[htbp]
    \centering
    \begin{subfigure}{0.48\textwidth}
        \centering
         \includegraphics[width=1\linewidth]{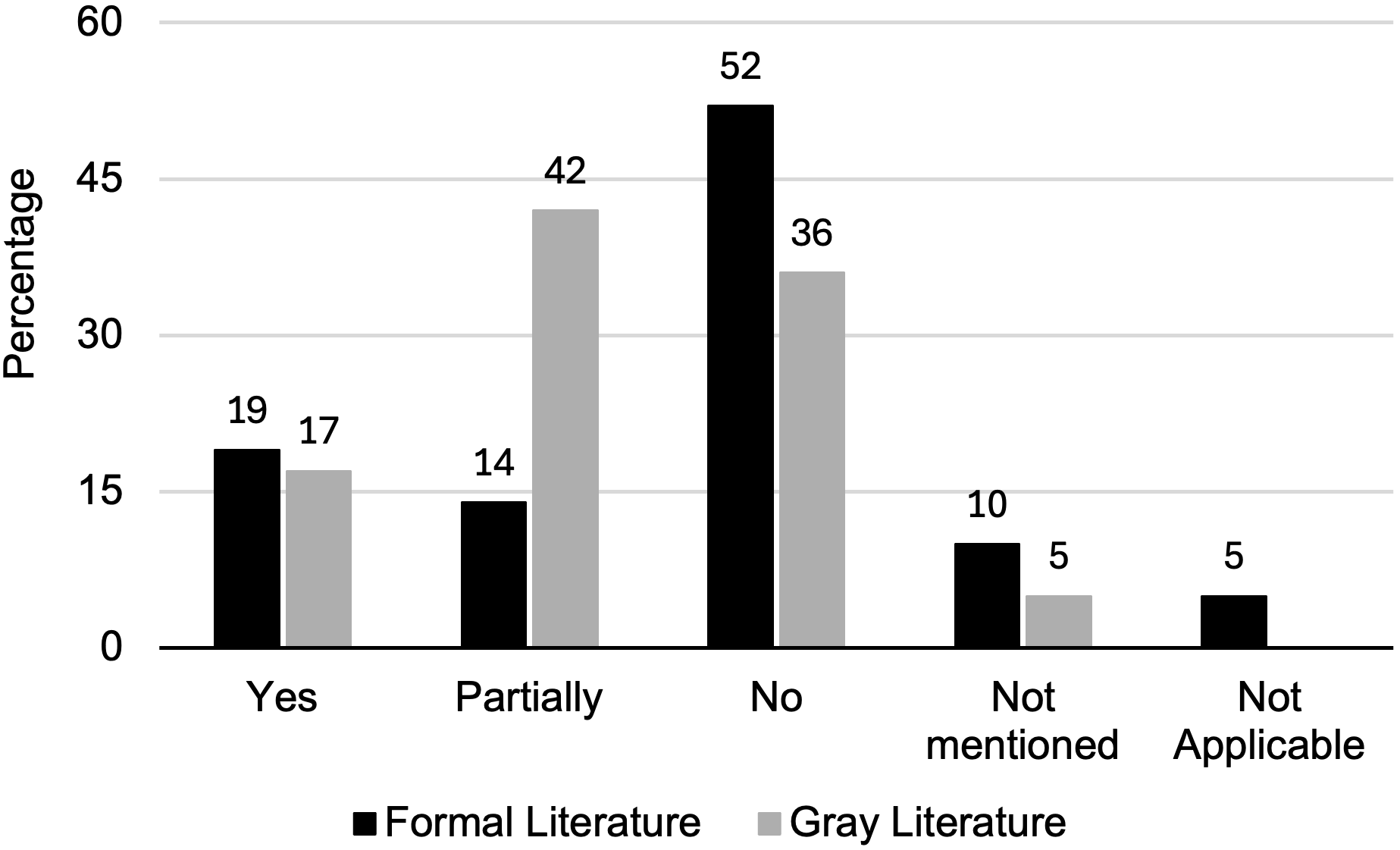}
        \caption{Ground Truth Dependency}
        \label{fig:labels}
        
    \end{subfigure}
    \hfill
    \begin{subfigure}{0.48\textwidth}
        \centering
         \includegraphics[width=1\linewidth]{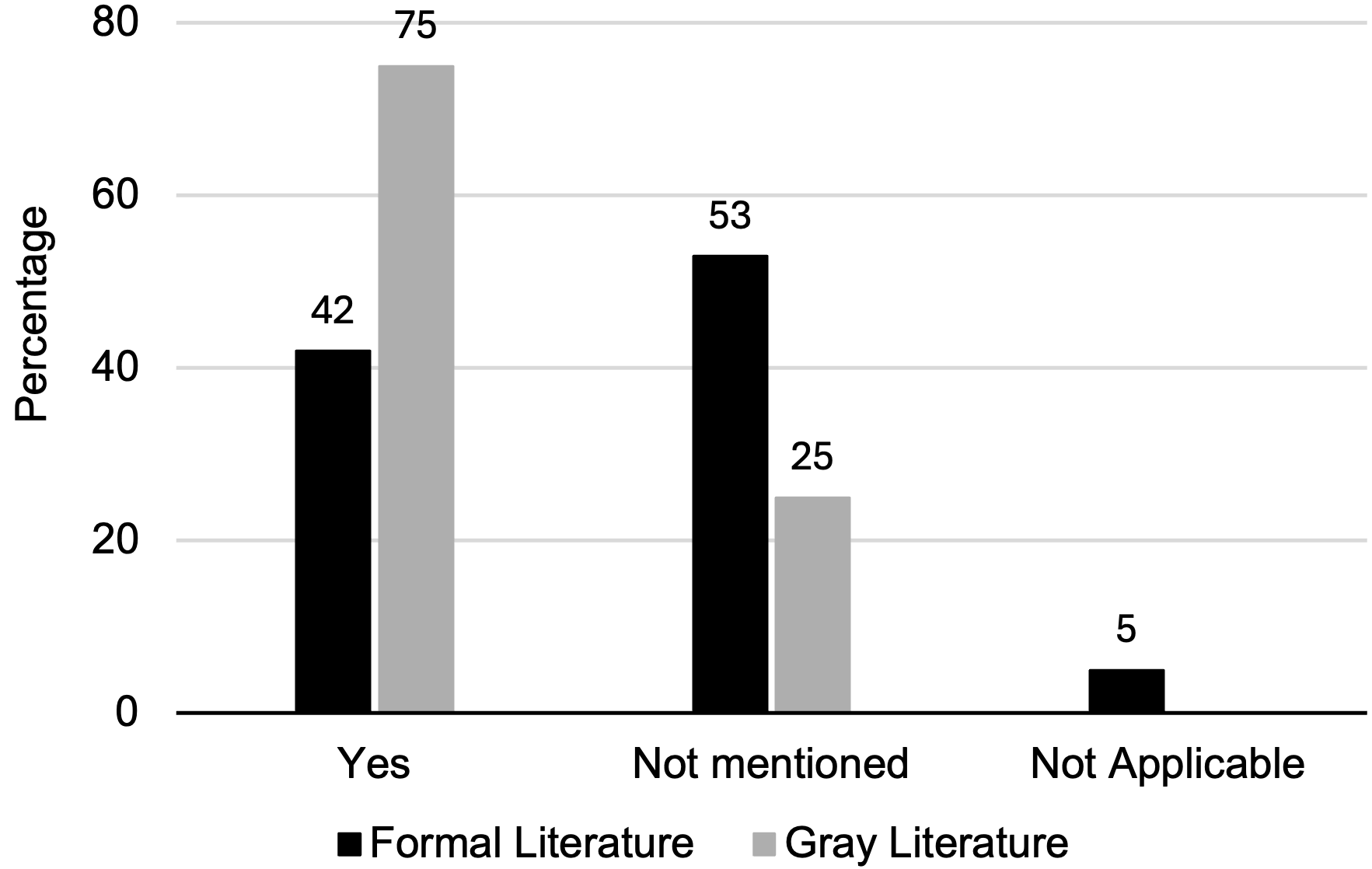}
        \caption{Logging Capabilities}
        \label{fig:logs}
    \end{subfigure}
    \caption{Characteristics of Monitors}
\end{figure}

\subsubsection{Ground Truth Dependency} Ground truth refers to the actual outcomes that an ML model attempts to predict. Certain monitoring techniques rely on ground truth labels at runtime to assess model performance or other aspects of the ML system. However, since ground truth labels are often not available in real time, we analyzed the included studies for such a dependency. Our findings are shown in Fig. \ref{fig:labels}. Interestingly, nearly half the studies \textit{do not} depend on ground truth for monitoring (P70, P124). Out of these, 52\% were from formal literature and 36\% from gray literature. A reasonable portion of studies, mostly gray literature (42\%), require ground truth \textit{partially}, meaning some labels but not all (P34, P46). In comparison, studies that require \textit{all labels} for monitoring were less common (P15, P136). A small portion of studies did \textit{not mention} clearly in the papers if ground truth data was necessary for their monitoring solutions (P101, P135). Lastly, there are some studies that do not propose a monitoring solution, instead, they focus on identifying current challenges in the field. We categorize these papers as \textit{not applicable} (P35, P43).

\subsubsection{Logging Capabilities} This section describes features of monitoring approaches to log data. As shown in Fig. \ref{fig:logs}, half of the studies \textit{support logging} of monitored data and events (P8, P110), nearly half of the studies do \textit{not mention} any such functionality (P13, P14), and a small portion of studies do not propose a monitoring approach, thus, we classify them as \textit{not applicable} (P31, P40). A significantly larger proportion of gray literature (+33\%) provided logging functionalities with their monitoring solutions, whereas this was missing in most formal literature papers. This might be because it was considered trivial in academic contexts that tend to focus more on complex monitoring problems; however, the significance of logs in an industrial context is well established.

\begin{center}
\begin{myframe}[width=48em,top=5pt,bottom=5pt,left=5pt,right=5pt,arc=10pt,auto outer arc,title=\centering\textbf{RQ2 Answer Summary}]

Frequently monitored aspects in ML systems include model performance, data drift, and anomalies, with similar distributions across formal and gray literature. Less frequent aspects, such as privacy, trust, and sustainability, were only found in formal literature. Statistical tests and distance metrics were the most prevalent techniques for comparing data distributions and detecting drifts. For model performance monitoring, ground truth labels are often leveraged to measure correctness, while system performance monitoring focuses on tracking resources and system operations. Learning-based methods were often employed to detect anomalies, safety violations, and model robustness. Frequently mentioned monitoring tools include Grafana, Prometheus, Evidently, and MLflow, all of which are open source and address specific aspects of ML monitoring. Cloud providers and enterprise monitoring tools typically have a broader scope, covering multiple aspects of ML monitoring, while open source tools are significantly narrower in scope. The majority of approaches do not require ground truth data for monitoring and provide functionality to log monitored data.
\end{myframe}
\end{center}

\subsection{RQ3. What are the benefits of the proposed monitoring solutions for ML systems?}
\subsubsection{Contributions}

\begin{figure}
    \centering
\includegraphics[width=0.9\linewidth]{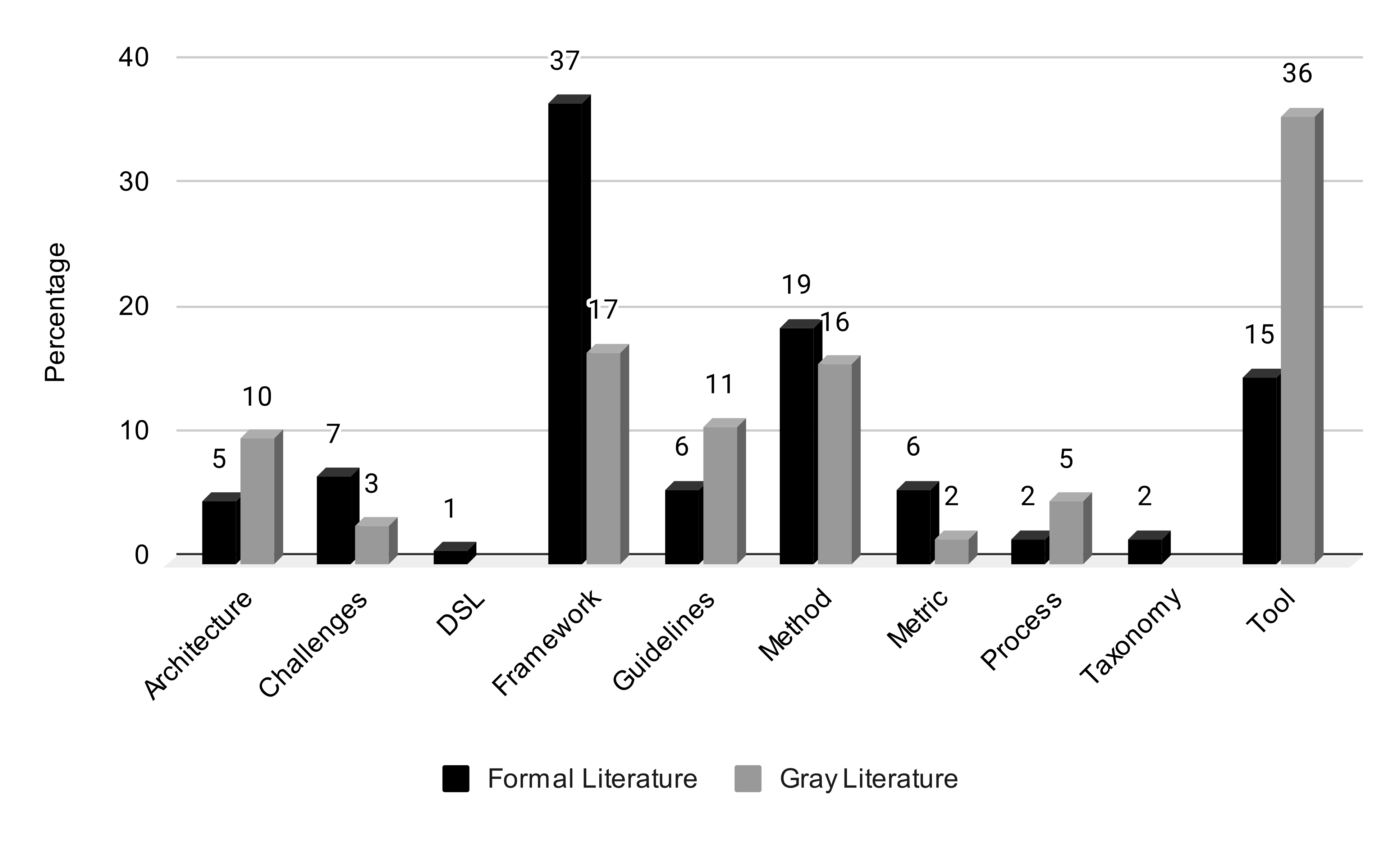}
    \caption{Study Contributions}
    \label{fig:contributions}
\end{figure}

We identified ten unique contributions from the included ML monitoring literature, shown in Fig. \ref{fig:contributions}. A monitoring \textit{framework} is by far the most common contribution, appearing significantly more in formal literature (+20\%) than gray literature. The opposite is observed for the second most common contribution, a monitoring \textit{tool}, which is more prevalent in gray literature (+21\%). A difference in priorities is evident between formal and gray literature, with the former focusing on high-level solutions (P13, P59) and the latter on concrete tools (P128, P129). Another frequent contribution is a monitoring \textit{method} describing specific steps to detect a runtime issue (P73, P106). A number of studies present monitoring \textit{guidelines} and \textit{challenges} based on previous experience, surveys and interviews of practitioners, and general observations (P31, P35). Few proposed an \textit{architecture} (P99, P117) or a novel monitoring \textit{metric} (P48, P86). The least common contributions were \textit{process} (P1, P2), \textit{taxonomy} (P24, P31), and \textit{domain-specific language (DSL)} (P5). A possible explanation for this is that most studies prioritize proposing generic, modular frameworks instead of workflows or processes. Taxonomies are only presented in works that aim to semantically structure ML monitoring literature, which is less common. DSLs, on the other hand, are designed to abstract and automate monitoring development, thereby reducing monitoring effort and technical expertise required for monitoring - an objective not shared by many studies.

\subsubsection{Study Type}
Following the classification scheme proposed in~\cite{petersen2008systematic}, we organize the selected studies into six categories based on their research approach. The results are tabulated in Table \ref{tab:study-type}, describing each research approach and the corresponding studies. 
\begin{table}
    \centering
    \small
    \caption{Study Type}
     \renewcommand{\arraystretch}{1.2}
    \begin{tabular}{>{\raggedright\arraybackslash}p{2.5cm}>{\raggedright\arraybackslash}p{3.5cm}>{\raggedright\arraybackslash}p{3cm}>{\raggedright\arraybackslash}p{3cm}c}
    \hline
        \textbf{Study Type} & \textbf{Examples} & \textbf{Formal Literature} & \textbf{Gray Literature} & \textbf{Percentage}\\
        \hline
        \textbf{Validation Research} & Lab experiments, academic case study (with students), simulation, prototyping, mathematical analysis, mathematical proof of properties, etc & P1, P2, P4, P8, P9, P11, P12, P14, P17, P18, P20, P21, P22, P25, P29, P32, P33, P36, P37, P38, P39, P41, P42, P44, P45, P46, P47, P48, P58, P59, P61, P69, P71, P72, P73, P75, P76, P77, P78, P79, P82, P83, P85, P86, P87, P88, P90, P91, P92, P94, P95, P96, P97, P98, P99, P102, P103, P104, P105, P106, P107, P112 & P52, P54, P67, P74, P84, P93, P109, P110, P134, P136 & 52.9\%\\
        \hline
        \textbf{Solution Proposal}  & No evaluation, proof-of-concept as small example or argument & P7, P13, P19, P24, P63, P66, P68, P70, P80, P111 & P55, P113, P114, P115, P116, P117, P118, P119, P120, P121, P122, P123, P124, P125, P126, P127, P128, P129, P130, P131, P132, P133 & 23.5\% \\
        \hline
        \textbf{Evaluation Research}  & Industrial evaluation, experiment/ survey with practitioners, ethnography, real-world project &  P3, P5, P6, P10, P15, P26, P30, P34, P35, P43, P49, P56, P57, P60, P89, P101, P108 & P65 & 13.2\%\\
        \hline
        \textbf{Philosophical Papers}  & Taxonomy or conceptual framework & P16, P23, P28, P40, P62, P100, P135 & P50, P51, P53 & 7.4\%\\
        \hline
        \textbf{Experience Papers}  & Some real-world experience & P27, P64, P81 & - & 2.2\%\\
        \hline
        \textbf{Opinion Papers} & Personal opinions & P31 & - & 0.7\%\\
        \hline
    \end{tabular}
     \renewcommand{\arraystretch}{1}
 
    \label{tab:study-type}
\end{table}

\textit{Validation research} is found to be the research type applied most frequently, particularly in formal literature (+32\%) since most researchers evaluate their proposed monitoring solutions through lab experiments and toy case studies, leaving industrial evaluation for future work. \textit{Solution proposal} is another prominent research approach, more so in gray literature (60\%) than formal literature (10\%). This disparity reflects the fast-paced, solution-oriented nature of the industry, where it is common to present practical monitoring tools or techniques without formal evaluation. For similar reasons, the majority of studies performing \textit{evaluation research} are also academic, despite the industrial focus of evaluation research. Overall, in comparison, fewer studies (\textasciitilde
13\%) perform industrial evaluations, perhaps due to the unavailability of real-world case studies. \textit{Philosophical papers} proposing conceptual frameworks or taxonomies are less common (\textasciitilde
7\%), while \textit{experience papers} and \textit{opinion papers} are scarce in formal literature and absent in gray literature.

\subsubsection{Benefits }
This section analyzes the benefits of ML monitoring approaches presented in the included studies. We categorize our findings under six broad themes shown in Table \ref{tab:benefits}. 

\begin{table}[htbp]
\centering
\caption{Benefits of Monitoring Approaches}
\label{tab:benefits}
\renewcommand{\arraystretch}{1.2}

\small
\begin{tabular}{>{\raggedright\arraybackslash}p{3cm}>{\raggedright\arraybackslash}p{3cm}p{3cm}p{3cm}c}
\hline
\textbf{Themes} & \textbf{Benefits} & \textbf{Formal Literature} & \textbf{Gray Literature} & \textbf{Percentage} \\
\hline

\multirow{2}{*}{\parbox{3cm}{\raggedright\textbf{Streamlined Operations}}}
& Ease of Implementation \& Integration & P1, P3, P4, P5, P8, P16, P19, P21, P27, P28, P32, P39, P47, P57, P60, P69, P72, P75, P91, P92, P108 & P67, P113, P115, P117, P118, P121, P122, P123, P124, P125, P130, P131, P132 & 11.9\% \\

& Cost Efficiency \& Resource Optimization & P4, P7, P14, P22, P26, P27, P36, P38, P39, P44, P46, P59, P61, P66, P73, P87, P88, P96, P102 & P74, P93, P116, P120, P128 & 8.4\% \\

& Fast Performance & P14, P15, P17, P36, P39, P46, P48, P56, P61, P63, P71, P85, P95, P102, P104 & P52, P110, P114, P115, P119, P123, P136 & 7.7\% \\

& Automation  & P2, P5, P8, P23, P59, P70, P75, P78, P80 & P53, P54 & 3.9\% \\
\hline
\multirow{2}{*}{\parbox{3cm}{\raggedright\textbf{Flexibility, Scalability, and Broad Applicability}} }
& Generalizable & P15, P21, P25, P32, P33, P39, P41, P46, P47, P72, P76, P83, P85, P86, P91, P92, P95, P98, P106, P112 & P114, P115, P117, P122, P134 & 8.8\% \\

& Scalability \& Robustness & P7, P12, P17, P18, P19, P25, P29, P30, P36, P56, P63, P94, P96, P99 & P93, P113, P116 & 6.0\% \\

& Comprehensive  & P1, P8, P11, P16, P17, P18, P21, P80, P81, P82, P101 & P114, P117, P124, P126, P130, P131 & 6.0\% \\

& Customizable & P8, P57, P70, P88 & P115, P119, P123, P128, P130, P131, P132 & 3.9\% \\
\hline
\multirow{2}{*}{\parbox{3cm}{\raggedright\textbf{Improved Monitor and Model Performance}}}
& Improved Accuracy  & P6, P13, P15, P25, P26, P27, P30, P33, P34, P36, P37, P38, P41, P42, P47, P57, P58, P63, P72, P73, P76, P81, P87, P91, P92, P95, P96, P97, P98, P103, P104, P105, P106, P112 & P52, P54, P74, P84, P120, P133 & 14.0\% \\
\hline

\multirow{2}{*}{\parbox{3cm}{\raggedright\textbf{Deeper Insights and Explainability}}}
& Enhanced Interpretability & P4, P10, P18, P20, P22, P37, P38, P68, P71, P87, P103 & P54, P65, P115, P118, P125, P127, P128, P129, P131, P133, P134 & 7.7\% \\

& Practical Implementation Guidance & P35, P40, P43, P45, P49, P62, P64, P89, P100, P101, P111, P135 & P50, P51, P55, P117, P126 & 6.0\% \\
\hline
\multirow{2}{*}{\parbox{3cm}{\raggedright\textbf{Increased Safety and Trust}}}
& Safety \& Risk Reduction & P9, P11, P12, P13, P23, P24, P28, P29, P31, P32, P58, P60, P61, P76, P77, P78, P79, P83, P90, P94, P107, P108 & P109 & 8.1\% \\
\hline
\multirow{2}{*}{\parbox{3cm}{\raggedright\textbf{Reduced Label Dependency}} }
& No Ground Truth Required & P22, P25, P33, P37, P42, P56, P70, P71, P82, P102, P111dis & P65, P67, P109, P110, P115, P124, P134 & 6.3\% \\

& Minimal Ground Truth Required & P36, P46, P69 & - & 1.1\% \\
\hline
\end{tabular}
\end{table}

\sectopic{Streamlined Operations} is the most prominent theme among studies. \textit{Ease of implementation and integration} of monitoring solutions is highlighted as a key benefit by several studies, with slightly greater emphasis in gray literature (+6\%). An example is seamless integration of monitors with existing ML systems without changing existing code (P16, P91). Other significant benefits found from both literatures are \textit{cost efficiency and resource optimization} and \textit{fast performance}. For instance, low latency using feasible computing resources for safety monitoring of autonomous systems with ML components (P61). Additionally, a few studies, mostly formal, propose automated techniques for developing ML monitors, such as transforming high-level specifications into code (P7).   

\sectopic{Flexibility, Scalability, and Broad Applicability} is the second most prominent theme among studies. An important benefit in this category is monitoring approaches being \textit{generalizable} to various ML models, platforms, and domains (P32, P39). \textit{Scalability and robustness} of ML monitors is another major benefit; techniques such as auto-scaling enable monitors to adapt to varying workloads (P93). A strength of some studies was \textit{comprehensive} monitoring solutions that cover several aspects of an ML system, for example, monitoring performance, drift, fairness, and explainability (P114). \textit{Customizable} monitoring approaches, such as custom-defined metrics (P115), were less common overall, but appeared more frequently in gray literature (+7\%).

\sectopic{Improved Monitor and Model Performance} refers to monitoring approaches that \textit{improve accuracy} of the monitors, thereby helping maintain or enhance the accuracy of the ML model. For example, by detecting novel classes with high precision or reducing false positives (P63, P103). This benefit is found twice as often in formal literature (16\%) compared to gray literature (8\%), perhaps due to the greater emphasis on theoretical rigor and evaluation metrics in academic research.

\sectopic{Deeper Insights and Explainability} covers aspects of improved understanding of the ML monitoring process and results. One of the benefits in this category is \textit{enhanced interpretability} through, for example, simple and easy to understand visualizations, correlations between model and data issues, and root cause analysis (P10). Interestingly, this benefit appeared more frequently in the gray literature (+10\%), which may reflect a greater demand for rapid diagnostic insight to enable faster mitigation in real-world ML systems. The other benefit in this theme is \textit{practical implementation guidance}, which refers to well-known challenges, recommendations, and desiderata for monitoring ML systems (P35, P49).

\sectopic{Increased Safety and Trust} includes monitoring approaches focusing on \textit{safety and risk reduction}, such as identifying misuse of ML systems (P24). This benefit is moderately represented in formal literature (11\%) while being nearly absent from gray literature (1\%). This discrepancy might be due to a significant portion of formal literature studies monitoring the safety of cyber-physical systems with ML components.

\sectopic{Reduced Label Dependency} is overall the least mentioned theme, including studies with benefits of \textit{no ground truth required} for monitoring or \textit{minimal ground truth required}. For ML systems, many monitoring approaches rely on ground truth or labels to calculate model performance. However, labels are often not available at runtime, making these monitoring approaches beneficial in such scenarios. An example is using a secondary ML model to detect deterioration in the performance of the primary model (P25).

\begin{figure}
    \centering
    \includegraphics[width=1\linewidth]{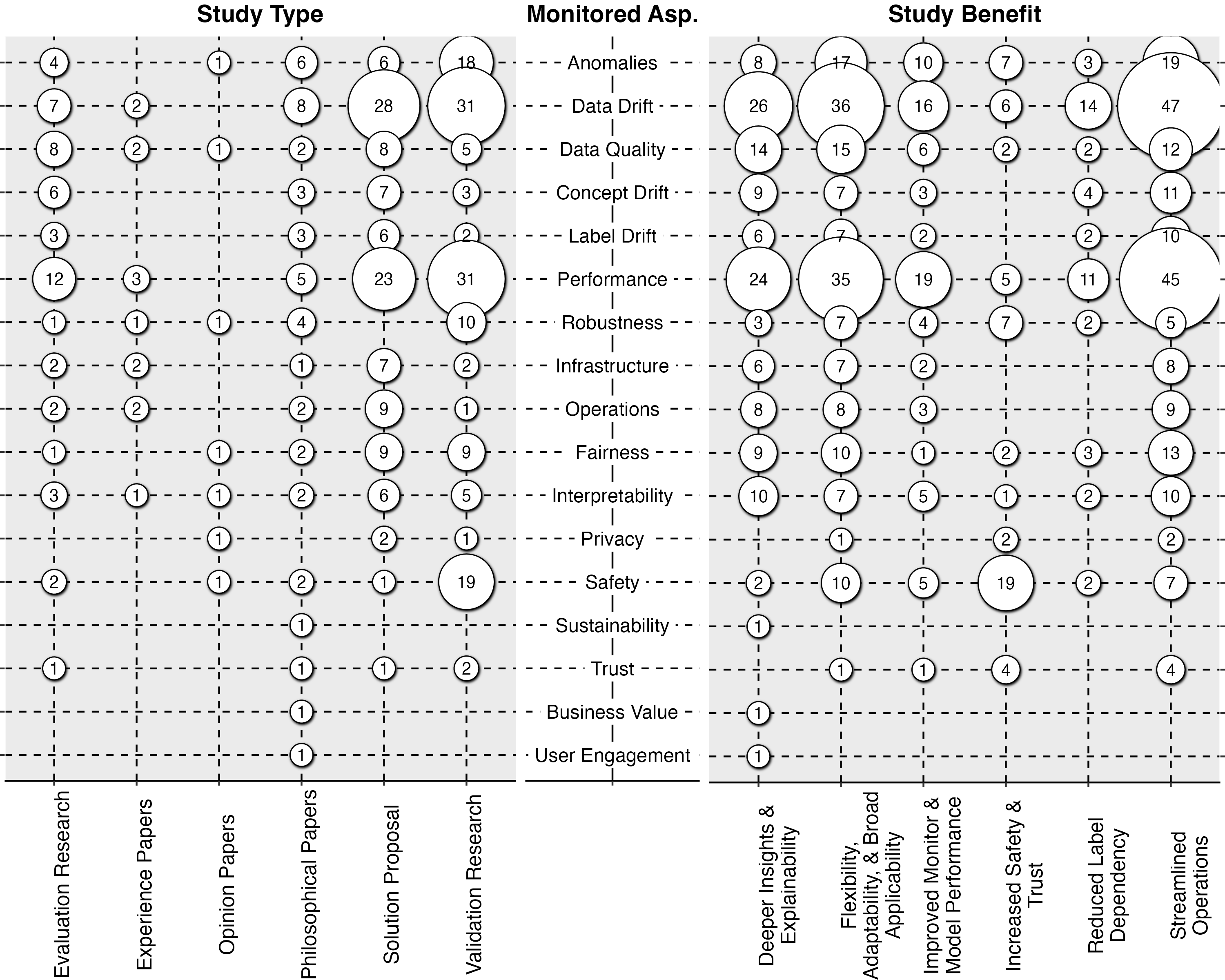}
    \caption{Bubble chart for study type, study benefit, and monitored aspects}
    \label{fig:bubblechart}
\end{figure}

Fig. \ref{fig:bubblechart} shows a bubble chart of the study type and benefit against their relevant monitored aspect. The size of the bubbles shows the frequency of studies in that category. For example, 18 validation research studies have been conducted to monitor \textit{anomalies} in ML systems, and the most common benefit reported is streamlined operations. Analysis of this figure revealed that studies that monitor model performance and data drift report a wide range of benefits, with the highest frequency. However, most of these studies are either not evaluated (solution proposal) or evaluated in an academic setting (validation research). Comparatively, studies evaluated in industrial settings (evaluation research) were much fewer and focused primarily on model performance, data drift, concept drift, and data quality. Additionally, a clear imbalance exists between monitoring solutions for technical aspects (e.g., performance and drift) and responsible ML aspects (e.g., privacy and trust). Technical aspects receive significantly greater attention in literature and benefit from more mature monitoring solutions than responsible ML aspects.

\begin{center}
\begin{myframe}[width=48em,top=5pt,bottom=5pt,left=5pt,right=5pt,arc=10pt,auto outer arc,title=\centering\textbf{RQ3 Answer Summary}]
The majority of studies present novel frameworks or tools for monitoring ML systems, while other less common contributions include monitoring methods, guidelines, and architectures. The most frequently cited benefits in the included studies were improved monitoring accuracy compared to baselines and ease of implementing and integrating monitors with existing ML systems. Comparatively, many fewer studies cited benefits related to automation and customization. In terms of study type, most monitoring approaches are either evaluated on academic case studies or are not evaluated at all, offering only a small proof of concept. Relatively, many fewer monitoring solutions were evaluated in industrial contexts.
\end{myframe}
\end{center}

\subsection{RQ4. What are the limitations of the proposed monitoring solutions for ML systems?}
\subsubsection{Limitations} \label{limitations}
The selected studies were examined for limitations in their proposed monitoring solutions. While 80 studies explicitly cite their limitations, the remaining 56 do not. 
We have classified the limitations reported in studies in six high-level categories shown in Table \ref{tab:limitations}. The studies that do not report any limitations are placed in the \textit{Not Mentioned} category. Overall, we note that nearly 70\% of gray literature studies do not report any limitations, compared to only 32\% in the formal literature.

\begin{table}[htbp]
\centering
\caption{Limitations of Monitoring Approaches}
\label{tab:limitations}
\renewcommand{\arraystretch}{1.2}

\small
\begin{tabular}{>{\raggedright\arraybackslash}p{4cm}p{4cm}p{4cm}c}
\hline
\textbf{Limitations} &  \textbf{Formal Literature} & \textbf{Gray Literature} & \textbf{Percentage} \\
\hline

\multirow{2}{*}{\parbox{3cm}{\raggedright\textbf{Method Limitations}}}
&  P2, P8, P21, P24, P28, P30, P32, P36, P37, P38, P39, P40, P42, P47, P48, P49, P56, P64, P68, P69, P71, P73, P76, P78, P80, P81, P86, P88, P89, P92, P94, P95, P103, P105, P108, P111, P135  & P52, P54, P65, P67, P84, P93, P123, P134, P136 & 24.3\% \\\hline

\multirow{2}{*}{\parbox{3cm}{\raggedright\textbf{Data Limitations}} }
& P8, P20, P21, P26, P30, P32, P33, P34, P36, P37, P44, P47, P48, P56, P61, P70, P71, P72, P76, P81, P83, P85, P86, P88, P92, P100, P112 & P52, P54, P93, P109, P118, P134 & 17.5\% \\\hline

\multirow{2}{*}{\parbox{3cm}{\raggedright\textbf{Evaluation Limitations}} }
& P2, P8, P9, P11, P25, P32, P35, P37, P38, P45, P47, P57, P61, P68, P80, P91, P95 & P53, P67, P134, P136 & 11.1\% \\\hline

\multirow{2}{*}{\parbox{3cm}{\raggedright\textbf{Operational Limitations}} }
& P2, P12, P19, P20, P21, P36, P38, P48, P58, P60, P63, P66, P72, P99, P100, P106 & P53, P134 & 9.5\% \\\hline
\multirow{2}{*}{\parbox{3cm}{\raggedright\textbf{Manual Effort}} }
& P2, P14, P20, P22, P30, P34, P36, P44, P66, P70, P79, P101, P105 & - & 6.9\% \\\hline
\multirow{1}{*}{\parbox{3cm}{\raggedright\textbf{Inherent Limitations}} }
& P28, P100 & - & 1.1\% \\\hline
\multirow{2}{*}{\parbox{3cm}{\raggedright\textbf{Not Mentioned}} }
& P1, P3, P4, P5, P6, P7, P10, P13, P15, P16, P17, P18, P23, P27, P29, P31, P41, P43, P46, P59, P62, P75, P77, P82, P87, P90, P96, P97, P98, P102, P104, P107 & P50, P51, P55, P74, P110, P113, P114, P115, P116, P117, P119, P120, P121, P122, P124, P125, P126, P127, P128, P129, P130, P131, P132, P133 & 29.6\% \\\hline
\end{tabular}
\end{table}

\sectopic{Method Limitations} were the most prevalent among formal and gray literature. These include weaknesses in the proposed framework or method, such as reduced effectiveness of monitors in certain conditions or for certain types of issues. It also includes studies that have a narrow scope in terms of the monitored ML technique or runtime issues, e.g., P21 only monitors supervised ML models, while P48 only monitors specific types of fairness properties.

\sectopic{Data Limitations} such as dependencies on large datasets, high quality data, and ground truth labels are frequently reported in studies. For instance, the accuracy of monitors in P32 relies heavily on the quality of training data since its used to develop a baseline for comparison. Approaches that utilize ML model-based monitors often require large datasets for training, some examples of this are P61, P72, and P86. Additionally, limitations in terms of data structure and data processing paradigm were also identified in some studies. For example, approaches proposed in P8, P21, P37, and P118 only support tabular data.

\sectopic{Evaluation Limitations} were reported mostly in formal literature, emphasizing that their current evaluations were not sufficient to demonstrate the effectiveness of the proposed work. This is primarily because the evaluation scenarios are simple (P53, P67, P80), restricted to small datasets (P2, P80), or lack industrial relevance (P8, P45). 

\sectopic{Operational Limitations} refer to challenges when using the monitoring solution in practice. Scalability limitations are often mentioned in studies due to the presented approaches being time-consuming or resource intensive. For example, the monitoring solutions of P20 and P38 become resource intensive as the volume of data increases. Other operational limitations arise from unrealistic assumptions made in studies, such as the lack of consideration for ground truth label delay in P36.

\sectopic{Manual Effort} is cited as a limitation for studies that require significant effort from ML engineers or domain experts. For ML engineers, this typically includes configuring monitoring thresholds, as discussed in P14, P20, P22, and P79, and analyzing reports to detect unwanted behavior, as shown in P2 and P30. For domain experts, the primary task is data annotation mentioned for instance in P36 and P70. Furthermore, these studies assume that expert labels are always correct.

\sectopic{Inherent Limitations} while rarely cited, refer to limitations that naturally exist in the domain or aspect being monitored. For example, in P28 the authors monitor trustworthiness of ML components, however, trust as a quality is difficult to quantify. Another example, mentioned in P100, is ML-induced confounding, where model predictions influence patient treatment decisions, making it difficult to evaluate the true performance of monitors in practice.

\begin{center}
\begin{myframe}[width=48em,top=5pt,bottom=5pt,left=5pt,right=5pt,arc=10pt,auto outer arc,title=\centering\textbf{RQ4 Answer Summary}]
The main limitations in ML monitoring solutions are weaknesses in the proposed method or a narrow solution scope. Data related limitations are also common, such as reliance on high quality data, large datasets, and ground truth labels. Furthermore, limitations in evaluation, scalability, and automation are also frequently highlighted in the selected studies.
\end{myframe}
\end{center}

\section{Threats to Validity}
This section presents the potential threats to the validity of our MLR and the mitigation strategies applied.

\subsection{Internal Validity}
Selection bias is the main threat to the internal validity of our study. To mitigate this, we followed the systematic literature review guidelines by Garousi~\cite{garousi2019guidelines} and Kitchenham~\cite{kitchenham2009systematic}. The MLR protocol was developed by the first author and reviewed by the other authors for further refinement before conducting the study. The search string was iteratively refined by assessing the relevance of results from six well-established databases. For gray literature, we relied on ArXiv and Google Search Engine, as they were the most suitable databases for our study~\cite{garousi2019guidelines}. The search string and inclusion/exclusion criteria were iteratively refined by assessing the relevance of results from six well-established databases. Following the guidelines, the search process was performed in several rounds as shown in Figure \ref{fig:search process}. The studies were filtered by the first author and validated by other authors. In the first round, studies were filtered based on title and abstract, in the second round by skimming the full text, and in the third round by reading the full text in detail. In case some potentially relevant studies were missed in the automated search, we also performed a complementary manual search. Additionally, to ensure the consistency of data extracted from the selected studies, all authors participated in a pilot test. Another threat to internal validity is the quality variability of studies, especially with gray literature. To reduce the impact of this threat, we have reported gray literature separately from formal literature and used relative percentages to do a fair comparison. 

\subsection{Construct Validity}
We performed a rigorous search of six well-established databases to find the most appropriate studies for this literature review. The inclusion and exclusion criteria were refined several times before the search for optimal results. After multiple rounds of filtering, the selected studies were highly relevant to monitoring ML systems and our research questions. The terminology used for monitoring concepts in the studies was inconsistent, which is a threat to our review. The impact of this threat was reduced by discussing all ambiguities with other authors to reach a consensus. Another threat was self-classification bias; we recorded results based on what was reported in the studies. While this risk cannot be entirely eliminated, we tried to interpret and categorize contributions as accurately as possible.

\subsection{Conclusion Validity}
To reduce threats to conclusion validity, we followed a well-planned search and data extraction process. All borderline cases during the search were discussed with other authors before making the final decision. For accurate data extraction, the first author extracted data for 9 studies and compared it with the data extracted by the other authors. A close match was found, after which the first author continued to extract data for the remaining studies. To reduce bias during data analysis and synthesis, all authors participated in several rounds of discussion to agree on how to categorize and represent the data. We used thematic analysis to organize and interpret the qualitative data consistently. For fair comparison while reporting, we share results from formal and gray literature separately with relative percentages.

\subsection{External Validity}
Our goal was to provide an overview of the current practices for monitoring ML systems by reviewing formal and gray literature. To find relevant studies, we conducted a systematic search following well-established guidelines~\cite{garousi2019guidelines, kitchenham2009systematic} and clearly defined inclusion and exclusion criteria. However, we cannot guarantee that we found all relevant studies, which is a threat to the generalizability of the study. We only included studies in the English language, with full text available, and at least four pages long. We did not include personal blogs, videos, and presentations due to the additional permission needed from authors to use them. For gray literature, we did not include articles from organizational websites where the terms and conditions restricted use in such a study or required special permission. Additionally, we only included the top 200 results from the Google Search Engine due to feasibility constraints, following the recommendation of~\cite{garousi2019guidelines}. While it is possible that some relevant studies were never found during the search process or were excluded during filtering, we believe the impact of this on our study would be minimal. Furthermore, to eliminate publication bias, we did not exclude any study based on quality or publication date.

\section{Recommendations for Future Research and Practice}
Based on the findings of this MLR, this section presents recommendations for future research and practice in monitoring ML systems.
\sectopic{(1) Enhance Automation in ML Monitoring.} 
Our study revealed that there are only a few studies that focus on automation for ML monitoring. This is surprising, since in several studies~\cite{shergadwala2022human, shankar2024we, bernardo2025continuous} practitioners report that monitoring ML systems is tedious due to the manual effort required during integration, configuration, issue diagnosis, and maintenance. The lack of automation in ML monitoring means that practitioners rely on manual setups that are time-consuming and difficult to scale, causing delays and inefficiencies in production~\cite{ghanta2019ml, zarour2025mlops}. A key direction for future work is to develop ML monitoring solutions and tools that require minimal human intervention. In particular, development methodologies like model-driven engineering (MDE)~\cite{kourouklidis2023domain, naveed2024towards} can be leveraged to create high-level abstractions of ML monitors that are automatically transformed into code~\cite{kourouklidis2023domain, naveed2024towards, wong2023mlguard}. This approach is also beneficial for the maintenance of ML monitors since they often require customizations and updates as the ML system and operating context evolve~\cite{shivashankar2025scalability}. Additionally, automating the configuration of monitors and selecting optimal alert thresholds is also crucial. Future work should investigate meta-learning~\cite{ghaderi2023threshold} and optimization algorithms~\cite{cummaudo2020threshy} to systematically select thresholds, tune parameters, and balance trade-offs such as sensitivity versus alert fatigue. Incorporating further automation in ML monitoring would improve practitioners' productivity, reduce technical barriers, reduce alert fatigue, and make setting up and maintaining ML monitors more efficient.\\

\sectopic{(2) Develop ML Monitoring Solutions for Real-World Adoption.}
For ML monitoring solutions to be adopted in real-world deployments, it is essential to address operational requirements such as scalability and industry-grade tooling~\cite{shivashankar2025scalability}. Our study found that many proposed solutions are resource-intensive and face scalability constraints. Academic works are often conceptual frameworks or prototypes, with industrial evaluations being scarce, while many gray literature tools assume the availability of ground truth. There is also a clear misalignment with practitioner needs, for example, while studies frequently report benefits such as ease of implementation and integration, practitioners still struggle with ML monitoring~\cite{shankar2024we}. These differences suggest a clear mismatch between literature and practice. Future work should bridge these gaps through closer collaboration between researchers and practitioners, focusing on usability studies, real-world evaluations, and co-development of tools~\cite{singer2022enhancing}. Rather than building solutions from scratch, approaches that extend existing tools through plugins or extensions can accelerate adoption. \\

\sectopic{(3) Consolidate ML Monitoring Guidelines.}
Despite the growing importance of ML monitoring, existing guidelines and best practices remain fragmented, incomplete, and often outdated, leaving practitioners without a unified reference point~\cite{breck2017ml, heyn2024empirical, klaise2020monitoring, zimelewicz2024ml, bhargavachallenges}. This fragmentation creates inconsistencies across organizations, increases workload for practitioners, and risks overlooking critical aspects of reliability. A potential direction for future work is to consolidate ML monitoring guidelines by systematically collecting, analyzing, and augmenting existing works. These guidelines can be organized under two broad categories: (1) general monitoring recommendations that apply to all ML systems, e.g., track data quality, integrate monitoring in CI/CD pipelines, and ensure operational reliability; and (2) context dependent monitoring recommendations that can vary based on various factors of the ML system, e.g., data characteristics, risk level, and integrations with other systems. These guidelines can also include examples of runtime issues, monitoring approaches, and successful resolution strategies. They can be iteratively refined through practitioners' feedback and eventually integrated into ML monitoring tools. Furthermore, with the rise of generative AI (genAI), these guidelines can be extended to include monitoring practices for genAI systems, or a similar approach can be adopted to develop monitoring guidelines specifically for such systems. A consolidated body of knowledge would not only streamline monitoring practices but also reduce overhead for practitioners and set a common standard for both academic research and industrial deployment. \\

\sectopic{(4) Standardize Requirements Specification for ML Monitoring.}
Specifying monitoring requirements for ML systems is challenging due to the absence of systematic methods~\cite{heyn2024empirical}. Practitioners typically follow a reactive approach, deciding what to monitor based on past incidents, intuition, or trial and error~\cite{shankar2024we, nannyml2024fromreactive}. This approach is error-prone because it leads to blind spots and delayed detection of critical issues. Without clear requirements, monitoring can become ineffective and misaligned with business objectives~\cite{nogare2025mlops}. Future work should focus on developing structured requirement specification methodologies for ML monitoring, such as requirement templates~\cite{darif2025requirements, mavin2009easy} and checklists~\cite{microsoft_ml_model_checklist}. These structured requirements can help practitioners clearly define \textit{what to monitor}, e.g., system aspects, metrics, \textit{how to monitor}, e.g., thresholds, frequency, triggers, and \textit{how to respond}, e.g., alerts, retraining, logging. Our study found that business aspects were the least monitored in ML systems. Linking these requirements to overall system goals or business objectives can provide traceability and tangible value to stakeholders. Furthermore, it would help organizations achieve compliance with regulatory requirements. Lastly, to streamline adoption, a domain-specific language (DSL) can be developed to facilitate defining and managing ML monitoring requirements~\cite{naveed2024towards}.\\

\sectopic{(5) Create Datasets to Evaluate ML Monitors.} 
To demonstrate the practical applicability and usefulness of an ML monitoring approach, the evaluation dataset must closely reflect the deployment environment and runtime conditions. Such datasets can be challenging to find, which weakens evaluation and results in ML monitors that may not be suitable for real-world applications. Evaluation limitations such as simple scenarios, small datasets, and a lack of industrial relevance are also cited in several studies detailed in Section \ref{limitations}. Future work can explore techniques to simulate realistic concept and data drift~\cite{lewis2022augur, lu2018learning}, inject anomalies~\cite{kim2024enhancing}, and generate adversarial inputs~\cite{kim2024enhancing}. Additionally, methods to create domain-specific synthetic datasets or augment existing datasets can help benchmark monitoring solutions more effectively and capture nuanced runtime issues. Practitioners can share the runtime issues encountered through real-world datasets or representative subsets. When real datasets cannot be shared due to privacy or organizational restrictions, synthetic data generation can be used to produce similar datasets. Even sharing the scenarios of issues in production ML systems and their resolution can further improve the evaluation and robustness of ML monitors~\cite{leest2024expert}.\\

\sectopic{(6) Incorporate Context in ML Monitoring. }
The operating context of ML systems is often specific to the application domain because of the unique risks, failure modes, data patterns, and usage patterns of each domain~\cite{shankar2024we, shergadwala2022human}. This context is critical for effective monitoring because generic solutions capture only surface-level issues, such as data drift and accuracy drops, while failing to detect nuanced domain-specific issues~\cite{leest2025tea, shankar2024we, shergadwala2022human}. This can lead to poorly calibrated thresholds, alert fatigue, difficulty in diagnosing issues, and over-reliance on domain experts~\cite{shankar2024we, xu2022dependency}. Yet, we found that most approaches monitor ML systems in isolation without much consideration of the broader context in which they operate. Future work should focus on integrating contextual information in ML monitoring and developing more domain-specific solutions. This can include monitoring the data sources and data processing pipelines; interactions between the ML system and other systems; and the broader real-world application of the ML system, such as human decision making~\cite{leest2025tea}. Additionally, the context can include important external events, such as policy changes or new market trends; shifts in user groups and interaction patterns; and other environmental variables relevant to the ML system, such as weather, seasonal effects, or infrastructure conditions~\cite{leest2025tea}. By incorporating these diverse contexts, ML monitoring approaches can detect nuanced issues, improve alert quality, and reduce practitioners' workload.\\

\sectopic{(7) Build Monitoring Solutions for Responsible ML. }
Responsible ML aspects are gaining traction due to increasing awareness of their role in building safe, ethical, and trustworthy ML systems. However, we found that aspects other than fairness and safety are often neglected in ML monitoring solutions. As is evident from several past events~\cite{wolf2017we, ross2018ibm, dastin2022amazon}, a lack of consideration for these aspects can cause serious reputational and financial damage to organizations. Further efforts are required in this area to build on top of high-level frameworks~\cite{oecd2019ai,australia2024aiethics, ey2024euai} and develop concrete metrics, approaches, and easy-to-use tools that can monitor these aspects. For sustainability monitoring, this involves tracking runtime energy consumption, retraining frequency, and the carbon footprint of ML systems for early detection of inefficiencies~\cite{jarvenpaa2024synthesis, cruz2025greening}. Accountability monitoring involves recording system operations and changes in usage patterns, model versions, ML and non-ML components, system configurations, and operational parameters~\cite{xia2024towards}. Regarding privacy monitoring, observing the ML system for predictions with extremely high confidence, unusual query patterns, or repeated probing of specific inputs can help detect attacks, enabling timely mitigation and safeguarding sensitive training data~\cite{rigaki2023survey}. For aspects that are inherently subjective and difficult to measure directly, such as trust and human values, proxy or composite metrics can be designed, and ML-based monitors can be developed and refined using human feedback to capture nuanced patterns. Better monitoring solutions for responsible ML would result in ML systems that align with ethical practices, maintain user trust, and maximize benefits for society.\\

\sectopic{(8) Leverage Large Language Models for ML Monitoring.}
Our review did not cover Large Language Models (LLMs), but given the increasing adoption of LLMs, there are several areas where the advanced capabilities of LLMs, such as reasoning, summarization, and generation can be valuable for ML monitoring, particularly when ground truth is unavailable and expert annotations are expensive. This data generation is also useful for evaluating monitors and training ML-based monitors. Another application area of LLMs is interpretability and root cause analysis of detected issues, a prominent challenge reported by practitioners~\cite{roy2024exploring, shankar2022towards, shergadwala2022human}. LLMs can analyze the detected issues, help diagnose underlying causes, and even recommend fixes. The explanations in natural language can significantly improve the mitigation time, especially for operations engineers who do not have ML expertise. Furthermore, domain expertise can be incorporated in LLMs through techniques such as Retrieval-Augmented Generation (RAG)~\cite{gao2023retrieval} to create domain specific ML monitors. Another emerging area is monitoring LLM-based systems, also known as LLM observability~\cite{ganesan2024llm, kale2025reliable}. Unlike traditional ML systems, LLMs require monitoring solutions tailored to their unique characteristics. Key aspects include tracking computational usage, cost, adversarial prompts, and hallucinations~\cite{ganesan2024llm}. While some solutions exist, they do not fully meet the requirements of practitioners, and there is a need for better LLM monitoring solutions~\cite{chen2025design}.

\section{Conclusion}
Monitoring ML systems is a complex task; it consists of several facets that need consideration. This includes deciding the appropriate monitoring aspect, technique, metric, and tool, which can be challenging given the increasing volume of literature on monitoring approaches for ML systems. To address this challenge, we conducted a multivocal literature review (MLR) on monitoring ML systems following the systematic guidelines by Garousi~\cite{garousi2019guidelines}. Our goal was to explore formal and gray literature studies to provide a structured and holistic overview of ML monitoring to improve understanding and practice. We found 9147 papers from our initial search that were carefully filtered in multiple iterations to reach a final pool of 136 studies, 100 from formal literature and 36 from gray literature. This is the first multivocal study in this area that presents insights from the entire ML monitoring landscape, including: 1) a categorization of monitored aspects in ML systems into five distinct categories; 2) an organization of monitoring techniques and metrics, grouped by aspect; 3) a comprehensive list of monitoring tools, including the aspects they monitor, distribution model, and their visualization, alerting, and logging capabilities; (4) the key motivations and goals that contextualize ML monitoring studies; (5) a categorization of ML techniques and domains where monitoring is applied; (6) the ground truth dependencies and logging capabilities of monitoring approaches; (7) the main contributions proposed in literature, such as frameworks and tools, along with their demonstrated benefits; and (8) the identification of six types of limitations that reveal current weaknesses and opportunities for advancing the field.

Furthermore, based on our findings we recommend the following directions for future research and practice: (1) enhance automation in ML monitoring; (2) develop ML monitoring solutions for real-world adoption; (3) consolidate ML monitoring guidelines; (4) standardize requirements specification for ML monitoring; (5) create datasets to evaluate ML monitors; (6) incorporate context in ML monitoring; (7) build monitoring solutions for responsible ML; and (8) leverage large language models for ML monitoring.

In summary, this study advances the understanding of monitoring ML systems by presenting a structured and unified view of techniques, metrics, and tools across diverse operational contexts. By systematically analyzing the capabilities, benefits and limitations of monitoring approaches, it provides practitioners with practical guidance for selecting appropriate monitoring techniques and tools, and highlights innovation opportunities for researchers and tool developers. Based on evidence from both industry and academia, the findings reveal misalignments, uncover gaps, and offer actionable insights to enhance the field of ML monitoring.



\section{Acknowledgments}
Naveed is supported by a Faculty of IT Post-graduate scholarship. Grundy and Haggag are supported by ARC Laureate Fellowship FL190100035. Haggag is also supported by a National Intelligence Post-doctoral Fellowship. This work is also
partly supported by ARC Discovery Project DP200100020.

\appendix
\section{Selected Studies} \label{StudiesList}

\begin{footnotesize}
\begin{itemize}
  \item[\textbf{P1:}] Anas Bodor, Meriem Hnida, and Daoudi Najima. From development to deployment: An approach to mlops monitoring for machine learning
model operationalization. In \textit{2023 14th International Conference on Intelligent Systems: Theories and Applications (SITA)}, pages 1–7. IEEE, 2023
\item[\textbf{P2:}] Grace A Lewis, Sebastián Echeverría, Lena Pons, and Jeffrey Chrabaszcz. Augur: A step towards realistic drift detection in production ml
systems. In \textit{Proceedings of the 1st Workshop on Software Engineering for Responsible AI}, pages 37–44, 2022
\item[\textbf{P3:}] Bradley Eck, Duygu Kabakci-Zorlu, Yan Chen, France Savard, and Xiaowei Bao. A monitoring framework for deployed machine learning
models with supply chain examples. In \textit{2022 IEEE International Conference on Big Data (Big Data)}, pages 2231–2238. IEEE, 2022
\item[\textbf{P4:}]Lucas Cardoso Silva, Fernando Rezende Zagatti, Bruno Silva Sette, Lucas Nildaimon dos Santos Silva, Daniel Lucrédio, Diego Furtado Silva, and
Helena de Medeiros Caseli. Benchmarking machine learning solutions in production. In \textit{2020 19th IEEE International Conference on Machine
Learning and Applications (ICMLA)}, pages 626–633. IEEE, 2020
\item[\textbf{P5:}] Panagiotis Kourouklidis, Dimitris Kolovos, Joost Noppen, and Nicholas Matragkas. A domain-specific language for monitoring ml model
performance. In \textit{2023 ACM/IEEE International Conference on Model Driven Engineering Languages and Systems Companion (MODELS-C)}, pages
266–275. IEEE, 2023
\item[\textbf{P6:}]Dimitrios Michael Manias, Ali Chouman, and Abdallah Shami. A model drift detection and adaptation framework for 5g core networks. In \textit{2022
IEEE International Mediterranean Conference on Communications and Networking (MeditCom)}, pages 197–202. IEEE, 2022
\item[\textbf{P7:}] Nikunj Parekh, Swathi Kurunji, and Alan Beck. Monitoring resources of machine learning engine in microservices architecture. In \textit{2018 IEEE
9th Annual Information Technology, Electronics and Mobile Communication Conference (IEMCON)}, pages 486–492. IEEE, 2018
\item[\textbf{P8:}] Yumo Luo, Mikko Raatikainen, and Jukka K Nurminen. Autonomously adaptive machine learning systems: Experimentation-driven open-source
pipeline. In \textit{2023 49th Euromicro Conference on Software Engineering and Advanced Applications (SEAA)}, pages 44–52. IEEE, 2023
\item[\textbf{P9:}] Jörg Grieser, Meng Zhang, Tim Warnecke, and Andreas Rausch. Assuring the safety of end-to-end learning-based autonomous driving through
runtime monitoring. In \textit{2020 23rd Euromicro Conference on Digital System Design (DSD)}, pages 476–483. IEEE, 2020
\item[\textbf{P10:}] Awalin Sopan and Konstantin Berlin. Ai total: Analyzing security ml models with imperfect data in production. In \textit{2021 IEEE Symposium on
Visualization for Cyber Security (VizSec)}, pages 10–14. IEEE, 2021
\item[\textbf{P11:}] Raul Sena Ferreira, Jean Arlat, Jérémie Guiochet, and Hélène Waeselynck. Benchmarking safety monitors for image classifiers with machine
learning. In \textit{2021 IEEE 26th Pacific Rim International Symposium on Dependable Computing (PRDC)}, pages 7–16. IEEE, 2021
\item[\textbf{P12:}] Houssem Guissouma, Moritz Zink, and Eric Sax. Continuous safety assessment of updated supervised learning models in shadow mode. In \textit{2023
IEEE 20th International Conference on Software Architecture Companion (ICSA-C)}, pages 301–308. IEEE, 2023
\item[\textbf{P13:}] Sophia Abraham, Zachariah Carmichael, Sreya Banerjee, Rosaura VidalMata, Ankit Agrawal, Md Nafee Al Islam, Walter Scheirer, and
Jane Cleland-Huang. Adaptive autonomy in human-on-the-loop vision-based robotics systems. In \textit{2021 IEEE/ACM 1st workshop on AI
engineering-software engineering for AI (WAIN)}, pages 113–120. IEEE, 2021
\item[\textbf{P14:}] Koustabh Dolui, Sam Michiels, Danny Hughes, and Hans Hallez. Context aware adaptive ml inference in mobile-cloud applications. In \textit{2022
IEEE 23rd International Symposium on a World of Wireless, Mobile and Multimedia Networks (WoWMoM)}, pages 90–99. IEEE, 2022
\item[\textbf{P15:}] Zhentao Xu, Ruoying Wang, Girish Balaji, Manas Bundele, Xiaofei Liu, Leo Liu, and Tie Wang. Alertiger: Deep learning for ai model health
monitoring at linkedin. In \textit{Proceedings of the 29th ACM SIGKDD Conference on Knowledge Discovery and Data Mining}, pages 5350–5359, 2023
\item[\textbf{P16:}] Shreya Shankar and Aditya G Parameswaran. Towards observability for production machine learning pipelines. In \textit{Proceedings of the VLDB
Endowment}, 15(13):4015–4022, 2022
\item[\textbf{P17:}] David Nigenda, Zohar Karnin, Muhammad Bilal Zafar, Raghu Ramesha, Alan Tan, Michele Donini, and Krishnaram Kenthapadi. Amazon
sagemaker model monitor: A system for real-time insights into deployed machine learning models. In \textit{Proceedings of the 28th ACM SIGKDD
Conference on Knowledge Discovery and Data Mining}, pages 3671–3681, 2022
\item[\textbf{P18:}] Michaela Hardt, Xiaoguang Chen, Xiaoyi Cheng, Michele Donini, Jason Gelman, Satish Gollaprolu, John He, Pedro Larroy, Xinyu Liu, Nick
McCarthy, et al. Amazon sagemaker clarify: Machine learning bias detection and explainability in the cloud. In \textit{Proceedings of the 27th ACM
SIGKDD conference on knowledge discovery \& data mining}, pages 2974–2983, 2021
\item[\textbf{P19:}] Anirban I Ghosh, Radhika Sharma, Karan Goyal, Balakarthikeyan Rajan, and Senthil Mani. Health assurance: Ai model monitoring platform. In
\textit{Proceedings of the Second International Conference on AI-ML Systems}, pages 1–7, 2022
\item[\textbf{P20:}] Hemadri Jayalath and Lakshmish Ramaswamy. Enhancing performance of operationalized machine learning models by analyzing user feedback.
In \textit{Proceedings of the 2022 4th International Conference on Image, Video and Signal Processing}, pages 197–203, 2022
\item[\textbf{P21:}] Arkadipta De, Satya Swaroop Gudipudi, Sourab Panchanan, and Maunendra Sankar Desarkar. Complai: Framework for multi-factor assessment
of black-box supervised machine learning models. In \textit{Proceedings of the 38th ACM/SIGAPP Symposium on Applied Computing}, pages 1096–1099,
2023
\item[\textbf{P22:}] Avijit Ghosh, Aalok Shanbhag, and Christo Wilson. Faircanary: Rapid continuous explainable fairness. In \textit{Proceedings of the 2022 AAAI/ACM
Conference on AI, Ethics, and Society}, pages 307–316, 2022
\item[\textbf{P23:}] Sheng Wong, Scott Barnett, Jessica Rivera-Villicana, Anj Simmons, Hala Abdelkader, Jean-Guy Schneider, and Rajesh Vasa. Mlguard: Defend
your machine learning model! In \textit{Proceedings of the 1st International Workshop on Dependability and Trustworthiness of Safety-Critical Systems
with Machine Learned Components}, pages 10–13, 2023
\item[\textbf{P24:}] Seyyed Ahmad Javadi, Chris Norval, Richard Cloete, and Jatinder Singh. Monitoring ai services for misuse. In \textit{Proceedings of the 2021 AAAI/ACM
Conference on AI, Ethics, and Society}, pages 597–607, 2021
\item[\textbf{P25:}] Sebastian Schelter, Tammo Rukat, and Felix Bießmann. Learning to validate the predictions of black box classifiers on unseen data. In \textit{Proceedings
of the 2020 ACM SIGMOD International Conference on Management of Data}, pages 1289–1299, 2020
\item[\textbf{P26:}] Shinan Liu, Francesco Bronzino, Paul Schmitt, Arjun Nitin Bhagoji, Nick Feamster, Hector Garcia Crespo, Timothy Coyle, and Brian Ward. Leaf:
Navigating concept drift in cellular networks. In \textit{Proceedings of the ACM on Networking}, 1(CoNEXT2):1–24, 2023
\item[\textbf{P27:}] Anes Bendimerad, Youcef Remil, Romain Mathonat, and Mehdi Kaytoue. On-premise aiops infrastructure for a software editor sme: an
experience report. In \textit{Proceedings of the 31st ACM Joint European Software Engineering Conference and Symposium on the Foundations of Software
Engineering}, pages 1820–1831, 2023
\item[\textbf{P28:}] Silverio Martínez-Fernández, Xavier Franch, Andreas Jedlitschka, Marc Oriol, and Adam Trendowicz. Developing and operating artificial
intelligence models in trustworthy autonomous systems. In \textit{International Conference on Research Challenges in Information Science}, pages
221–229. Springer, 2021
\item[\textbf{P29:}]  Saddek Bensalem, Panagiotis Katsaros, Dejan Ničković, Brian Hsuan-Cheng Liao, Ricardo Ruiz Nolasco, Mohamed Abd El Salam Ahmed,
Tewodros A Beyene, Filip Cano, Antoine Delacourt, Hasan Esen, et al. Continuous engineering for trustworthy learning-enabled autonomous
systems. In \textit{International Conference on Bridging the Gap between AI and Reality}, pages 256–278. Springer, 2023
\item[\textbf{P30:}] Lisa Ehrlinger, Verena Haunschmid, Davide Palazzini, and Christian Lettner. A daql to monitor data quality in machine learning applications. In \textit{
Database and Expert Systems Applications: 30th International Conference}, DEXA 2019, Linz, Austria, August 26–29, 2019, Proceedings, Part I 30,
pages 227–237. Springer, 2019
\item[\textbf{P31:}] Roman V Yampolskiy. On monitorability of ai. In \textit{AI and Ethics}, pages 1–19, 2024
\item[\textbf{P32:}] Al-Harith Farhad, Ioannis Sorokos, Andreas Schmidt, Mohammed Naveed Akram, Koorosh Aslansefat, and Daniel Schneider. Keep your
distance: determining sampling and distance thresholds in machine learning monitoring. In \textit{International Symposium on Model-Based Safety and
Assessment}, pages 219–234. Springer, 2022
\item[\textbf{P33:}] Sindhu Ghanta, Sriram Subramanian, Lior Khermosh, Swaminathan Sundararaman, Harshil Shah, Yakov Goldberg, Drew Roselli, and Nisha
Talagala. Ml health monitor: taking the pulse of machine learning algorithms in production. In \textit{Applications of Machine Learning}, volume 11139,
pages 191–202. SPIE, 2019
\item[\textbf{P34:}] Christopher Ré, Feng Niu, Pallavi Gudipati, and Charles Srisuwananukorn. Overton: A data system for monitoring and improving machine-learned
products. In \textit{Conference on Innovative Data Systems Research (CIDR)}, 2019
\item[\textbf{P35:}] Murtuza N Shergadwala, Himabindu Lakkaraju, and Krishnaram Kenthapadi. A human-centric perspective on model monitoring. In \textit{Proceedings
of the AAAI Conference on Human Computation and Crowdsourcing}, volume 10, pages 173–183, 2022
\item[\textbf{P36:}] Tony Ginart, Martin Jinye Zhang, and James Zou. Mldemon: Deployment monitoring for machine learning systems. In \textit{International conference
on artificial intelligence and statistics}, pages 3962–3997. PMLR, 2022
\item[\textbf{P37:}] Carlos Mougan and Dan Saattrup Nielsen. Monitoring model deterioration with explainable uncertainty estimation via non-parametric bootstrap.
In \textit{Proceedings of the AAAI Conference on Artificial Intelligence}, volume 37, pages 15037–15045, 2023
\item[\textbf{P38:}] Da Song, Zhijie Wang, Yuheng Huang, Lei Ma, and Tianyi Zhang. Deeplens: interactive out-of-distribution data detection in nlp models. In
\textit{Proceedings of the 2023 CHI Conference on Human Factors in Computing Systems}, pages 1–17, 2023
\item[\textbf{P39:}] Myeongseob Ko, Xinyu Yang, Zhengjie Ji, Hoang Anh Just, Peng Gao, Anoop Kumar, and Ruoxi Jia. Privmon: A stream-based system for
real-time privacy attack detection for machine learning models. In \textit{Proceedings of the 26th International Symposium on Research in Attacks},
Intrusions and Defenses, pages 264–281, 2023
\item[\textbf{P40:}]Tim Schröder and Michael Schulz. Monitoring machine learning models: a categorization of challenges and methods. In \textit{Data Science and
Management}, 5(3):105–116, 2022
\item[\textbf{P41:}] Gusseppe Bravo-Rocca, Peini Liu, Jordi Guitart, Ajay Dholakia, David Ellison, Jeffrey Falkanger, and Miroslav Hodak. Scanflow: A multi-graph
framework for machine learning workflow management, supervision, and debugging. In \textit{Expert Systems with Applications}, 202:117232, 2022
\item[\textbf{P42:}] Anna Malinovskaya, Pavlo Mozharovskyi, and Philipp Otto. Statistical process monitoring of artificial neural networks. In \textit{Technometrics}, 66(1):104–117, 2024
\item[\textbf{P43:}] Eduardo Zimelewicz, Marcos Kalinowski, Daniel Mendez, Görkem Giray, Antonio Pedro Santos Alves, Niklas Lavesson, Kelly Azevedo,
Hugo Villamizar, Tatiana Escovedo, Helio Lopes, et al. Ml-enabled systems model deployment and monitoring: Status quo and problems. In
\textit{International Conference on Software Quality}, pages 112–131. Springer, 2024
\item[\textbf{P44:}] Conor K Corbin, Rob Maclay, Aakash Acharya, Sreedevi Mony, Soumya Punnathanam, Rahul Thapa, Nikesh Kotecha, Nigam H Shah, and
Jonathan H Chen. Deployr: a technical framework for deploying custom real-time machine learning models into the electronic medical record. In \textit{
Journal of the American Medical Informatics Association}, 30(9):1532–1542, 2023
\item[\textbf{P45:}] Jean Feng, Adarsh Subbaswamy, Alexej Gossmann, Harvineet Singh, Berkman Sahiner, Mi-Ok Kim, Gene Anthony Pennello, Nicholas Petrick,
Romain Pirracchio, and Fan Xia. Designing monitoring strategies for deployed machine learning algorithms: navigating performativity through
a causal lens. In \textit{Causal Learning and Reasoning}, pages 587–608. PMLR, 2024
\item[\textbf{P46:}] Huong Ha. An efficient framework for monitoring subgroup performance of machine learning systems. In \textit{NeurIPS ML Safety Workshop}
\item[\textbf{P47:}] Daniel Kang, Deepti Raghavan, Peter Bailis, and Matei Zaharia. Model assertions for monitoring and improving ml models. In \textit{Proceedings of
Machine Learning and Systems}, 2:481–496, 2020
\item[\textbf{P48:}] homas Henzinger, Mahyar Karimi, Konstantin Kueffner, and Kaushik Mallik. Runtime monitoring of dynamic fairness properties. In \textit{Proceedings
of the 2023 ACM Conference on Fairness, Accountability, and Transparency}, pages 604–614, 2023
\item[\textbf{P49:}] Shreya Shankar, Rolando Garcia, Joseph M Hellerstein, and Aditya G Parameswaran. " we have no idea how models will behave in production
until production": How engineers operationalize machine learning. In \textit{Proceedings of the ACM on Human-Computer Interaction}, 8(CSCW1):1–34,
2024
\item[\textbf{P50:}] Janis Klaise, Arnaud Van Looveren, Clive Cox, Giovanni Vacanti, and Alexandru Coca. Monitoring and explainability of models in production. \textit{
arXiv preprint arXiv:2007.06299}, 2020
\item[\textbf{P51:}] an Hendrycks, Nicholas Carlini, John Schulman, and Jacob Steinhardt. Unsolved problems in ml safety. \textit{arXiv preprint arXiv:2109.13916}, 2021
\item[\textbf{P52:}] Nader Karayanni, Robert J Shahla, and Chieh-Lien Hsiao. Distributed monitoring for data distribution shifts in edge-ml fraud detection. \textit{arXiv
preprint arXiv:2401.05219}, 2024
\item[\textbf{P53:}] Tom Diethe, Tom Borchert, Eno Thereska, Borja Balle, and Neil Lawrence. Continual learning in practice. \textit{arXiv preprint arXiv:1903.05202}, 2019
\item[\textbf{P54:}] Wei Hao, Zixi Wang, Lauren Hong, Lingxiao Li, Nader Karayanni, Chengzhi Mao, Junfeng Yang, and Asaf Cidon. Monitoring and adapting ml
models on mobile devices. \textit{arXiv preprint arXiv:2305.07772}, 2023
\item[\textbf{P55:}] Mahed Abroshan, Michael Burkhart, Oscar Giles, Sam Greenbury, Zoe Kourtzi, Jack Roberts, Mihaela van der Schaar, Jannetta S Steyn, Alan
Wilson, and May Yong. Safe ai for health and beyond–monitoring to transform a health service. \textit{arXiv preprint arXiv:2303.01513}, 2023
\item[\textbf{P56:}] Xianzhe Zhou, Wally Lo Faro, Xiaoying Zhang, and Ravi Santosh Arvapally. A framework to monitor machine learning systems using concept
drift detection. In \textit{Business Information Systems: 22nd International Conference, BIS 2019, Seville, Spain, June 26–28, 2019, Proceedings, Part I 22},
pages 218–231. Springer, 2019
\item[\textbf{P57:}] Rosa Candela, Pietro Michiardi, Maurizio Filippone, and Maria A Zuluaga. Model monitoring and dynamic model selection in travel time-series
forecasting. In \textit{Machine Learning and Knowledge Discovery in Databases: Applied Data Science Track: European Conference, ECML PKDD 2020,
Ghent, Belgium, September 14–18, 2020, Proceedings, Part IV}, pages 513–529. Springer, 2021
\item[\textbf{P58:}] Joris Guerin, Kevin Delmas, and Jérémie Guiochet. Evaluation of runtime monitoring for uav emergency landing. In \textit{2022 International Conference
on Robotics and Automation (ICRA)}, pages 9703–9709. IEEE, 2022
\item[\textbf{P59:}] Natnaree Jubju and Natawut Nupairoj. End-to-end data characteristics monitoring for mlops. In \textit{2024 21st International Joint Conference on
Computer Science and Software Engineering (JCSSE)}, pages 280–285. IEEE, 2024
\item[\textbf{P60:}] Abdul-Rasheed Ottun, Rasinthe Marasinghe, Toluwani Elemosho, Mohan Liyanage, Mohamad Ragab, Prachi Bagave, Marcus Westberg, Mehrdad
Asadi, Michell Boerger, Chamara Sandeepa, et al. The spatial architecture: Design and development experiences from gauging and monitoring
the ai inference capabilities of modern applications. In \textit{2024 IEEE 44th International Conference on Distributed Computing Systems (ICDCS)}, pages
947–959. IEEE, 2024
\item[\textbf{P61:}] Sepehr Sharifi, Andrea Stocco, and Lionel C Briand. System safety monitoring of learned components using temporal metric forecasting. In \textit{ACM
Transactions on Software Engineering and Methodology}, 2024
\item[\textbf{P62:}] Surabhi Bhargava and Shubham Singhal. Challenges, solutions, and best practices in post-deployment monitoring of machine learning models
\item[\textbf{P63:}] Mithun Kumar Pusukuri. Designing observable machine learning pipelines for real-time credit risk detection: A scalable approach. In \textit{Journal of
Computer Engineering and Technology (IJCET)}, 15(6):1482–1491, 2024
\item[\textbf{P64:}] Louis Faust, Patrick Wilson, Shusaku Asai, Sunyang Fu, Hongfang Liu, Xiaoyang Ruan, Curt Storlie, et al. Considerations for quality control
monitoring of machine learning models in clinical practice. In \textit{JMIR Medical Informatics}, 12(1):e50437, 2024
\item[\textbf{P65:}] Fábio Pinto, Marco OP Sampaio, and Pedro Bizarro. Automatic model monitoring for data streams. \textit{arXiv preprint arXiv:1908.04240}, 2019
\item[\textbf{P66:}] Jose Morales, Luiz Antunes, Patrick Earl, Robert Edman, Jeffrey Hamed, Douglas Reynolds, Katherine R Maffey, Joseph Yankel, and Hasan Yasar.
Insights on implementing a metrics baseline for post-deployment ai container monitoring. In \textit{Proceedings of the 2024 International Conference on
Software and Systems Processes}, pages 46–55, 2024
\item[\textbf{P67:}] Iqra Aslam, Abhishek Buragohain, Daniel Bamal, Adina Aniculaesei, Meng Zhang, and Andreas Rausch. A method for the runtime validation of
ai-based environment perception in automated driving system. \textit{arXiv preprint arXiv:2412.16762}, 2024
\item[\textbf{P68:}] Jean Feng, Adarsh Subbaswamy, Alexej Gossmann, Harvineet Singh, Berkman Sahiner, Mi-Ok Kim, Gene Pennello, Nicholas Petrick, Romain
Pirracchio, and Fan Xia. Towards a post-market monitoring framework for machine learning-based medical devices: A case study. In \textit{NeurIPS
2023 Workshop on Regulatable ML}, 2023
\item[\textbf{P69:}] Zhihui Shao and Jianyi Yang. Increasing the trustworthiness of deep neural networks via accuracy monitoring. In \textit{Workshop on Artificial
Intelligence Safety 2020 (co-located with IJCAI-PRICAI 2020)}, 2020
\item[\textbf{P70:}] Joran Leest, Claudia Raibulet, Ilias Gerostathopoulos, and Patricia Lago. Expert monitoring: Human-centered concept drift detection in machine
learning operations. In \textit{Proceedings of the 2024 ACM/IEEE 44th International Conference on Software Engineering: New Ideas and Emerging Results},
pages 1–5, 2024
\item[\textbf{P71:}] Gou Tan, Pengfei Chen, and Min Li. Online data drift detection for anomaly detection services based on deep learning towards multivariate
time series. In \textit{2023 IEEE 23rd International Conference on Software Quality, Reliability, and Security (QRS)}, pages 1–11. IEEE, 2023
\item[\textbf{P72:}]Michael Weiss and Paolo Tonella. Fail-safe execution of deep learning based systems through uncertainty monitoring. In \textit{2021 14th IEEE
conference on software testing, verification and validation (ICST)}, pages 24–35. IEEE, 2021
\item[\textbf{P73:}] Jon Ayerdi, Asier Iriarte, Pablo Valle, Ibai Roman, Miren Illarramendi, and Aitor Arrieta. Marmot: Metamorphic runtime monitoring of
autonomous driving systems. In \textit{ACM Transactions on Software Engineering and Methodology}, 34(1):1–35, 2024
\item[\textbf{P74:}] Ioannis Mavromatis, Stefano De Feo, and Aftab Khan. Flame: Adaptive and reactive concept drift mitigation for federated learning deployments. \textit{
arXiv preprint arXiv:2410.01386}, 202
\item[\textbf{P75:}] Muqsit Azeem, Marta Grobelna, Sudeep Kanav, Jan Křetínsk `y, Stefanie Mohr, and Sabine Rieder. Monitizer: automating design and evaluation
of neural network monitors. In \textit{International Conference on Computer Aided Verification}, pages 265–279. Springer, 2024
\item[\textbf{P76:}] Amirhossein Zolfagharian, Manel Abdellatif, Lionel C Briand, and S Ramesh. Smarla: A safety monitoring approach for deep reinforcement
learning agents. In \textit{IEEE Transactions on Software Engineering}, 2024
\item[\textbf{P77:}] Mohd Hafeez Osman, Stefan Kugele, and Sina Shafaei. Run-time safety monitoring framework for ai-based systems: Automated driving cases.
In \textit{2019 26th Asia-Pacific Software Engineering Conference (APSEC)}, pages 442–449. IEEE, 2019
\item[\textbf{P78:}] Michael Austin Langford, Kenneth H Chan, Jonathon Emil Fleck, Philip K McKinley, and Betty HC Cheng. Modalas: Model-driven assurance
for learning-enabled autonomous systems. In \textit{2021 ACM/IEEE 24th International Conference on Model Driven Engineering Languages and Systems
(MODELS)}, pages 182–193. IEEE, 2021
\item[\textbf{P79:}] Hazem Torfah and Sanjit A Seshia. Runtime monitors for operational design domains of black-box ml-models. In \textit{NeurIPS ML Safety Workshop}, 2022
\item[\textbf{P80:}] Hira Naveed, John Grundy, Chetan Arora, Hourieh Khalajzadeh, and Omar Haggag. Towards runtime monitoring for responsible machine
learning using model-driven engineering. In \textit{Proceedings of the ACM/IEEE 27th International Conference on Model Driven Engineering Languages
and Systems}, pages 195–202, 2024
\item[\textbf{P81:}] Eric Breck, Shanqing Cai, Eric Nielsen, Michael Salib, and D Sculley. The ml test score: A rubric for ml production readiness and technical debt
reduction. In \textit{2017 IEEE international conference on big data (big data)}, pages 1123–1132. IEEE, 2017
\item[\textbf{P82:}] David A Cieslak and Nitesh V Chawla. A framework for monitoring classifiers’ performance: when and why failure occurs? In \textit{Knowledge and
Information Systems}, 18(1):83–108, 2009
\item[\textbf{P83:}] Piergiuseppe Mallozzi, Ezequiel Castellano, Patrizio Pelliccione, Gerardo Schneider, and Kenji Tei. A runtime monitoring framework to enforce
invariants on reinforcement learning agents exploring complex environments. In \textit{2019 IEEE/ACM 2nd International Workshop on Robotics
Software Engineering (RoSE)}, pages 5–12. IEEE, 2019
\item[\textbf{P84:}] Florian Heinrichs. Monitoring machine learning models: Online detection of relevant deviations. \textit{arXiv preprint arXiv:2309.15187}, 2023
\item[\textbf{P85:}] Yuhang Chen, Chih-Hong Cheng, Jun Yan, and Rongjie Yan. Monitoring object detection abnormalities via data-label and post-algorithm
abstractions. In \textit{2021 IEEE/RSJ International Conference on Intelligent Robots and Systems (IROS)}, pages 6688–6693. IEEE, 2021
\item[\textbf{P86:}] Changshun Wu, Yliès Falcone, and Saddek Bensalem. Customizable reference runtime monitoring of neural networks using resolution boxes.
In \textit{International Conference on Runtime Verification}, pages 23–41. Springer, 2023
\item[\textbf{P87:}] Florian Geissler, Syed Qutub, Michael Paulitsch, and Karthik Pattabiraman. A low-cost strategic monitoring approach for scalable and
interpretable error detection in deep neural networks. In \textit{International Conference on Computer Safety, Reliability, and Security}, pages 75–88.
Springer, 2023
\item[\textbf{P88:}] Thomas A Henzinger, Anna Lukina, and Christian Schilling. Outside the box: Abstraction-based monitoring of neural networks. In \textit{24th
European Conference on Artificial Intelligence}, volume 325, 2020
\item[\textbf{P89:}] Hans-Martin Heyn, Eric Knauss, Iswarya Malleswaran, and Shruthi Dinakaran. An empirical investigation of challenges of specifying training
data and runtime monitors for critical software with machine learning and their relation to architectural decisions. In \textit{Requirements Engineering},
29(1):97–117, 2024
\item[\textbf{P90:}] Joris Guerin, Raul Sena Ferreira, Kevin Delmas, and Jérémie Guiochet. Unifying evaluation of machine learning safety monitors. In \textit{2022 IEEE
33rd International Symposium on Software Reliability Engineering (ISSRE)}, pages 414–422. IEEE, 2022
\item[\textbf{P91:}] Manzoor Hussain, Nazakat Ali, and Jang-Eui Hong. Deepguard: A framework for safeguarding autonomous driving systems from inconsistent
behaviour. In \textit{Automated Software Engineering}, 29(1):1, 2022
\item[\textbf{P92:}]Yan Xiao, Ivan Beschastnikh, David S Rosenblum, Changsheng Sun, Sebastian Elbaum, Yun Lin, and Jin Song Dong. Self-checking deep neural
networks in deployment. In \textit{2021 IEEE/ACM 43rd International Conference on Software Engineering (ICSE)}, pages 372–384. IEEE, 2021
\item[\textbf{P93:}] Javier de la Rúa Martínez. Scalable architecture for automating machine learning model monitoring, Master’s thesis, "KTH Royal Institute of Technology", 2020
\item[\textbf{P94:}]Darren Cofer, Isaac Amundson, Ramachandra Sattigeri, Arjun Passi, Christopher Boggs, Eric Smith, Limei Gilham, Taejoon Byun, and Sanjai
Rayadurgam. Run-time assurance for learning-based aircraft taxiing. In \textit{2020 AIAA/IEEE 39th Digital Avionics Systems Conference (DASC)}, pages
1–9. IEEE, 2020
\item[\textbf{P95:}] Hariharan Arunachalam, Zhuoling Huang, Marc Hanheide, and Leonardo Guevara. Runtime anomaly monitoring of human perception models
for robotic systems. In \textit{2024 IEEE 20th International Conference on Automation Science and Engineering (CASE)}, pages 723–729. IEEE, 2024
\item[\textbf{P96:}] Vahid Hashemi, Jan Křetínsk `y, Sabine Rieder, Torsten Schön, and Jan Vorhoff. Gaussian-based and outside-the-box runtime monitoring join
forces. In \textit{International Conference on Runtime Verification}, pages 218–228. Springer, 2024
\item[\textbf{P97:}] Vahid Hashemi, Jan Křetínsk `y, Sabine Rieder, and Jessica Schmidt. Runtime monitoring for out-of-distribution detection in object detection
neural networks. In \textit{International Symposium on Formal Methods}, pages 622–634. Springer, 2023
\item[\textbf{P98:}] Yi Cai, Xiaohui Wan, Zhihao Liu, and Zheng Zheng. Drlfailuremonitor: A dynamic failure monitoring approach for deep reinforcement learning
system. In \textit{2024 IEEE 35th International Symposium on Software Reliability Engineering (ISSRE)}, pages 487–498. IEEE, 2024
\item[\textbf{P99:}] A Cummaudo, S Barnett, R Vasa, J Grundy. Beware the evolving 'intelligent'web service! An integration architecture tactic to guard AI-first components. In \textit{28th ACM Joint Meeting on European Software Engineering Conference and Symposium on the Foundations of Software Engineering}, 2020
\item[\textbf{P100:}] Jean Feng, Rachael V Phillips, Ivana Malenica, Andrew Bishara, Alan E Hubbard, Leo A Celi, and Romain Pirracchio. Clinical artificial intelligence
quality improvement: towards continual monitoring and updating of ai algorithms in healthcare. \textit{NPJ digital medicine}, 5(1):66, 2022
\item[\textbf{P101:}] Adrian-Ioan ARGESANU and Gheorghe-Daniel ANDREESCU. From data to decisions: The importance of monitoring ml systems in industrial
settings. \textit{Acta Technica Napocensis-Series: Applied Mathematics, Mechanics, and Engineering}, 66(3), 2023
\item[\textbf{P102:}] Quazi Marufur Rahman, Niko Sünderhauf, and Feras Dayoub. Online monitoring of object detection performance during deployment. In \textit{2021
IEEE/RSJ International Conference on Intelligent Robots and Systems (IROS)}, pages 4839–4845. IEEE, 2021
\item[\textbf{P103:}] Konstantin Kueffner, Anna Lukina, Christian Schilling, and Thomas A Henzinger. Into the unknown: active monitoring of neural networks
(extended version). In \textit{International Journal on Software Tools for Technology Transfer}, 25(4):575–592, 2023
\item[\textbf{P104:}] Arjun Gupta and Luca Carlone. Online monitoring for neural network based monocular pedestrian pose estimation. In \textit{2020 IEEE 23rd
International Conference on Intelligent Transportation Systems (ITSC)}, pages 1–8. IEEE, 2020
\item[\textbf{P105:}] Fateh Boudardara, Abderraouf Boussif, Pierre-Jean Meyer, and Mohamed Ghazel. Monitoring of neural network classifiers using neuron
activation paths. In \textit{International Conference on Verification and Evaluation of Computer and Communication Systems}, pages 81–96. Springer, 2024
\item[\textbf{P106:}] Yan Xiao, Ivan Beschastnikh, Yun Lin, Rajdeep Singh Hundal, Xiaofei Xie, David S Rosenblum, and Jin Song Dong. Self-checking deep neural
networks for anomalies and adversaries in deployment. In \textit{IEEE Transactions on Dependable and Secure Computing}, 2022
\item[\textbf{P107:}] Shengduo Chen, Yaowei Sun, Dachuan Li, Qiang Wang, Qi Hao, and Joseph Sifakis. Runtime safety assurance for learning-enabled control of
autonomous driving vehicles. In \textit{2022 International Conference on Robotics and Automation (ICRA)}, pages 8978–8984. IEEE, 2022
\item[\textbf{P108:}] Darren Cofer, Ramachandra Sattigeri, Isaac Amundson, Junaid Babar, Saqib Hasan, Eric W Smith, Karthik Nukala, Denis Osipychev, Matthew A
Moser, James L Paunicka, et al. Flight test of a collision avoidance neural network with run-time assurance. In \textit{2022 IEEE/AIAA 41st Digital
Avionics Systems Conference (DASC)}, pages 1–10. IEEE, 2022
\item[\textbf{P109:}] Ozan Vardal, Richard Hawkins, Colin Paterson, Chiara Picardi, Daniel Omeiza, Lars Kunze, and Ibrahim Habli. Learning run-time safety
monitors for machine learning components. \textit{arXiv preprint arXiv:2406.16220}, 2024
\item[\textbf{P110:}] Vasantha Kumar Venugopal, Abhishek Gupta, Rohit Takhar, and Vidur Mahajan. New epochs in ai supervision: Design and implementation of
an autonomous radiology ai monitoring system. \textit{arXiv preprint arXiv:2311.14305}, 2023
\item[\textbf{P111:}] Patrick Baier and Stanimir Dragiev. Challenges in live monitoring of machine learning systems. \textit{APPLICATION IN LIFE SCIENCES AND BEYOND},
page 1, 2021
\item[\textbf{P112:}] Wenbo Shao, Boqi Li, Wenhao Yu, Jiahui Xu, and Hong Wang. When is it likely to fail? performance monitor for black-box trajectory prediction
model. In \textit{IEEE Transactions on Automation Science and Engineering}, 2024
\item[\textbf{P113:}] Amazon Web Services. Mlops: Continuous delivery for machine learning on aws. Technical report, Amazon Web Services, 2020
\item[\textbf{P114:}] Keegan Hines. How to build a production-ready model monitoring system for your enterprise, June 2020. Accessed: 2025-04-28
\item[\textbf{P115:}] Comet ML. Production monitoring for the complete ml lifecycle, 2021. Accessed: 2025-04-28
\item[\textbf{P116:}] Shun Mao and Oleksandr Saienko. Building a scalable machine learning model monitoring system with datarobot, June 2023. Accessed:
2025-04-28
\item[\textbf{P117:}]Evidently AI. Model monitoring for ml in production: A comprehensive guide, January 2025. Accessed: 2025-04-28
\item[\textbf{P118:}] Google Cloud. Vertex ai model monitoring overview, 2025. Accessed: 2025-04-28
\item[\textbf{P119:}] H2O.ai. H2o mlops: Operate ai models with transparency and scale, 2025. Accessed: 2025-04-28
\item[\textbf{P120:}] IBM. Ibm watson studio, 2025. Accessed: 2025-04-28
\item[\textbf{P121:}] Iguazio. Model monitoring, 2025. Accessed: 2025-04-28
\item[\textbf{P122:}] KNIME. Model monitoring in a data science context, April 2021. Accessed: 2025-04-28
\item[\textbf{P123:}] Microsoft. Model monitoring in production, February 2025. Accessed: 2025-04-28
\item[\textbf{P124:}] Maciej Balawejder. How to set up a machine learning monitoring system with nannyml and grafana, April 2023. Accessed: 2025-04-28
\item[\textbf{P125:}] Oracle. Oracle machine learning monitoring, 2025. Accessed: 2025-04-28
\item[\textbf{P126:}] Valohai. What is model monitoring?, 2025. Accessed: 2025-04-28
\item[\textbf{P127:}] Vertier. Mlops with vetiver - monitor, 2025. Accessed: 2025-04-28
\item[\textbf{P128:}] Censius. Monitor machine learning models using censius ai observability platform, 2025. Accessed: 2025-04-28
\item[\textbf{P129:}] Mona Labs. Mona: Ai model performance insights platform, 2025. Accessed: 2025-04-28
\item[\textbf{P130:}] Datatron. Ai model monitoring and governance, 2025. Accessed: 2025-04-28
\item[\textbf{P131:}] Superwise. Ai monitoring, 2025. Accessed: 2025-04-28
\item[\textbf{P132:}] MLRun. Model monitoring, 2025. Accessed: 2025-04-28
\item[\textbf{P133:}] Hao Xu, Can Liu, and Chiranjeet Chetia. Explainable feature drift monitoring system for predictive machine learning models. 2024
\item[\textbf{P134:}] Carlos Mougan. ]Model monitoring in the absence of labeled data via feature attributions distributions. PhD thesis, University of Southampton,
2025
\item[\textbf{P135:}] Kai-Kristian Kemell, Jukka K Nurminen, and Ville Vakkuri. Monitoring machine learning systems from the point of view of ai ethics. In
\textit{Proceedings of the Conference on Technology Ethics 2024 (Tethics 2024)}. RWTH Aachen, 2024
\item[\textbf{P136:}] Adam Ekblom. Monitoring with mlops for clinical decision support. Master’s thesis, "Lund University", 2024

\end{itemize}
\end{footnotesize}

\bibliographystyle{ACM-Reference-Format}
\bibliography{sample-base}


\begin{thebibliography}{143}


\ifx \showCODEN    \undefined \def \showCODEN     #1{\unskip}     \fi
\ifx \showISBNx    \undefined \def \showISBNx     #1{\unskip}     \fi
\ifx \showISBNxiii \undefined \def \showISBNxiii  #1{\unskip}     \fi
\ifx \showISSN     \undefined \def \showISSN      #1{\unskip}     \fi
\ifx \showLCCN     \undefined \def \showLCCN      #1{\unskip}     \fi
\ifx \shownote     \undefined \def \shownote      #1{#1}          \fi
\ifx \showarticletitle \undefined \def \showarticletitle #1{#1}   \fi
\ifx \showURL      \undefined \def \showURL       {\relax}        \fi
\providecommand\bibfield[2]{#2}
\providecommand\bibinfo[2]{#2}
\providecommand\natexlab[1]{#1}
\providecommand\showeprint[2][]{arXiv:#2}

\bibitem[gre({[n.\,d.]})]%
        {greatexpectations}
 \bibinfo{year}{[n.\,d.]}\natexlab{}.
\newblock \bibinfo{title}{Great Expectations Core}.
\newblock
\urldef\tempurl%
\url{https://greatexpectations.io/gx-core/}
\showURL{%
\tempurl}
\newblock
\shownote{Accessed: 2025-07-02}.


\bibitem[kib({[n.\,d.]})]%
        {kibana}
 \bibinfo{year}{[n.\,d.]}\natexlab{}.
\newblock \bibinfo{title}{Kibana: Explore, visualize, and share insights from your data}.
\newblock
\urldef\tempurl%
\url{https://www.elastic.co/kibana}
\showURL{%
\tempurl}
\newblock
\shownote{Accessed: 2025-07-02}.


\bibitem[pro({[n.\,d.]})]%
        {prometheus}
 \bibinfo{year}{[n.\,d.]}\natexlab{}.
\newblock \bibinfo{title}{Prometheus: Monitoring system \& time series database}.
\newblock
\urldef\tempurl%
\url{https://prometheus.io/}
\showURL{%
\tempurl}
\newblock
\shownote{Accessed: 2025-07-02}.


\bibitem[Adamopoulou and Moussiades(2020)]%
        {adamopoulou2020chatbots}
\bibfield{author}{\bibinfo{person}{Eleni Adamopoulou} {and} \bibinfo{person}{Lefteris Moussiades}.} \bibinfo{year}{2020}\natexlab{}.
\newblock \showarticletitle{Chatbots: History, technology, and applications}.
\newblock \bibinfo{journal}{\emph{Machine Learning with applications}}  \bibinfo{volume}{2} (\bibinfo{year}{2020}), \bibinfo{pages}{100006}.
\newblock


\bibitem[{Amazon Web Services}(2024a)]%
        {awscloudwatch}
\bibfield{author}{\bibinfo{person}{{Amazon Web Services}}.} \bibinfo{year}{2024}\natexlab{a}.
\newblock \bibinfo{title}{Amazon CloudWatch}.
\newblock
\urldef\tempurl%
\url{https://aws.amazon.com/cloudwatch/}
\showURL{%
\tempurl}
\newblock
\shownote{Accessed: 2025-07-01}.


\bibitem[{Amazon Web Services}(2024b)]%
        {awssagemaker}
\bibfield{author}{\bibinfo{person}{{Amazon Web Services}}.} \bibinfo{year}{2024}\natexlab{b}.
\newblock \bibinfo{title}{Amazon SageMaker}.
\newblock
\urldef\tempurl%
\url{https://aws.amazon.com/sagemaker/}
\showURL{%
\tempurl}
\newblock
\shownote{Accessed: 2025-07-01}.


\bibitem[Amershi et~al\mbox{.}(2019a)]%
        {8804457}
\bibfield{author}{\bibinfo{person}{Saleema Amershi}, \bibinfo{person}{Andrew Begel}, \bibinfo{person}{Christian Bird}, \bibinfo{person}{Robert DeLine}, \bibinfo{person}{Harald Gall}, \bibinfo{person}{Ece Kamar}, \bibinfo{person}{Nachiappan Nagappan}, \bibinfo{person}{Besmira Nushi}, {and} \bibinfo{person}{Thomas Zimmermann}.} \bibinfo{year}{2019}\natexlab{a}.
\newblock \showarticletitle{Software Engineering for Machine Learning: A Case Study}. In \bibinfo{booktitle}{\emph{2019 IEEE/ACM 41st International Conference on Software Engineering: Software Engineering in Practice (ICSE-SEIP)}}. \bibinfo{pages}{291--300}.
\newblock
\href{https://doi.org/10.1109/ICSE-SEIP.2019.00042}{doi:\nolinkurl{10.1109/ICSE-SEIP.2019.00042}}


\bibitem[Amershi et~al\mbox{.}(2019b)]%
        {amershi2019software}
\bibfield{author}{\bibinfo{person}{Saleema Amershi}, \bibinfo{person}{Andrew Begel}, \bibinfo{person}{Christian Bird}, \bibinfo{person}{Robert DeLine}, \bibinfo{person}{Harald Gall}, \bibinfo{person}{Ece Kamar}, \bibinfo{person}{Nachiappan Nagappan}, \bibinfo{person}{Besmira Nushi}, {and} \bibinfo{person}{Thomas Zimmermann}.} \bibinfo{year}{2019}\natexlab{b}.
\newblock \showarticletitle{Software engineering for machine learning: A case study}. In \bibinfo{booktitle}{\emph{2019 IEEE/ACM 41st International Conference on Software Engineering: Software Engineering in Practice (ICSE-SEIP)}}. IEEE, \bibinfo{pages}{291--300}.
\newblock


\bibitem[Andersen et~al\mbox{.}(2024)]%
        {andersen2024monitoring}
\bibfield{author}{\bibinfo{person}{Eline~Sandvig Andersen}, \bibinfo{person}{Johan~Baden Birk-Korch}, \bibinfo{person}{Rasmus~S{\o}gaard Hansen}, \bibinfo{person}{Line~Haugaard Fly}, \bibinfo{person}{Richard R{\"o}ttger}, \bibinfo{person}{Diana Maria~Cespedes Arcani}, \bibinfo{person}{Claus~Lohman Brasen}, \bibinfo{person}{Ivan Brandslund}, {and} \bibinfo{person}{Jonna~Skov Madsen}.} \bibinfo{year}{2024}\natexlab{}.
\newblock \showarticletitle{Monitoring performance of clinical artificial intelligence in health care: a scoping review}.
\newblock \bibinfo{journal}{\emph{JBI evidence synthesis}} \bibinfo{volume}{22}, \bibinfo{number}{12} (\bibinfo{year}{2024}), \bibinfo{pages}{2423--2446}.
\newblock


\bibitem[{Apache Software Foundation}(2024)]%
        {apachejmeter}
\bibfield{author}{\bibinfo{person}{{Apache Software Foundation}}.} \bibinfo{year}{2024}\natexlab{}.
\newblock \bibinfo{title}{Apache JMeter}.
\newblock
\urldef\tempurl%
\url{https://jmeter.apache.org/}
\showURL{%
\tempurl}
\newblock
\shownote{Accessed: 2025-07-01}.


\bibitem[{Arthur AI}(2025)]%
        {arthur_ai_observability}
\bibfield{author}{\bibinfo{person}{{Arthur AI}}.} \bibinfo{year}{2025}\natexlab{}.
\newblock \bibinfo{title}{Observability}.
\newblock
\urldef\tempurl%
\url{https://www.arthur.ai/solution/observability}
\showURL{%
\tempurl}
\newblock
\shownote{Accessed: 2025-07-01}.


\bibitem[Awoyemi et~al\mbox{.}(2017)]%
        {awoyemi2017credit}
\bibfield{author}{\bibinfo{person}{John~O Awoyemi}, \bibinfo{person}{Adebayo~O Adetunmbi}, {and} \bibinfo{person}{Samuel~A Oluwadare}.} \bibinfo{year}{2017}\natexlab{}.
\newblock \showarticletitle{Credit card fraud detection using machine learning techniques: A comparative analysis}. In \bibinfo{booktitle}{\emph{2017 international conference on computing networking and informatics (ICCNI)}}. IEEE, \bibinfo{pages}{1--9}.
\newblock


\bibitem[Azeem et~al\mbox{.}(2024)]%
        {azeem2024monitizer}
\bibfield{author}{\bibinfo{person}{Muqsit Azeem}, \bibinfo{person}{Marta Grobelna}, \bibinfo{person}{Sudeep Kanav}, \bibinfo{person}{Jan K{\v{r}}et{\'\i}nsk{\`y}}, \bibinfo{person}{Stefanie Mohr}, {and} \bibinfo{person}{Sabine Rieder}.} \bibinfo{year}{2024}\natexlab{}.
\newblock \showarticletitle{Monitizer: automating design and evaluation of neural network monitors}. In \bibinfo{booktitle}{\emph{International Conference on Computer Aided Verification}}. Springer, \bibinfo{pages}{265--279}.
\newblock


\bibitem[{Berkeley LearnVerify Group}(2024)]%
        {verifai}
\bibfield{author}{\bibinfo{person}{{Berkeley LearnVerify Group}}.} \bibinfo{year}{2024}\natexlab{}.
\newblock \bibinfo{title}{VerifAI: A Toolkit for the Formal Design and Analysis of AI-Based Systems}.
\newblock
\urldef\tempurl%
\url{https://github.com/BerkeleyLearnVerify/VerifAI}
\showURL{%
\tempurl}
\newblock
\shownote{Accessed: 2025-07-01}.


\bibitem[Bernardi et~al\mbox{.}(2019)]%
        {bernardi2019150}
\bibfield{author}{\bibinfo{person}{Lucas Bernardi}, \bibinfo{person}{Themistoklis Mavridis}, {and} \bibinfo{person}{Pablo Estevez}.} \bibinfo{year}{2019}\natexlab{}.
\newblock \showarticletitle{150 successful machine learning models: 6 lessons learned at booking. com}. In \bibinfo{booktitle}{\emph{Proceedings of the 25th ACM SIGKDD international conference on knowledge discovery \& data mining}}. \bibinfo{pages}{1743--1751}.
\newblock


\bibitem[Bernardo et~al\mbox{.}(2025)]%
        {bernardo2025continuous}
\bibfield{author}{\bibinfo{person}{Jo{\~a}o~Helis Bernardo}, \bibinfo{person}{Daniel~Alencar da Costa}, \bibinfo{person}{Filipe~Roseiro Cogo}, \bibinfo{person}{S{\'e}rgio~Queir{\'o}z de Medeiros}, {and} \bibinfo{person}{Uir{\'a} Kulesza}.} \bibinfo{year}{2025}\natexlab{}.
\newblock \showarticletitle{Continuous Integration Practices in Machine Learning Projects: The PractitionersPerspective}.
\newblock \bibinfo{journal}{\emph{arXiv preprint arXiv:2502.17378}} (\bibinfo{year}{2025}).
\newblock


\bibitem[Bhargava and Singhal({[n.\,d.]})]%
        {bhargavachallenges}
\bibfield{author}{\bibinfo{person}{Surabhi Bhargava} {and} \bibinfo{person}{Shubham Singhal}.} \bibinfo{year}{[n.\,d.]}\natexlab{}.
\newblock \showarticletitle{Challenges, Solutions, and Best Practices in Post-Deployment Monitoring of Machine Learning Models}.
\newblock  (\bibinfo{year}{[n.\,d.]}).
\newblock


\bibitem[Bosch et~al\mbox{.}(2021)]%
        {bosch2021engineering}
\bibfield{author}{\bibinfo{person}{Jan Bosch}, \bibinfo{person}{Helena~Holmstr{\"o}m Olsson}, {and} \bibinfo{person}{Ivica Crnkovic}.} \bibinfo{year}{2021}\natexlab{}.
\newblock \showarticletitle{Engineering ai systems: A research agenda}.
\newblock \bibinfo{journal}{\emph{Artificial intelligence paradigms for smart cyber-physical systems}} (\bibinfo{year}{2021}), \bibinfo{pages}{1--19}.
\newblock


\bibitem[Breck et~al\mbox{.}(2017)]%
        {breck2017ml}
\bibfield{author}{\bibinfo{person}{Eric Breck}, \bibinfo{person}{Shanqing Cai}, \bibinfo{person}{Eric Nielsen}, \bibinfo{person}{Michael Salib}, {and} \bibinfo{person}{D Sculley}.} \bibinfo{year}{2017}\natexlab{}.
\newblock \showarticletitle{The ML test score: A rubric for ML production readiness and technical debt reduction}. In \bibinfo{booktitle}{\emph{2017 IEEE international conference on big data (big data)}}. IEEE, \bibinfo{pages}{1123--1132}.
\newblock


\bibitem[C. and collaborators(2023)]%
        {skshift}
\bibfield{author}{\bibinfo{person}{Sébastien~M. C.} {and} \bibinfo{person}{collaborators}.} \bibinfo{year}{2023}\natexlab{}.
\newblock \bibinfo{title}{skshift: A Python Toolbox for Shift Detection}.
\newblock
\urldef\tempurl%
\url{https://skshift.readthedocs.io/en/latest/}
\showURL{%
\tempurl}
\newblock
\shownote{Accessed: 2025-07-02}.


\bibitem[Carvalho et~al\mbox{.}(2019)]%
        {carvalho2019systematic}
\bibfield{author}{\bibinfo{person}{Thyago~P Carvalho}, \bibinfo{person}{Fabr{\'\i}zzio~AAMN Soares}, \bibinfo{person}{Roberto Vita}, \bibinfo{person}{Roberto da~P Francisco}, \bibinfo{person}{Jo{\~a}o~P Basto}, {and} \bibinfo{person}{Symone~GS Alcal{\'a}}.} \bibinfo{year}{2019}\natexlab{}.
\newblock \showarticletitle{A systematic literature review of machine learning methods applied to predictive maintenance}.
\newblock \bibinfo{journal}{\emph{Computers \& Industrial Engineering}}  \bibinfo{volume}{137} (\bibinfo{year}{2019}), \bibinfo{pages}{106024}.
\newblock


\bibitem[{Censius AI}(2025)]%
        {censius_ai_monitoring}
\bibfield{author}{\bibinfo{person}{{Censius AI}}.} \bibinfo{year}{2025}\natexlab{}.
\newblock \bibinfo{title}{Monitor Machine Learning Models using Censius AI Observability Platform}.
\newblock
\urldef\tempurl%
\url{https://censius.ai/monitoring}
\showURL{%
\tempurl}
\newblock
\shownote{Accessed: 2025-07-01}.


\bibitem[Chen et~al\mbox{.}(2022)]%
        {chen2022estimating}
\bibfield{author}{\bibinfo{person}{Lingjiao Chen}, \bibinfo{person}{Matei Zaharia}, {and} \bibinfo{person}{James~Y Zou}.} \bibinfo{year}{2022}\natexlab{}.
\newblock \showarticletitle{Estimating and explaining model performance when both covariates and labels shift}.
\newblock \bibinfo{journal}{\emph{Advances in Neural Information Processing Systems}}  \bibinfo{volume}{35} (\bibinfo{year}{2022}), \bibinfo{pages}{11467--11479}.
\newblock


\bibitem[Chen et~al\mbox{.}(2025)]%
        {chen2025design}
\bibfield{author}{\bibinfo{person}{Xin Chen}, \bibinfo{person}{Yan Li}, {and} \bibinfo{person}{Xiaoming Wang}.} \bibinfo{year}{2025}\natexlab{}.
\newblock \showarticletitle{Design Principles and Guidelines for LLM Observability: Insights from Developers}. In \bibinfo{booktitle}{\emph{Proceedings of the Extended Abstracts of the CHI Conference on Human Factors in Computing Systems}}. \bibinfo{pages}{1--9}.
\newblock


\bibitem[Clarke and Braun(2017)]%
        {clarke2017thematic}
\bibfield{author}{\bibinfo{person}{Victoria Clarke} {and} \bibinfo{person}{Virginia Braun}.} \bibinfo{year}{2017}\natexlab{}.
\newblock \showarticletitle{Thematic analysis}.
\newblock \bibinfo{journal}{\emph{The journal of positive psychology}} \bibinfo{volume}{12}, \bibinfo{number}{3} (\bibinfo{year}{2017}), \bibinfo{pages}{297--298}.
\newblock


\bibitem[{Comet ML, Inc.}(2025)]%
        {comet_model_production_monitoring}
\bibfield{author}{\bibinfo{person}{{Comet ML, Inc.}}} \bibinfo{year}{2025}\natexlab{}.
\newblock \bibinfo{title}{Model Production Monitoring}.
\newblock
\urldef\tempurl%
\url{https://www.comet.com/site/products/model-production-monitoring/}
\showURL{%
\tempurl}
\newblock
\shownote{Accessed: 2025-07-01}.


\bibitem[Community(2018)]%
        {heapster}
\bibfield{author}{\bibinfo{person}{Kubernetes Community}.} \bibinfo{year}{2018}\natexlab{}.
\newblock \bibinfo{title}{Heapster: Performance Monitoring and Metrics Collection for Kubernetes}.
\newblock
\urldef\tempurl%
\url{https://github.com/kubernetes-retired/heapster}
\showURL{%
\tempurl}
\newblock
\shownote{Project deprecated and archived. Accessed: 2025-07-01}.


\bibitem[Cruz et~al\mbox{.}(2025)]%
        {cruz2025greening}
\bibfield{author}{\bibinfo{person}{Lu{\'\i}s Cruz}, \bibinfo{person}{Jo{\~a}o~Paulo Fernandes}, \bibinfo{person}{Maja~H Kirkeby}, \bibinfo{person}{Silverio Mart{\'\i}nez-Fern{\'a}ndez}, \bibinfo{person}{June Sallou}, \bibinfo{person}{Hina Anwar}, \bibinfo{person}{Enrique Barba~Roque}, \bibinfo{person}{Justus Bogner}, \bibinfo{person}{Joel Casta{\~n}o}, \bibinfo{person}{Fernando Castor}, {et~al\mbox{.}}} \bibinfo{year}{2025}\natexlab{}.
\newblock \showarticletitle{Greening ai-enabled systems with software engineering: A research agenda for environmentally sustainable ai practices}.
\newblock \bibinfo{journal}{\emph{ACM SIGSOFT Software Engineering Notes}} \bibinfo{volume}{50}, \bibinfo{number}{3} (\bibinfo{year}{2025}), \bibinfo{pages}{14--23}.
\newblock


\bibitem[Cummaudo et~al\mbox{.}(2020)]%
        {cummaudo2020threshy}
\bibfield{author}{\bibinfo{person}{Alex Cummaudo}, \bibinfo{person}{Scott Barnett}, \bibinfo{person}{Rajesh Vasa}, {and} \bibinfo{person}{John Grundy}.} \bibinfo{year}{2020}\natexlab{}.
\newblock \showarticletitle{Threshy: Supporting safe usage of intelligent web services}. In \bibinfo{booktitle}{\emph{Proceedings of the 28th ACM Joint Meeting on European Software Engineering Conference and Symposium on the Foundations of Software Engineering}}. \bibinfo{pages}{1645--1649}.
\newblock


\bibitem[Darif et~al\mbox{.}(2025)]%
        {darif2025requirements}
\bibfield{author}{\bibinfo{person}{Ikram Darif}, \bibinfo{person}{Ghizlane El~Boussaidi}, {and} \bibinfo{person}{S{\`e}gla Kpodjedo}.} \bibinfo{year}{2025}\natexlab{}.
\newblock \showarticletitle{Requirements specification using templates: a model-driven approach}.
\newblock \bibinfo{journal}{\emph{Software and Systems Modeling}} (\bibinfo{year}{2025}), \bibinfo{pages}{1--38}.
\newblock


\bibitem[Dastin(2022)]%
        {dastin2022amazon}
\bibfield{author}{\bibinfo{person}{Jeffrey Dastin}.} \bibinfo{year}{2022}\natexlab{}.
\newblock \showarticletitle{Amazon scraps secret AI recruiting tool that showed bias against women}.
\newblock In \bibinfo{booktitle}{\emph{Ethics of data and analytics}}. \bibinfo{publisher}{Auerbach Publications}, \bibinfo{pages}{296--299}.
\newblock


\bibitem[{Datadog, Inc.}(2025)]%
        {datadog_product_overview}
\bibfield{author}{\bibinfo{person}{{Datadog, Inc.}}} \bibinfo{year}{2025}\natexlab{}.
\newblock \bibinfo{title}{Infrastructure \& Application Monitoring as a Service}.
\newblock
\urldef\tempurl%
\url{https://www.datadoghq.com/product/}
\showURL{%
\tempurl}
\newblock
\shownote{Accessed: 2025-07-01}.


\bibitem[{DataRobot, Inc.}(2025)]%
        {datarobot_ai_observability}
\bibfield{author}{\bibinfo{person}{{DataRobot, Inc.}}} \bibinfo{year}{2025}\natexlab{}.
\newblock \bibinfo{title}{AI Observability}.
\newblock
\urldef\tempurl%
\url{https://www.datarobot.com/product/ai-observability/}
\showURL{%
\tempurl}
\newblock
\shownote{Accessed: 2025-07-01}.


\bibitem[{Datatron, Inc.}(2025)]%
        {datatron_ai_monitoring_governance}
\bibfield{author}{\bibinfo{person}{{Datatron, Inc.}}} \bibinfo{year}{2025}\natexlab{}.
\newblock \bibinfo{title}{AI Model Monitoring and Governance}.
\newblock
\urldef\tempurl%
\url{https://datatron.com/ai-monitoring-ai-governance/}
\showURL{%
\tempurl}
\newblock
\shownote{Accessed: 2025-07-01}.


\bibitem[{Department of Industry, Science and Resources}(2024)]%
        {australia2024aiethics}
\bibfield{author}{\bibinfo{person}{{Department of Industry, Science and Resources}}.} \bibinfo{year}{2024}\natexlab{}.
\newblock \bibinfo{title}{Australia’s Artificial Intelligence Ethics Principles}.
\newblock
\urldef\tempurl%
\url{https://www.industry.gov.au/publications/australias-artificial-intelligence-ethics-principles}
\showURL{%
\tempurl}
\newblock
\shownote{Accessed: 2025-09-05}.


\bibitem[{Docker, Inc.}(2024)]%
        {dockerstats}
\bibfield{author}{\bibinfo{person}{{Docker, Inc.}}} \bibinfo{year}{2024}\natexlab{}.
\newblock \bibinfo{title}{docker container stats}.
\newblock
\urldef\tempurl%
\url{https://docs.docker.com/reference/cli/docker/container/stats/}
\showURL{%
\tempurl}
\newblock
\shownote{Accessed: 2025-07-01}.


\bibitem[Dong et~al\mbox{.}(2021)]%
        {dong2021survey}
\bibfield{author}{\bibinfo{person}{Shi Dong}, \bibinfo{person}{Ping Wang}, {and} \bibinfo{person}{Khushnood Abbas}.} \bibinfo{year}{2021}\natexlab{}.
\newblock \showarticletitle{A survey on deep learning and its applications}.
\newblock \bibinfo{journal}{\emph{Computer Science Review}}  \bibinfo{volume}{40} (\bibinfo{year}{2021}), \bibinfo{pages}{100379}.
\newblock


\bibitem[Dressel and Farid(2018)]%
        {dressel2018accuracy}
\bibfield{author}{\bibinfo{person}{Julia Dressel} {and} \bibinfo{person}{Hany Farid}.} \bibinfo{year}{2018}\natexlab{}.
\newblock \showarticletitle{The accuracy, fairness, and limits of predicting recidivism}.
\newblock \bibinfo{journal}{\emph{Science advances}} \bibinfo{volume}{4}, \bibinfo{number}{1} (\bibinfo{year}{2018}), \bibinfo{pages}{eaao5580}.
\newblock


\bibitem[Ehrlinger et~al\mbox{.}(2019)]%
        {ehrlinger2019daql}
\bibfield{author}{\bibinfo{person}{Lisa Ehrlinger}, \bibinfo{person}{Verena Haunschmid}, \bibinfo{person}{Davide Palazzini}, {and} \bibinfo{person}{Christian Lettner}.} \bibinfo{year}{2019}\natexlab{}.
\newblock \showarticletitle{A DaQL to monitor data quality in machine learning applications}. In \bibinfo{booktitle}{\emph{Database and Expert Systems Applications: 30th International Conference, DEXA 2019, Linz, Austria, August 26--29, 2019, Proceedings, Part I 30}}. Springer, \bibinfo{pages}{227--237}.
\newblock


\bibitem[Eken et~al\mbox{.}(2024)]%
        {eken2024multivocal}
\bibfield{author}{\bibinfo{person}{Beyza Eken}, \bibinfo{person}{Samodha Pallewatta}, \bibinfo{person}{Nguyen~Khoi Tran}, \bibinfo{person}{Ayse Tosun}, {and} \bibinfo{person}{Muhammad~Ali Babar}.} \bibinfo{year}{2024}\natexlab{}.
\newblock \showarticletitle{A Multivocal Review of MLOps Practices, Challenges and Open Issues}.
\newblock \bibinfo{journal}{\emph{arXiv preprint arXiv:2406.09737}} (\bibinfo{year}{2024}).
\newblock


\bibitem[{Elastic NV}(2024)]%
        {logstash}
\bibfield{author}{\bibinfo{person}{{Elastic NV}}.} \bibinfo{year}{2024}\natexlab{}.
\newblock \bibinfo{title}{Logstash: Collect, Parse, and Enrich Your Data}.
\newblock
\urldef\tempurl%
\url{https://www.elastic.co/logstash}
\showURL{%
\tempurl}
\newblock
\shownote{Accessed: 2025-07-01}.


\bibitem[{Ernst \& Young (EY)}(2024)]%
        {ey2024euai}
\bibfield{author}{\bibinfo{person}{{Ernst \& Young (EY)}}.} \bibinfo{year}{2024}\natexlab{}.
\newblock \bibinfo{title}{The EU AI Act: What it means for your business}.
\newblock
\urldef\tempurl%
\url{https://www.ey.com/en_ch/insights/forensic-integrity-services/the-eu-ai-act-what-it-means-for-your-business}
\showURL{%
\tempurl}
\newblock
\shownote{Accessed: 2025-09-05}.


\bibitem[{Evidently AI}(2025)]%
        {evidently2025monitoring}
\bibfield{author}{\bibinfo{person}{{Evidently AI}}.} \bibinfo{year}{2025}\natexlab{}.
\newblock \bibinfo{title}{Model Monitoring for ML in Production: A Comprehensive Guide}.
\newblock
\urldef\tempurl%
\url{https://www.evidentlyai.com/ml-in-production/model-monitoring}
\showURL{%
\tempurl}
\newblock
\shownote{Accessed: 2025-04-28}.


\bibitem[Ferreira et~al\mbox{.}(2021)]%
        {ferreira2021benchmarking}
\bibfield{author}{\bibinfo{person}{Raul~Sena Ferreira}, \bibinfo{person}{Jean Arlat}, \bibinfo{person}{J{\'e}r{\'e}mie Guiochet}, {and} \bibinfo{person}{H{\'e}l{\`e}ne Waeselynck}.} \bibinfo{year}{2021}\natexlab{}.
\newblock \showarticletitle{Benchmarking safety monitors for image classifiers with machine learning}. In \bibinfo{booktitle}{\emph{2021 IEEE 26th Pacific Rim International Symposium on Dependable Computing (PRDC)}}. IEEE, \bibinfo{pages}{7--16}.
\newblock


\bibitem[Ferreira et~al\mbox{.}(2024)]%
        {ferreira2024safety}
\bibfield{author}{\bibinfo{person}{Raul~Sena Ferreira}, \bibinfo{person}{Joris Gu{\'e}rin}, \bibinfo{person}{Kevin Delmas}, \bibinfo{person}{J{\'e}r{\'e}mie Guiochet}, {and} \bibinfo{person}{H{\'e}l{\`e}ne Waeselynck}.} \bibinfo{year}{2024}\natexlab{}.
\newblock \showarticletitle{Safety Monitoring of Machine Learning Perception Functions: a Survey}.
\newblock \bibinfo{journal}{\emph{arXiv preprint arXiv:2412.06869}} (\bibinfo{year}{2024}).
\newblock


\bibitem[Ganesan(2024)]%
        {ganesan2024llm}
\bibfield{author}{\bibinfo{person}{P Ganesan}.} \bibinfo{year}{2024}\natexlab{}.
\newblock \showarticletitle{LLM-Powered Observability Enhancing Monitoring and Diagnostics}.
\newblock \bibinfo{journal}{\emph{J Artif Intell Mach Learn \& Data Sci}} \bibinfo{volume}{2}, \bibinfo{number}{2} (\bibinfo{year}{2024}), \bibinfo{pages}{1329--1336}.
\newblock


\bibitem[Gao et~al\mbox{.}(2023)]%
        {gao2023retrieval}
\bibfield{author}{\bibinfo{person}{Yunfan Gao}, \bibinfo{person}{Yun Xiong}, \bibinfo{person}{Xinyu Gao}, \bibinfo{person}{Kangxiang Jia}, \bibinfo{person}{Jinliu Pan}, \bibinfo{person}{Yuxi Bi}, \bibinfo{person}{Yixin Dai}, \bibinfo{person}{Jiawei Sun}, \bibinfo{person}{Haofen Wang}, {and} \bibinfo{person}{Haofen Wang}.} \bibinfo{year}{2023}\natexlab{}.
\newblock \showarticletitle{Retrieval-augmented generation for large language models: A survey}.
\newblock \bibinfo{journal}{\emph{arXiv preprint arXiv:2312.10997}} \bibinfo{volume}{2}, \bibinfo{number}{1} (\bibinfo{year}{2023}).
\newblock


\bibitem[Garousi et~al\mbox{.}(2019)]%
        {garousi2019guidelines}
\bibfield{author}{\bibinfo{person}{Vahid Garousi}, \bibinfo{person}{Michael Felderer}, {and} \bibinfo{person}{Mika~V M{\"a}ntyl{\"a}}.} \bibinfo{year}{2019}\natexlab{}.
\newblock \showarticletitle{Guidelines for including grey literature and conducting multivocal literature reviews in software engineering}.
\newblock \bibinfo{journal}{\emph{Information and software technology}}  \bibinfo{volume}{106} (\bibinfo{year}{2019}), \bibinfo{pages}{101--121}.
\newblock


\bibitem[G{\'e}ron(2022)]%
        {geron2022hands}
\bibfield{author}{\bibinfo{person}{Aur{\'e}lien G{\'e}ron}.} \bibinfo{year}{2022}\natexlab{}.
\newblock \bibinfo{booktitle}{\emph{Hands-on machine learning with Scikit-Learn, Keras, and TensorFlow}}.
\newblock \bibinfo{publisher}{" O'Reilly Media, Inc."}.
\newblock


\bibitem[Ghaderi~Zefrehi et~al\mbox{.}(2023)]%
        {ghaderi2023threshold}
\bibfield{author}{\bibinfo{person}{Hossein Ghaderi~Zefrehi}, \bibinfo{person}{Ghazaal Sheikhi}, {and} \bibinfo{person}{Hakan Alt{\i}n{\c{c}}ay}.} \bibinfo{year}{2023}\natexlab{}.
\newblock \showarticletitle{Threshold prediction for detecting rare positive samples using a meta-learner}.
\newblock \bibinfo{journal}{\emph{Pattern Analysis and Applications}} \bibinfo{volume}{26}, \bibinfo{number}{1} (\bibinfo{year}{2023}), \bibinfo{pages}{289--306}.
\newblock


\bibitem[Ghanta et~al\mbox{.}(2019)]%
        {ghanta2019ml}
\bibfield{author}{\bibinfo{person}{Sindhu Ghanta}, \bibinfo{person}{Sriram Subramanian}, \bibinfo{person}{Lior Khermosh}, \bibinfo{person}{Swaminathan Sundararaman}, \bibinfo{person}{Harshil Shah}, \bibinfo{person}{Yakov Goldberg}, \bibinfo{person}{Drew Roselli}, {and} \bibinfo{person}{Nisha Talagala}.} \bibinfo{year}{2019}\natexlab{}.
\newblock \showarticletitle{ML health monitor: taking the pulse of machine learning algorithms in production}. In \bibinfo{booktitle}{\emph{Applications of Machine Learning}}, Vol.~\bibinfo{volume}{11139}. SPIE, \bibinfo{pages}{191--202}.
\newblock


\bibitem[Ghosh et~al\mbox{.}(2022)]%
        {ghosh2022faircanary}
\bibfield{author}{\bibinfo{person}{Avijit Ghosh}, \bibinfo{person}{Aalok Shanbhag}, {and} \bibinfo{person}{Christo Wilson}.} \bibinfo{year}{2022}\natexlab{}.
\newblock \showarticletitle{Faircanary: Rapid continuous explainable fairness}. In \bibinfo{booktitle}{\emph{Proceedings of the 2022 AAAI/ACM Conference on AI, Ethics, and Society}}. \bibinfo{pages}{307--316}.
\newblock


\bibitem[Ginart et~al\mbox{.}(2022)]%
        {ginart2022mldemon}
\bibfield{author}{\bibinfo{person}{Tony Ginart}, \bibinfo{person}{Martin~Jinye Zhang}, {and} \bibinfo{person}{James Zou}.} \bibinfo{year}{2022}\natexlab{}.
\newblock \showarticletitle{Mldemon: Deployment monitoring for machine learning systems}. In \bibinfo{booktitle}{\emph{International conference on artificial intelligence and statistics}}. PMLR, \bibinfo{pages}{3962--3997}.
\newblock


\bibitem[{Google Cloud}(2024)]%
        {googlecloud}
\bibfield{author}{\bibinfo{person}{{Google Cloud}}.} \bibinfo{year}{2024}\natexlab{}.
\newblock \bibinfo{title}{Google Cloud Platform}.
\newblock
\urldef\tempurl%
\url{https://cloud.google.com/?hl=en}
\showURL{%
\tempurl}
\newblock
\shownote{Accessed: 2025-07-29}.


\bibitem[{Google LLC}(2025)]%
        {google_vertex_ai_model_monitoring}
\bibfield{author}{\bibinfo{person}{{Google LLC}}.} \bibinfo{year}{2025}\natexlab{}.
\newblock \bibinfo{title}{Vertex AI Model Monitoring Overview}.
\newblock
\urldef\tempurl%
\url{https://cloud.google.com/vertex-ai/docs/model-monitoring/overview}
\showURL{%
\tempurl}
\newblock
\shownote{Accessed: 2025-07-01}.


\bibitem[{Grafana Labs}(2024)]%
        {grafanaobservability}
\bibfield{author}{\bibinfo{person}{{Grafana Labs}}.} \bibinfo{year}{2024}\natexlab{}.
\newblock \bibinfo{title}{Grafana Cloud Application Observability}.
\newblock
\urldef\tempurl%
\url{https://grafana.com/products/cloud/application-observability/}
\showURL{%
\tempurl}
\newblock
\shownote{Accessed: 2025-07-01}.


\bibitem[Grieser et~al\mbox{.}(2020)]%
        {grieser2020assuring}
\bibfield{author}{\bibinfo{person}{J{\"o}rg Grieser}, \bibinfo{person}{Meng Zhang}, \bibinfo{person}{Tim Warnecke}, {and} \bibinfo{person}{Andreas Rausch}.} \bibinfo{year}{2020}\natexlab{}.
\newblock \showarticletitle{Assuring the safety of end-to-end learning-based autonomous driving through runtime monitoring}. In \bibinfo{booktitle}{\emph{2020 23rd Euromicro Conference on Digital System Design (DSD)}}. IEEE, \bibinfo{pages}{476--483}.
\newblock


\bibitem[{H2O.ai}(2025)]%
        {h2o_mlops}
\bibfield{author}{\bibinfo{person}{{H2O.ai}}.} \bibinfo{year}{2025}\natexlab{}.
\newblock \bibinfo{title}{H2O MLOps: Operate AI Models with Transparency and Scale}.
\newblock
\urldef\tempurl%
\url{https://h2o.ai/platform/ai-cloud/operate/h2o-mlops/}
\showURL{%
\tempurl}
\newblock
\shownote{Accessed: 2025-07-01}.


\bibitem[Hardt et~al\mbox{.}(2021)]%
        {hardt2021amazon}
\bibfield{author}{\bibinfo{person}{Michaela Hardt}, \bibinfo{person}{Xiaoguang Chen}, \bibinfo{person}{Xiaoyi Cheng}, \bibinfo{person}{Michele Donini}, \bibinfo{person}{Jason Gelman}, \bibinfo{person}{Satish Gollaprolu}, \bibinfo{person}{John He}, \bibinfo{person}{Pedro Larroy}, \bibinfo{person}{Xinyu Liu}, \bibinfo{person}{Nick McCarthy}, {et~al\mbox{.}}} \bibinfo{year}{2021}\natexlab{}.
\newblock \showarticletitle{Amazon sagemaker clarify: Machine learning bias detection and explainability in the cloud}. In \bibinfo{booktitle}{\emph{Proceedings of the 27th ACM SIGKDD conference on knowledge discovery \& data mining}}. \bibinfo{pages}{2974--2983}.
\newblock


\bibitem[Heyn et~al\mbox{.}(2024)]%
        {heyn2024empirical}
\bibfield{author}{\bibinfo{person}{Hans-Martin Heyn}, \bibinfo{person}{Eric Knauss}, \bibinfo{person}{Iswarya Malleswaran}, {and} \bibinfo{person}{Shruthi Dinakaran}.} \bibinfo{year}{2024}\natexlab{}.
\newblock \showarticletitle{An empirical investigation of challenges of specifying training data and runtime monitors for critical software with machine learning and their relation to architectural decisions}.
\newblock \bibinfo{journal}{\emph{Requirements Engineering}} \bibinfo{volume}{29}, \bibinfo{number}{1} (\bibinfo{year}{2024}), \bibinfo{pages}{97--117}.
\newblock


\bibitem[{IBM Corporation}(2025)]%
        {ibm_watson_studio}
\bibfield{author}{\bibinfo{person}{{IBM Corporation}}.} \bibinfo{year}{2025}\natexlab{}.
\newblock \bibinfo{title}{IBM Watson Studio}.
\newblock
\urldef\tempurl%
\url{https://www.ibm.com/products/watson-studio}
\showURL{%
\tempurl}
\newblock
\shownote{Accessed: 2025-07-01}.


\bibitem[IEEE(1990)]%
        {ieee1990ieee}
\bibfield{author}{\bibinfo{person}{RS IEEE}.} \bibinfo{year}{1990}\natexlab{}.
\newblock \showarticletitle{IEEE Standard Glossary of Software Engineering Terminology. IEEE Std 610.12-1990}.
\newblock \bibinfo{journal}{\emph{CA: IEEE Comput. Societ.}}  \bibinfo{volume}{169} (\bibinfo{year}{1990}).
\newblock


\bibitem[{Iguazio}(2025)]%
        {iguazio_model_monitoring}
\bibfield{author}{\bibinfo{person}{{Iguazio}}.} \bibinfo{year}{2025}\natexlab{}.
\newblock \bibinfo{title}{Model Monitoring}.
\newblock
\urldef\tempurl%
\url{https://www.iguazio.com/solutions/model-monitoring/}
\showURL{%
\tempurl}
\newblock
\shownote{Accessed: 2025-04-28}.


\bibitem[{Iguazio Ltd.}(2024)]%
        {mlrun}
\bibfield{author}{\bibinfo{person}{{Iguazio Ltd.}}} \bibinfo{year}{2024}\natexlab{}.
\newblock \bibinfo{title}{MLRun - An Open Source MLOps Framework}.
\newblock
\urldef\tempurl%
\url{https://www.mlrun.org/}
\showURL{%
\tempurl}
\newblock
\shownote{Accessed: 2025-07-01}.


\bibitem[{InfluxData}(2024)]%
        {telegraf}
\bibfield{author}{\bibinfo{person}{{InfluxData}}.} \bibinfo{year}{2024}\natexlab{}.
\newblock \bibinfo{title}{Telegraf: Server Agent for Collecting and Reporting Metrics}.
\newblock
\urldef\tempurl%
\url{https://www.influxdata.com/time-series-platform/telegraf/}
\showURL{%
\tempurl}
\newblock
\shownote{Accessed: 2025-07-01}.


\bibitem[Jain et~al\mbox{.}(2023)]%
        {jain2023survey}
\bibfield{author}{\bibinfo{person}{Yash Jain}, \bibinfo{person}{Sidhant Khamankar}, \bibinfo{person}{Husain Fatepurwala}, \bibinfo{person}{Suruchi Dedgaonkar}, \bibinfo{person}{Priya Shelke}, {and} \bibinfo{person}{Riddhi Mirajkar}.} \bibinfo{year}{2023}\natexlab{}.
\newblock \showarticletitle{Survey on Model Monitoring in NLP.}
\newblock \bibinfo{journal}{\emph{Grenze International Journal of Engineering \& Technology (GIJET)}} \bibinfo{volume}{9}, \bibinfo{number}{2} (\bibinfo{year}{2023}).
\newblock


\bibitem[Jarrett and collaborators(2023)]%
        {explanationspace}
\bibfield{author}{\bibinfo{person}{Daniel Jarrett} {and} \bibinfo{person}{collaborators}.} \bibinfo{year}{2023}\natexlab{}.
\newblock \bibinfo{title}{Explanation Space: A Toolkit for Explaining Machine Learning Models}.
\newblock
\urldef\tempurl%
\url{https://explanationspace.readthedocs.io/en/latest/}
\showURL{%
\tempurl}
\newblock
\shownote{Accessed: 2025-07-02}.


\bibitem[J{\"a}rvenp{\"a}{\"a} et~al\mbox{.}(2024)]%
        {jarvenpaa2024synthesis}
\bibfield{author}{\bibinfo{person}{Heli J{\"a}rvenp{\"a}{\"a}}, \bibinfo{person}{Patricia Lago}, \bibinfo{person}{Justus Bogner}, \bibinfo{person}{Grace Lewis}, \bibinfo{person}{Henry Muccini}, {and} \bibinfo{person}{Ipek Ozkaya}.} \bibinfo{year}{2024}\natexlab{}.
\newblock \showarticletitle{A synthesis of green architectural tactics for ml-enabled systems}. In \bibinfo{booktitle}{\emph{Proceedings of the 46th International Conference on Software Engineering: Software Engineering in Society}}. \bibinfo{pages}{130--141}.
\newblock


\bibitem[Jayalath and Ramaswamy(2022)]%
        {jayalath2022enhancing}
\bibfield{author}{\bibinfo{person}{Hemadri Jayalath} {and} \bibinfo{person}{Lakshmish Ramaswamy}.} \bibinfo{year}{2022}\natexlab{}.
\newblock \showarticletitle{Enhancing performance of operationalized machine learning models by analyzing user feedback}. In \bibinfo{booktitle}{\emph{Proceedings of the 2022 4th International Conference on Image, Video and Signal Processing}}. \bibinfo{pages}{197--203}.
\newblock


\bibitem[Kale et~al\mbox{.}(2025)]%
        {kale2025reliable}
\bibfield{author}{\bibinfo{person}{Neil Kale}, \bibinfo{person}{Chen Bo~Calvin Zhang}, \bibinfo{person}{Kevin Zhu}, \bibinfo{person}{Ankit Aich}, \bibinfo{person}{Paula Rodriguez}, \bibinfo{person}{Scale~Red Team}, \bibinfo{person}{Christina~Q Knight}, {and} \bibinfo{person}{Zifan Wang}.} \bibinfo{year}{2025}\natexlab{}.
\newblock \showarticletitle{Reliable Weak-to-Strong Monitoring of LLM Agents}.
\newblock \bibinfo{journal}{\emph{arXiv preprint arXiv:2508.19461}} (\bibinfo{year}{2025}).
\newblock


\bibitem[Karval and Singh(2023)]%
        {karval2023catching}
\bibfield{author}{\bibinfo{person}{Ritesh Karval} {and} \bibinfo{person}{Kamakhya~Narain Singh}.} \bibinfo{year}{2023}\natexlab{}.
\newblock \showarticletitle{Catching Silent Failures: A Machine Learning Model Monitoring and Explainability Survey}. In \bibinfo{booktitle}{\emph{2023 OITS International Conference on Information Technology (OCIT)}}. IEEE, \bibinfo{pages}{526--532}.
\newblock


\bibitem[Khanal et~al\mbox{.}(2020)]%
        {khanal2020systematic}
\bibfield{author}{\bibinfo{person}{Shristi~Shakya Khanal}, \bibinfo{person}{PWC Prasad}, \bibinfo{person}{Abeer Alsadoon}, {and} \bibinfo{person}{Angelika Maag}.} \bibinfo{year}{2020}\natexlab{}.
\newblock \showarticletitle{A systematic review: machine learning based recommendation systems for e-learning}.
\newblock \bibinfo{journal}{\emph{Education and Information Technologies}} \bibinfo{volume}{25}, \bibinfo{number}{4} (\bibinfo{year}{2020}), \bibinfo{pages}{2635--2664}.
\newblock


\bibitem[Khomh et~al\mbox{.}(2018)]%
        {khomh2018software}
\bibfield{author}{\bibinfo{person}{Foutse Khomh}, \bibinfo{person}{Bram Adams}, \bibinfo{person}{Jinghui Cheng}, \bibinfo{person}{Marios Fokaefs}, {and} \bibinfo{person}{Giuliano Antoniol}.} \bibinfo{year}{2018}\natexlab{}.
\newblock \showarticletitle{Software engineering for machine-learning applications: The road ahead}.
\newblock \bibinfo{journal}{\emph{IEEE Software}} \bibinfo{volume}{35}, \bibinfo{number}{5} (\bibinfo{year}{2018}), \bibinfo{pages}{81--84}.
\newblock


\bibitem[Kim and Lee(2024)]%
        {kim2024enhancing}
\bibfield{author}{\bibinfo{person}{Hyuntae Kim} {and} \bibinfo{person}{Changhee Lee}.} \bibinfo{year}{2024}\natexlab{}.
\newblock \showarticletitle{Enhancing anomaly detection via generating diversified and hard-to-distinguish synthetic anomalies}. In \bibinfo{booktitle}{\emph{Proceedings of the 33rd ACM International Conference on Information and Knowledge Management}}. \bibinfo{pages}{1089--1098}.
\newblock


\bibitem[Kitchenham et~al\mbox{.}(2009)]%
        {kitchenham2009systematic}
\bibfield{author}{\bibinfo{person}{Barbara Kitchenham}, \bibinfo{person}{O~Pearl Brereton}, \bibinfo{person}{David Budgen}, \bibinfo{person}{Mark Turner}, \bibinfo{person}{John Bailey}, {and} \bibinfo{person}{Stephen Linkman}.} \bibinfo{year}{2009}\natexlab{}.
\newblock \showarticletitle{Systematic literature reviews in software engineering--a systematic literature review}.
\newblock \bibinfo{journal}{\emph{Information and software technology}} \bibinfo{volume}{51}, \bibinfo{number}{1} (\bibinfo{year}{2009}), \bibinfo{pages}{7--15}.
\newblock


\bibitem[Klaise et~al\mbox{.}(2020)]%
        {klaise2020monitoring}
\bibfield{author}{\bibinfo{person}{Janis Klaise}, \bibinfo{person}{Arnaud Van~Looveren}, \bibinfo{person}{Clive Cox}, \bibinfo{person}{Giovanni Vacanti}, {and} \bibinfo{person}{Alexandru Coca}.} \bibinfo{year}{2020}\natexlab{}.
\newblock \showarticletitle{Monitoring and explainability of models in production}.
\newblock \bibinfo{journal}{\emph{arXiv preprint arXiv:2007.06299}} (\bibinfo{year}{2020}).
\newblock


\bibitem[Klein(2021)]%
        {klein2021mlops}
\bibfield{author}{\bibinfo{person}{Bruno Klein}.} \bibinfo{year}{2021}\natexlab{}.
\newblock \bibinfo{title}{Planning for Successful MLOps}.
\newblock \bibinfo{howpublished}{AWS Prescriptive Guidance}.
\newblock
\urldef\tempurl%
\url{https://docs.aws.amazon.com/prescriptive-guidance/latest/ml-operations-planning/ml-operations-planning.pdf}
\showURL{%
\tempurl}
\newblock
\shownote{Amazon Web Services, December 2021}.


\bibitem[{KNIME}(2021)]%
        {knime2021monitoring}
\bibfield{author}{\bibinfo{person}{{KNIME}}.} \bibinfo{year}{2021}\natexlab{}.
\newblock \bibinfo{title}{Model Monitoring in a Data Science Context}.
\newblock
\urldef\tempurl%
\url{https://www.knime.com/blog/model-monitoring-in-data-science-context}
\showURL{%
\tempurl}
\newblock
\shownote{Accessed: 2025-04-28}.


\bibitem[Ko et~al\mbox{.}(2023)]%
        {ko2023privmon}
\bibfield{author}{\bibinfo{person}{Myeongseob Ko}, \bibinfo{person}{Xinyu Yang}, \bibinfo{person}{Zhengjie Ji}, \bibinfo{person}{Hoang~Anh Just}, \bibinfo{person}{Peng Gao}, \bibinfo{person}{Anoop Kumar}, {and} \bibinfo{person}{Ruoxi Jia}.} \bibinfo{year}{2023}\natexlab{}.
\newblock \showarticletitle{Privmon: A stream-based system for real-time privacy attack detection for machine learning models}. In \bibinfo{booktitle}{\emph{Proceedings of the 26th International Symposium on Research in Attacks, Intrusions and Defenses}}. \bibinfo{pages}{264--281}.
\newblock


\bibitem[Kourouklidis et~al\mbox{.}(2023)]%
        {kourouklidis2023domain}
\bibfield{author}{\bibinfo{person}{Panagiotis Kourouklidis}, \bibinfo{person}{Dimitris Kolovos}, \bibinfo{person}{Joost Noppen}, {and} \bibinfo{person}{Nicholas Matragkas}.} \bibinfo{year}{2023}\natexlab{}.
\newblock \showarticletitle{A domain-specific language for monitoring ML model performance}. In \bibinfo{booktitle}{\emph{2023 ACM/IEEE International Conference on Model Driven Engineering Languages and Systems Companion (MODELS-C)}}. IEEE, \bibinfo{pages}{266--275}.
\newblock


\bibitem[Lee and Shin(2020)]%
        {lee2020machine}
\bibfield{author}{\bibinfo{person}{In Lee} {and} \bibinfo{person}{Yong~Jae Shin}.} \bibinfo{year}{2020}\natexlab{}.
\newblock \showarticletitle{Machine learning for enterprises: Applications, algorithm selection, and challenges}.
\newblock \bibinfo{journal}{\emph{Business Horizons}} \bibinfo{volume}{63}, \bibinfo{number}{2} (\bibinfo{year}{2020}), \bibinfo{pages}{157--170}.
\newblock


\bibitem[Leest et~al\mbox{.}(2024)]%
        {leest2024expert}
\bibfield{author}{\bibinfo{person}{Joran Leest}, \bibinfo{person}{Claudia Raibulet}, \bibinfo{person}{Ilias Gerostathopoulos}, {and} \bibinfo{person}{Patricia Lago}.} \bibinfo{year}{2024}\natexlab{}.
\newblock \showarticletitle{Expert Monitoring: Human-Centered Concept Drift Detection in Machine Learning Operations}. In \bibinfo{booktitle}{\emph{Proceedings of the 2024 ACM/IEEE 44th International Conference on Software Engineering: New Ideas and Emerging Results}}. \bibinfo{pages}{1--5}.
\newblock


\bibitem[Leest et~al\mbox{.}(2025)]%
        {leest2025tea}
\bibfield{author}{\bibinfo{person}{Joran Leest}, \bibinfo{person}{Claudia Raibulet}, \bibinfo{person}{Patricia Lago}, {and} \bibinfo{person}{Ilias Gerostathopoulos}.} \bibinfo{year}{2025}\natexlab{}.
\newblock \showarticletitle{From Tea Leaves to System Maps: A Survey and Framework on Context-aware Machine Learning Monitoring}.
\newblock \bibinfo{journal}{\emph{IEEE Transactions on Software Engineering}} (\bibinfo{year}{2025}).
\newblock


\bibitem[Lewis et~al\mbox{.}(2022)]%
        {lewis2022augur}
\bibfield{author}{\bibinfo{person}{Grace~A Lewis}, \bibinfo{person}{Sebasti{\'a}n Echeverr{\'\i}a}, \bibinfo{person}{Lena Pons}, {and} \bibinfo{person}{Jeffrey Chrabaszcz}.} \bibinfo{year}{2022}\natexlab{}.
\newblock \showarticletitle{Augur: A step towards realistic drift detection in production ml systems}. In \bibinfo{booktitle}{\emph{Proceedings of the 1st Workshop on Software Engineering for Responsible AI}}. \bibinfo{pages}{37--44}.
\newblock


\bibitem[Lipton et~al\mbox{.}(2018)]%
        {lipton2018detecting}
\bibfield{author}{\bibinfo{person}{Zachary Lipton}, \bibinfo{person}{Yu-Xiang Wang}, {and} \bibinfo{person}{Alexander Smola}.} \bibinfo{year}{2018}\natexlab{}.
\newblock \showarticletitle{Detecting and correcting for label shift with black box predictors}. In \bibinfo{booktitle}{\emph{International conference on machine learning}}. PMLR, \bibinfo{pages}{3122--3130}.
\newblock


\bibitem[Ltd(2020)]%
        {alibidetect}
\bibfield{author}{\bibinfo{person}{Seldon~Technologies Ltd}.} \bibinfo{year}{2020}\natexlab{}.
\newblock \bibinfo{title}{Alibi Detect: Algorithms for Outlier, Adversarial and Drift Detection}.
\newblock
\urldef\tempurl%
\url{https://github.com/SeldonIO/alibi-detect}
\showURL{%
\tempurl}
\newblock
\shownote{Accessed: 2025-07-02}.


\bibitem[Lu et~al\mbox{.}(2018)]%
        {lu2018learning}
\bibfield{author}{\bibinfo{person}{Jie Lu}, \bibinfo{person}{Anjin Liu}, \bibinfo{person}{Fan Dong}, \bibinfo{person}{Feng Gu}, \bibinfo{person}{Joao Gama}, {and} \bibinfo{person}{Guangquan Zhang}.} \bibinfo{year}{2018}\natexlab{}.
\newblock \showarticletitle{Learning under concept drift: A review}.
\newblock \bibinfo{journal}{\emph{IEEE transactions on knowledge and data engineering}} \bibinfo{volume}{31}, \bibinfo{number}{12} (\bibinfo{year}{2018}), \bibinfo{pages}{2346--2363}.
\newblock


\bibitem[Mart{\'\i}nez-Fern{\'a}ndez et~al\mbox{.}(2021)]%
        {martinez2021developing}
\bibfield{author}{\bibinfo{person}{Silverio Mart{\'\i}nez-Fern{\'a}ndez}, \bibinfo{person}{Xavier Franch}, \bibinfo{person}{Andreas Jedlitschka}, \bibinfo{person}{Marc Oriol}, {and} \bibinfo{person}{Adam Trendowicz}.} \bibinfo{year}{2021}\natexlab{}.
\newblock \showarticletitle{Developing and operating artificial intelligence models in trustworthy autonomous systems}. In \bibinfo{booktitle}{\emph{International Conference on Research Challenges in Information Science}}. Springer, \bibinfo{pages}{221--229}.
\newblock


\bibitem[Mavin et~al\mbox{.}(2009)]%
        {mavin2009easy}
\bibfield{author}{\bibinfo{person}{Alistair Mavin}, \bibinfo{person}{Philip Wilkinson}, \bibinfo{person}{Adrian Harwood}, {and} \bibinfo{person}{Mark Novak}.} \bibinfo{year}{2009}\natexlab{}.
\newblock \showarticletitle{Easy approach to requirements syntax (EARS)}. In \bibinfo{booktitle}{\emph{2009 17th IEEE international requirements engineering conference}}. IEEE, \bibinfo{pages}{317--322}.
\newblock


\bibitem[{Microsoft Corporation}(2024)]%
        {azuremlmonitoring}
\bibfield{author}{\bibinfo{person}{{Microsoft Corporation}}.} \bibinfo{year}{2024}\natexlab{}.
\newblock \bibinfo{title}{Model monitoring in Azure Machine Learning}.
\newblock
\urldef\tempurl%
\url{https://learn.microsoft.com/en-us/azure/machine-learning/concept-model-monitoring?view=azureml-api-2}
\showURL{%
\tempurl}
\newblock
\shownote{Accessed: 2025-07-01}.


\bibitem[{Microsoft Corporation}(2025)]%
        {azure_monitor_overview}
\bibfield{author}{\bibinfo{person}{{Microsoft Corporation}}.} \bibinfo{year}{2025}\natexlab{}.
\newblock \bibinfo{title}{Azure Monitor overview}.
\newblock
\urldef\tempurl%
\url{https://learn.microsoft.com/en-us/azure/azure-monitor/fundamentals/overview}
\showURL{%
\tempurl}
\newblock
\shownote{Accessed: 2025-07-01}.


\bibitem[{Microsoft Industry Solutions Engineering}(2024)]%
        {microsoft_ml_model_checklist}
\bibfield{author}{\bibinfo{person}{{Microsoft Industry Solutions Engineering}}.} \bibinfo{year}{2024}\natexlab{}.
\newblock \bibinfo{title}{ML Model Production Checklist}.
\newblock
\urldef\tempurl%
\url{https://microsoft.github.io}
\showURL{%
\tempurl}
\newblock
\shownote{Accessed: 2025-09-07}.


\bibitem[{MLflow Contributors}(2024)]%
        {mlflowtracking}
\bibfield{author}{\bibinfo{person}{{MLflow Contributors}}.} \bibinfo{year}{2024}\natexlab{}.
\newblock \bibinfo{title}{MLflow Tracking}.
\newblock
\urldef\tempurl%
\url{https://mlflow.org/docs/latest/ml/tracking}
\showURL{%
\tempurl}
\newblock
\shownote{Accessed: 2025-07-01}.


\bibitem[{Mona Labs}(2025)]%
        {mona_ml_monitoring}
\bibfield{author}{\bibinfo{person}{{Mona Labs}}.} \bibinfo{year}{2025}\natexlab{}.
\newblock \bibinfo{title}{Intelligent Monitoring Solution for Machine Learning}.
\newblock
\urldef\tempurl%
\url{https://www.monalabs.io/machine-learning-ai}
\showURL{%
\tempurl}
\newblock
\shownote{Accessed: 2025-07-01}.


\bibitem[Moreno-Torres et~al\mbox{.}(2012)]%
        {moreno2012unifying}
\bibfield{author}{\bibinfo{person}{Jose~G Moreno-Torres}, \bibinfo{person}{Troy Raeder}, \bibinfo{person}{Roc{\'\i}o Alaiz-Rodr{\'\i}guez}, \bibinfo{person}{Nitesh~V Chawla}, {and} \bibinfo{person}{Francisco Herrera}.} \bibinfo{year}{2012}\natexlab{}.
\newblock \showarticletitle{A unifying view on dataset shift in classification}.
\newblock \bibinfo{journal}{\emph{Pattern recognition}} \bibinfo{volume}{45}, \bibinfo{number}{1} (\bibinfo{year}{2012}), \bibinfo{pages}{521--530}.
\newblock


\bibitem[Mueller and Massaron(2021)]%
        {mueller2021machine}
\bibfield{author}{\bibinfo{person}{John~Paul Mueller} {and} \bibinfo{person}{Luca Massaron}.} \bibinfo{year}{2021}\natexlab{}.
\newblock \bibinfo{booktitle}{\emph{Machine learning for dummies}}.
\newblock \bibinfo{publisher}{John Wiley \& Sons}.
\newblock


\bibitem[{NannyML}(2024)]%
        {nannyml}
\bibfield{author}{\bibinfo{person}{{NannyML}}.} \bibinfo{year}{2024}\natexlab{}.
\newblock \bibinfo{title}{NannyML - Monitor and Detect Silent Model Failure in Production}.
\newblock
\urldef\tempurl%
\url{https://www.nannyml.com/}
\showURL{%
\tempurl}
\newblock
\shownote{Accessed: 2025-07-01}.


\bibitem[Naveed et~al\mbox{.}(2024)]%
        {naveed2024towards}
\bibfield{author}{\bibinfo{person}{Hira Naveed}, \bibinfo{person}{John Grundy}, \bibinfo{person}{Chetan Arora}, \bibinfo{person}{Hourieh Khalajzadeh}, {and} \bibinfo{person}{Omar Haggag}.} \bibinfo{year}{2024}\natexlab{}.
\newblock \showarticletitle{Towards Runtime Monitoring for Responsible Machine Learning using Model-driven Engineering}. In \bibinfo{booktitle}{\emph{Proceedings of the ACM/IEEE 27th International Conference on Model Driven Engineering Languages and Systems}}. \bibinfo{pages}{195--202}.
\newblock


\bibitem[Nogare and Silveira(2025)]%
        {nogare2025mlops}
\bibfield{author}{\bibinfo{person}{Diego Nogare} {and} \bibinfo{person}{Ismar~Frango Silveira}.} \bibinfo{year}{2025}\natexlab{}.
\newblock \showarticletitle{MLOps for Machine Learning Model Lifecycle Automation-A Systematic Literature Review}.
\newblock \bibinfo{journal}{\emph{Authorea Preprints}} (\bibinfo{year}{2025}).
\newblock


\bibitem[{Oracle Corporation}(2025)]%
        {oracle_ml_monitoring}
\bibfield{author}{\bibinfo{person}{{Oracle Corporation}}.} \bibinfo{year}{2025}\natexlab{}.
\newblock \bibinfo{title}{Monitoring Oracle Machine Learning Models}.
\newblock
\urldef\tempurl%
\url{https://docs.oracle.com/en/database/oracle/machine-learning/oml-monitoring/}
\showURL{%
\tempurl}
\newblock
\shownote{Accessed: 2025-07-01}.


\bibitem[{Organisation for Economic Co-operation and Development (OECD)}(2019)]%
        {oecd2019ai}
\bibfield{author}{\bibinfo{person}{{Organisation for Economic Co-operation and Development (OECD)}}.} \bibinfo{year}{2019}\natexlab{}.
\newblock \bibinfo{title}{OECD Principles on Artificial Intelligence}.
\newblock
\urldef\tempurl%
\url{https://oecd.ai/en/ai-principles}
\showURL{%
\tempurl}
\newblock
\shownote{Accessed: 2025-09-05}.


\bibitem[Ovadia et~al\mbox{.}(2019)]%
        {ovadia2019can}
\bibfield{author}{\bibinfo{person}{Yaniv Ovadia}, \bibinfo{person}{Emily Fertig}, \bibinfo{person}{Jie Ren}, \bibinfo{person}{Zachary Nado}, \bibinfo{person}{David Sculley}, \bibinfo{person}{Sebastian Nowozin}, \bibinfo{person}{Joshua Dillon}, \bibinfo{person}{Balaji Lakshminarayanan}, {and} \bibinfo{person}{Jasper Snoek}.} \bibinfo{year}{2019}\natexlab{}.
\newblock \showarticletitle{Can you trust your model's uncertainty? evaluating predictive uncertainty under dataset shift}.
\newblock \bibinfo{journal}{\emph{Advances in neural information processing systems}}  \bibinfo{volume}{32} (\bibinfo{year}{2019}).
\newblock


\bibitem[Pattan and Smitha(2020)]%
        {pattan2020survey}
\bibfield{author}{\bibinfo{person}{Yogita~M Pattan} {and} \bibinfo{person}{GR Smitha}.} \bibinfo{year}{2020}\natexlab{}.
\newblock \showarticletitle{A Survey on Machine Learning Model Monitoring for Performance Assessment}.
\newblock  (\bibinfo{year}{2020}).
\newblock


\bibitem[Petersen et~al\mbox{.}(2008)]%
        {petersen2008systematic}
\bibfield{author}{\bibinfo{person}{Kai Petersen}, \bibinfo{person}{Robert Feldt}, \bibinfo{person}{Shahid Mujtaba}, {and} \bibinfo{person}{Michael Mattsson}.} \bibinfo{year}{2008}\natexlab{}.
\newblock \showarticletitle{Systematic mapping studies in software engineering}. In \bibinfo{booktitle}{\emph{12th international conference on evaluation and assessment in software engineering (EASE)}}. BCS Learning \& Development.
\newblock


\bibitem[Protschky et~al\mbox{.}(2025)]%
        {protschky2025gets}
\bibfield{author}{\bibinfo{person}{Dominik Protschky}, \bibinfo{person}{Luis L{\"a}mmermann}, \bibinfo{person}{Peter Hofmann}, {and} \bibinfo{person}{Nils Urbach}.} \bibinfo{year}{2025}\natexlab{}.
\newblock \showarticletitle{What Gets Measured Gets Improved: Monitoring Machine Learning Applications in Their Production Environments}.
\newblock \bibinfo{journal}{\emph{IEEE Access}} (\bibinfo{year}{2025}).
\newblock


\bibitem[{Qiamo (Luca) Zheng / NannyML}(2024)]%
        {nannyml2024fromreactive}
\bibfield{author}{\bibinfo{person}{{Qiamo (Luca) Zheng / NannyML}}.} \bibinfo{year}{2024}\natexlab{}.
\newblock \bibinfo{title}{From Reactive to Proactive: Shift your ML Monitoring Approach}.
\newblock
\urldef\tempurl%
\url{https://www.nannyml.com/blog/proactive-ml-monitoring-workflow/}
\showURL{%
\tempurl}
\newblock
\shownote{Accessed: 2025-09-08}.


\bibitem[Rackspace(2023)]%
        {rackspace2023report}
\bibfield{author}{\bibinfo{person}{Rackspace}.} \bibinfo{year}{2023}\natexlab{}.
\newblock \bibinfo{booktitle}{\emph{2023 AI/ML Report: AI/ML adoption surges despite challenges}}.
\newblock
\urldef\tempurl%
\url{https://www.rackspace.com/en-au/blog/2023-ai-ml-report}
\showURL{%
\tempurl}


\bibitem[Rahman et~al\mbox{.}(2021)]%
        {rahman2021run}
\bibfield{author}{\bibinfo{person}{Quazi~Marufur Rahman}, \bibinfo{person}{Peter Corke}, {and} \bibinfo{person}{Feras Dayoub}.} \bibinfo{year}{2021}\natexlab{}.
\newblock \showarticletitle{Run-time monitoring of machine learning for robotic perception: A survey of emerging trends}.
\newblock \bibinfo{journal}{\emph{IEEE Access}}  \bibinfo{volume}{9} (\bibinfo{year}{2021}), \bibinfo{pages}{20067--20075}.
\newblock


\bibitem[{Redash}(2024)]%
        {redash}
\bibfield{author}{\bibinfo{person}{{Redash}}.} \bibinfo{year}{2024}\natexlab{}.
\newblock \bibinfo{title}{Redash - Data Visualization and Dashboarding Tool}.
\newblock
\urldef\tempurl%
\url{https://redash.io/product/}
\showURL{%
\tempurl}
\newblock
\shownote{Accessed: 2025-07-01}.


\bibitem[Rigaki and Garcia(2023)]%
        {rigaki2023survey}
\bibfield{author}{\bibinfo{person}{Maria Rigaki} {and} \bibinfo{person}{Sebastian Garcia}.} \bibinfo{year}{2023}\natexlab{}.
\newblock \showarticletitle{A survey of privacy attacks in machine learning}.
\newblock \bibinfo{journal}{\emph{Comput. Surveys}} \bibinfo{volume}{56}, \bibinfo{number}{4} (\bibinfo{year}{2023}), \bibinfo{pages}{1--34}.
\newblock


\bibitem[Ross and Swetlitz(2018)]%
        {ross2018ibm}
\bibfield{author}{\bibinfo{person}{Casey Ross} {and} \bibinfo{person}{Ike Swetlitz}.} \bibinfo{year}{2018}\natexlab{}.
\newblock \showarticletitle{IBM’s Watson supercomputer recommended ‘unsafe and incorrect’cancer treatments, internal documents show}.
\newblock \bibinfo{journal}{\emph{Stat}}  \bibinfo{volume}{25} (\bibinfo{year}{2018}), \bibinfo{pages}{1--10}.
\newblock


\bibitem[Roy et~al\mbox{.}(2024)]%
        {roy2024exploring}
\bibfield{author}{\bibinfo{person}{Devjeet Roy}, \bibinfo{person}{Xuchao Zhang}, \bibinfo{person}{Rashi Bhave}, \bibinfo{person}{Chetan Bansal}, \bibinfo{person}{Pedro Las-Casas}, \bibinfo{person}{Rodrigo Fonseca}, {and} \bibinfo{person}{Saravan Rajmohan}.} \bibinfo{year}{2024}\natexlab{}.
\newblock \showarticletitle{Exploring llm-based agents for root cause analysis}. In \bibinfo{booktitle}{\emph{Companion proceedings of the 32nd ACM international conference on the foundations of software engineering}}. \bibinfo{pages}{208--219}.
\newblock


\bibitem[SaaSworthy(2024)]%
        {saasworthy2024mlstats}
\bibfield{author}{\bibinfo{person}{SaaSworthy}.} \bibinfo{year}{2024}\natexlab{}.
\newblock \bibinfo{title}{Machine Learning Statistics in 2024}.
\newblock
\urldef\tempurl%
\url{https://www.saasworthy.com/blog/machine-learning-statistics-2?utm_source=chatgpt.com}
\showURL{%
\tempurl}
\newblock
\shownote{Accessed: 2025-08-18}.


\bibitem[Salih et~al\mbox{.}(2025)]%
        {salih2025perspective}
\bibfield{author}{\bibinfo{person}{Ahmed~M Salih}, \bibinfo{person}{Zahra Raisi-Estabragh}, \bibinfo{person}{Ilaria~Boscolo Galazzo}, \bibinfo{person}{Petia Radeva}, \bibinfo{person}{Steffen~E Petersen}, \bibinfo{person}{Karim Lekadir}, {and} \bibinfo{person}{Gloria Menegaz}.} \bibinfo{year}{2025}\natexlab{}.
\newblock \showarticletitle{A perspective on explainable artificial intelligence methods: SHAP and LIME}.
\newblock \bibinfo{journal}{\emph{Advanced Intelligent Systems}} \bibinfo{volume}{7}, \bibinfo{number}{1} (\bibinfo{year}{2025}), \bibinfo{pages}{2400304}.
\newblock


\bibitem[Schr{\"o}der and Schulz(2022)]%
        {schroder2022monitoring}
\bibfield{author}{\bibinfo{person}{Tim Schr{\"o}der} {and} \bibinfo{person}{Michael Schulz}.} \bibinfo{year}{2022}\natexlab{}.
\newblock \showarticletitle{Monitoring machine learning models: a categorization of challenges and methods}.
\newblock \bibinfo{journal}{\emph{Data Science and Management}} \bibinfo{volume}{5}, \bibinfo{number}{3} (\bibinfo{year}{2022}), \bibinfo{pages}{105--116}.
\newblock


\bibitem[Shankar et~al\mbox{.}(2024)]%
        {shankar2024we}
\bibfield{author}{\bibinfo{person}{Shreya Shankar}, \bibinfo{person}{Rolando Garcia}, \bibinfo{person}{Joseph~M Hellerstein}, {and} \bibinfo{person}{Aditya~G Parameswaran}.} \bibinfo{year}{2024}\natexlab{}.
\newblock \showarticletitle{" We Have No Idea How Models will Behave in Production until Production": How Engineers Operationalize Machine Learning}.
\newblock \bibinfo{journal}{\emph{Proceedings of the ACM on Human-Computer Interaction}} \bibinfo{volume}{8}, \bibinfo{number}{CSCW1} (\bibinfo{year}{2024}), \bibinfo{pages}{1--34}.
\newblock


\bibitem[Shankar and Parameswaran(2022)]%
        {shankar2022towards}
\bibfield{author}{\bibinfo{person}{Shreya Shankar} {and} \bibinfo{person}{Aditya~G Parameswaran}.} \bibinfo{year}{2022}\natexlab{}.
\newblock \showarticletitle{Towards Observability for Production Machine Learning Pipelines}.
\newblock \bibinfo{journal}{\emph{Proceedings of the VLDB Endowment}} \bibinfo{volume}{15}, \bibinfo{number}{13} (\bibinfo{year}{2022}), \bibinfo{pages}{4015--4022}.
\newblock


\bibitem[Shergadwala et~al\mbox{.}(2022)]%
        {shergadwala2022human}
\bibfield{author}{\bibinfo{person}{Murtuza~N Shergadwala}, \bibinfo{person}{Himabindu Lakkaraju}, {and} \bibinfo{person}{Krishnaram Kenthapadi}.} \bibinfo{year}{2022}\natexlab{}.
\newblock \showarticletitle{A human-centric perspective on model monitoring}. In \bibinfo{booktitle}{\emph{Proceedings of the AAAI Conference on Human Computation and Crowdsourcing}}, Vol.~\bibinfo{volume}{10}. \bibinfo{pages}{173--183}.
\newblock


\bibitem[Shivashankar et~al\mbox{.}(2025)]%
        {shivashankar2025scalability}
\bibfield{author}{\bibinfo{person}{Karthik Shivashankar}, \bibinfo{person}{Ghadi S~Al Hajj}, {and} \bibinfo{person}{Antonio Martini}.} \bibinfo{year}{2025}\natexlab{}.
\newblock \showarticletitle{Scalability and Maintainability Challenges and Solutions in Machine Learning: Systematic Literature Review}.
\newblock \bibinfo{journal}{\emph{arXiv preprint arXiv:2504.11079}} (\bibinfo{year}{2025}).
\newblock


\bibitem[Singer et~al\mbox{.}(2022)]%
        {singer2022enhancing}
\bibfield{author}{\bibinfo{person}{Sara~J Singer}, \bibinfo{person}{Katherine~C Kellogg}, \bibinfo{person}{Ari~B Galper}, {and} \bibinfo{person}{Deborah Viola}.} \bibinfo{year}{2022}\natexlab{}.
\newblock \showarticletitle{Enhancing the value to users of machine learning-based clinical decision support tools: A framework for iterative, collaborative development and implementation}.
\newblock \bibinfo{journal}{\emph{Health Care Management Review}} \bibinfo{volume}{47}, \bibinfo{number}{2} (\bibinfo{year}{2022}), \bibinfo{pages}{E21--E31}.
\newblock


\bibitem[Song et~al\mbox{.}(2023)]%
        {song2023deeplens}
\bibfield{author}{\bibinfo{person}{Da Song}, \bibinfo{person}{Zhijie Wang}, \bibinfo{person}{Yuheng Huang}, \bibinfo{person}{Lei Ma}, {and} \bibinfo{person}{Tianyi Zhang}.} \bibinfo{year}{2023}\natexlab{}.
\newblock \showarticletitle{DeepLens: interactive out-of-distribution data detection in NLP models}. In \bibinfo{booktitle}{\emph{Proceedings of the 2023 CHI Conference on Human Factors in Computing Systems}}. \bibinfo{pages}{1--17}.
\newblock


\bibitem[Strickland(2019)]%
        {strickland2019ibm}
\bibfield{author}{\bibinfo{person}{Eliza Strickland}.} \bibinfo{year}{2019}\natexlab{}.
\newblock \showarticletitle{IBM Watson, heal thyself: How IBM overpromised and underdelivered on AI health care}.
\newblock \bibinfo{journal}{\emph{IEEE spectrum}} \bibinfo{volume}{56}, \bibinfo{number}{4} (\bibinfo{year}{2019}), \bibinfo{pages}{24--31}.
\newblock


\bibitem[{Superwise}(2025)]%
        {superwise_ml_monitoring}
\bibfield{author}{\bibinfo{person}{{Superwise}}.} \bibinfo{year}{2025}\natexlab{}.
\newblock \bibinfo{title}{ML Monitoring | Superwise ML Observability}.
\newblock
\urldef\tempurl%
\url{https://platform.superwise.ai/ml-monitoring/}
\showURL{%
\tempurl}
\newblock
\shownote{Accessed: 2025-07-01}.


\bibitem[{TensorFlow Authors}(2024)]%
        {tensorboard}
\bibfield{author}{\bibinfo{person}{{TensorFlow Authors}}.} \bibinfo{year}{2024}\natexlab{}.
\newblock \bibinfo{title}{TensorBoard: Visualization Toolkit for Machine Learning Experimentation}.
\newblock
\urldef\tempurl%
\url{https://www.tensorflow.org/tensorboard}
\showURL{%
\tempurl}
\newblock
\shownote{Accessed: 2025-07-01}.


\bibitem[{The Fluent Bit Authors}(2024)]%
        {fluentbit}
\bibfield{author}{\bibinfo{person}{{The Fluent Bit Authors}}.} \bibinfo{year}{2024}\natexlab{}.
\newblock \bibinfo{title}{Fluent Bit - Fast and Lightweight Log Processor and Forwarder}.
\newblock
\urldef\tempurl%
\url{https://fluentbit.io/}
\showURL{%
\tempurl}
\newblock
\shownote{Accessed: 2025-07-01}.


\bibitem[Ulus(2024)]%
        {doubt}
\bibfield{author}{\bibinfo{person}{Dogan Ulus}.} \bibinfo{year}{2024}\natexlab{}.
\newblock \bibinfo{title}{doubt: A Python package for non-parametric uncertainty estimation}.
\newblock
\urldef\tempurl%
\url{https://pypi.org/project/doubt/}
\showURL{%
\tempurl}
\newblock
\shownote{Accessed: 2025-07-01}.


\bibitem[{Uncertainty Wizard Contributors}(2024)]%
        {uncertaintywizard}
\bibfield{author}{\bibinfo{person}{{Uncertainty Wizard Contributors}}.} \bibinfo{year}{2024}\natexlab{}.
\newblock \bibinfo{title}{Uncertainty Wizard: Uncertainty-Aware Machine Learning with Keras}.
\newblock
\urldef\tempurl%
\url{https://uncertainty-wizard.readthedocs.io/en/latest/}
\showURL{%
\tempurl}
\newblock
\shownote{Accessed: 2025-07-01}.


\bibitem[{Valohai}(2025)]%
        {valohai_model_monitoring}
\bibfield{author}{\bibinfo{person}{{Valohai}}.} \bibinfo{year}{2025}\natexlab{}.
\newblock \bibinfo{title}{What Is Model Monitoring?}
\newblock
\urldef\tempurl%
\url{https://valohai.com/model-monitoring/}
\showURL{%
\tempurl}
\newblock
\shownote{Accessed: 2025-07-01}.


\bibitem[Van~Engelen and Hoos(2020)]%
        {van2020survey}
\bibfield{author}{\bibinfo{person}{Jesper~E Van~Engelen} {and} \bibinfo{person}{Holger~H Hoos}.} \bibinfo{year}{2020}\natexlab{}.
\newblock \showarticletitle{A survey on semi-supervised learning}.
\newblock \bibinfo{journal}{\emph{Machine learning}} \bibinfo{volume}{109}, \bibinfo{number}{2} (\bibinfo{year}{2020}), \bibinfo{pages}{373--440}.
\newblock


\bibitem[{Vertier}(2025)]%
        {vetiver2025monitoring}
\bibfield{author}{\bibinfo{person}{{Vertier}}.} \bibinfo{year}{2025}\natexlab{}.
\newblock \bibinfo{title}{MLOps with vetiver - Monitor}.
\newblock
\urldef\tempurl%
\url{https://vetiver.rstudio.com/get-started/monitor.html}
\showURL{%
\tempurl}
\newblock
\shownote{Accessed: 2025-04-28}.


\bibitem[Wohlin(2014)]%
        {wohlin2014guidelines}
\bibfield{author}{\bibinfo{person}{Claes Wohlin}.} \bibinfo{year}{2014}\natexlab{}.
\newblock \showarticletitle{Guidelines for snowballing in systematic literature studies and a replication in software engineering}. In \bibinfo{booktitle}{\emph{Proceedings of the 18th international conference on evaluation and assessment in software engineering}}. \bibinfo{pages}{1--10}.
\newblock


\bibitem[Wolf et~al\mbox{.}(2017)]%
        {wolf2017we}
\bibfield{author}{\bibinfo{person}{Marty~J Wolf}, \bibinfo{person}{Keith Miller}, {and} \bibinfo{person}{Frances~S Grodzinsky}.} \bibinfo{year}{2017}\natexlab{}.
\newblock \showarticletitle{Why we should have seen that coming: comments on Microsoft's tay" experiment," and wider implications}.
\newblock \bibinfo{journal}{\emph{Acm Sigcas Computers and Society}} \bibinfo{volume}{47}, \bibinfo{number}{3} (\bibinfo{year}{2017}), \bibinfo{pages}{54--64}.
\newblock


\bibitem[Wong et~al\mbox{.}(2023)]%
        {wong2023mlguard}
\bibfield{author}{\bibinfo{person}{Sheng Wong}, \bibinfo{person}{Scott Barnett}, \bibinfo{person}{Jessica Rivera-Villicana}, \bibinfo{person}{Anj Simmons}, \bibinfo{person}{Hala Abdelkader}, \bibinfo{person}{Jean-Guy Schneider}, {and} \bibinfo{person}{Rajesh Vasa}.} \bibinfo{year}{2023}\natexlab{}.
\newblock \showarticletitle{Mlguard: Defend your machine learning model!}. In \bibinfo{booktitle}{\emph{Proceedings of the 1st International Workshop on Dependability and Trustworthiness of Safety-Critical Systems with Machine Learned Components}}. \bibinfo{pages}{10--13}.
\newblock


\bibitem[Xia et~al\mbox{.}(2024)]%
        {xia2024towards}
\bibfield{author}{\bibinfo{person}{Boming Xia}, \bibinfo{person}{Qinghua Lu}, \bibinfo{person}{Liming Zhu}, \bibinfo{person}{Sung~Une Lee}, \bibinfo{person}{Yue Liu}, {and} \bibinfo{person}{Zhenchang Xing}.} \bibinfo{year}{2024}\natexlab{}.
\newblock \showarticletitle{Towards a responsible ai metrics catalogue: A collection of metrics for ai accountability}. In \bibinfo{booktitle}{\emph{Proceedings of the IEEE/ACM 3rd International Conference on AI Engineering-Software Engineering for AI}}. \bibinfo{pages}{100--111}.
\newblock


\bibitem[Xu et~al\mbox{.}(2022)]%
        {xu2022dependency}
\bibfield{author}{\bibinfo{person}{Xiwei Xu}, \bibinfo{person}{Chen Wang}, \bibinfo{person}{Zhen Wang}, \bibinfo{person}{Qinghua Lu}, {and} \bibinfo{person}{Liming Zhu}.} \bibinfo{year}{2022}\natexlab{}.
\newblock \showarticletitle{Dependency tracking for risk mitigation in machine learning (ML) systems}. In \bibinfo{booktitle}{\emph{Proceedings of the 44th International Conference on Software Engineering: Software Engineering in Practice}}. \bibinfo{pages}{145--146}.
\newblock


\bibitem[Xu et~al\mbox{.}(2023)]%
        {xu2023alertiger}
\bibfield{author}{\bibinfo{person}{Zhentao Xu}, \bibinfo{person}{Ruoying Wang}, \bibinfo{person}{Girish Balaji}, \bibinfo{person}{Manas Bundele}, \bibinfo{person}{Xiaofei Liu}, \bibinfo{person}{Leo Liu}, {and} \bibinfo{person}{Tie Wang}.} \bibinfo{year}{2023}\natexlab{}.
\newblock \showarticletitle{Alertiger: Deep learning for ai model health monitoring at linkedin}. In \bibinfo{booktitle}{\emph{Proceedings of the 29th ACM SIGKDD Conference on Knowledge Discovery and Data Mining}}. \bibinfo{pages}{5350--5359}.
\newblock
\urldef\tempurl%
\url{https://github.com/linkedin/AlerTiger}
\showURL{%
\tempurl}


\bibitem[Yao et~al\mbox{.}(2017)]%
        {yao2017complexity}
\bibfield{author}{\bibinfo{person}{Yuanshun Yao}, \bibinfo{person}{Zhujun Xiao}, \bibinfo{person}{Bolun Wang}, \bibinfo{person}{Bimal Viswanath}, \bibinfo{person}{Haitao Zheng}, {and} \bibinfo{person}{Ben~Y Zhao}.} \bibinfo{year}{2017}\natexlab{}.
\newblock \showarticletitle{Complexity vs. performance: empirical analysis of machine learning as a service}. In \bibinfo{booktitle}{\emph{Proceedings of the 2017 Internet Measurement Conference}}. \bibinfo{pages}{384--397}.
\newblock


\bibitem[{Yue Zhao and Zain Nasrullah and Zheng Li}(2024)]%
        {pyod}
\bibfield{author}{\bibinfo{person}{{Yue Zhao and Zain Nasrullah and Zheng Li}}.} \bibinfo{year}{2024}\natexlab{}.
\newblock \bibinfo{title}{PyOD: A Python Toolbox for Scalable Outlier Detection}.
\newblock
\urldef\tempurl%
\url{https://pyod.readthedocs.io/en/latest/}
\showURL{%
\tempurl}
\newblock
\shownote{Accessed: 2025-07-01}.


\bibitem[Zamani et~al\mbox{.}(2021)]%
        {zamani2021machine}
\bibfield{author}{\bibinfo{person}{Kareshna Zamani}, \bibinfo{person}{Didar Zowghi}, {and} \bibinfo{person}{Chetan Arora}.} \bibinfo{year}{2021}\natexlab{}.
\newblock \showarticletitle{Machine learning in requirements engineering: A mapping study}. In \bibinfo{booktitle}{\emph{2021 IEEE 29th International Requirements Engineering Conference Workshops (REW)}}. IEEE, \bibinfo{pages}{116--125}.
\newblock


\bibitem[Zarour et~al\mbox{.}(2025)]%
        {zarour2025mlops}
\bibfield{author}{\bibinfo{person}{Mohammad Zarour}, \bibinfo{person}{Hamza Alzabut}, {and} \bibinfo{person}{Khalid~T Al-Sarayreh}.} \bibinfo{year}{2025}\natexlab{}.
\newblock \showarticletitle{MLOps best practices, challenges and maturity models: A systematic literature review}.
\newblock \bibinfo{journal}{\emph{Information and Software Technology}}  \bibinfo{volume}{183} (\bibinfo{year}{2025}), \bibinfo{pages}{107733}.
\newblock


\bibitem[Zeileis et~al\mbox{.}(2024)]%
        {strucchange}
\bibfield{author}{\bibinfo{person}{Achim Zeileis}, \bibinfo{person}{Friedrich Leisch}, {and} \bibinfo{person}{Kurt Hornik}.} \bibinfo{year}{2024}\natexlab{}.
\newblock \bibinfo{booktitle}{\emph{strucchange: Testing for Structural Change in Linear Regression Models}}.
\newblock Comprehensive R Archive Network (CRAN), https://CRAN.R-project.org.
\newblock
\urldef\tempurl%
\url{https://cran.r-project.org/web/packages/strucchange/index.html}
\showURL{%
\tempurl}
\newblock
\shownote{R package version 1.5-3}.


\bibitem[Zhang and Tsai(2003)]%
        {zhang2003machine}
\bibfield{author}{\bibinfo{person}{Du Zhang} {and} \bibinfo{person}{Jeffrey~JP Tsai}.} \bibinfo{year}{2003}\natexlab{}.
\newblock \showarticletitle{Machine learning and software engineering}.
\newblock \bibinfo{journal}{\emph{Software Quality Journal}}  \bibinfo{volume}{11} (\bibinfo{year}{2003}), \bibinfo{pages}{87--119}.
\newblock


\bibitem[Zimelewicz et~al\mbox{.}(2024)]%
        {zimelewicz2024ml}
\bibfield{author}{\bibinfo{person}{Eduardo Zimelewicz}, \bibinfo{person}{Marcos Kalinowski}, \bibinfo{person}{Daniel Mendez}, \bibinfo{person}{G{\"o}rkem Giray}, \bibinfo{person}{Antonio~Pedro Santos~Alves}, \bibinfo{person}{Niklas Lavesson}, \bibinfo{person}{Kelly Azevedo}, \bibinfo{person}{Hugo Villamizar}, \bibinfo{person}{Tatiana Escovedo}, \bibinfo{person}{Helio Lopes}, {et~al\mbox{.}}} \bibinfo{year}{2024}\natexlab{}.
\newblock \showarticletitle{Ml-enabled systems model deployment and monitoring: Status quo and problems}. In \bibinfo{booktitle}{\emph{International Conference on Software Quality}}. Springer, \bibinfo{pages}{112--131}.
\newblock


\end{thebibliography}

\appendix

\end{document}